\documentclass[11pt]{article}
\usepackage{amsfonts}
\usepackage{amssymb}
\usepackage{mathrsfs}
\usepackage{graphicx}
\usepackage{amsmath,amsthm,amssymb,amscd}
\usepackage[all]{xy}
\usepackage{subfig}
\usepackage{hyperref}
\usepackage{multirow}
\usepackage{color}
\usepackage{amsfonts}
\usepackage{float}
\usepackage{mathtools,slashed}
\usepackage[utf8]{inputenc}
\usepackage[T1]{fontenc}

\usepackage{dsfont}
\usepackage{slashed}
\usepackage{empheq}
\usepackage{multirow}
\usepackage{bbm}
\usepackage{fancyhdr}
\usepackage{tikz}
\usepackage[utf8]{inputenc}
\usepackage[T1]{fontenc}

\def\D{\prod\limits_{i=1}^n\langle i ,i+1\rangle}

\newcommand{\cd}{{\cal D}}

\newcommand{\cn}{{\cal N}}

\newcommand{\cl}{{\cal L}}

\newcommand{\co}{{\cal O}}

\newcommand{\cq}{{\cal Q}}

\def\eqa{\begin{eqnarray}}
\def\eqae{\end{eqnarray}}
\def\eq{\begin{equation}}
\def\eqe{\end{equation}}
\def\be{\begin{equation}}
\def\ee{\end{equation}}
\def\bea{\begin{eqnarray}}
\def\eea{\end{eqnarray}}
\def\ba{\begin{array}}
\def\ea{\end{array}}
\def\bd{\begin{displaymath}}
\def\ed{\end{displaymath}}

\def\tr{{\rm tr}}
\def\>{\rangle}
\def\<{\langle}
\def\a{\alpha}
\def\b{\beta}

\def\del{\delta}
\def\e{\epsilon}
\def\f{\phi}

\def\j{\psi}

\def\l{\lambda}
\def\m{\mu}
\def\n{\nu}

\def\q{\theta}
\def\r{\rho}
\def\s{\sigma}
\def\t{\tau}

\def\z{\zeta}
\def\D{\Delta}
\def\F{\Phi}
\def\G{\Gamma}

\def\L{\Lambda}

\def\X{\Xi}

\def\pa{\partial}

\newcommand{\red}{\textcolor[rgb]{0.9,0.00,0.00}}

\numberwithin{equation}{section}

\begin{document}

\begin{titlepage}
\hfill MCTP-17-01

\begin{center}

\vskip .7 cm

{\large \bf  Microstate Counting of $AdS_4$ Hyperbolic Black Hole Entropy}

\vskip .6 cm

{\large \bf   via the Topologically Twisted Index}

\end{center}
\vskip 1 cm
\begin{center}
{\bf \small Alejandro Cabo-Bizet${}^a$, Victor I. Giraldo-Rivera${}^{b}$ and Leopoldo A. Pando Zayas${}^c$ }
\end{center}

\vskip .4cm 
\centerline{\it ${}^a$ Instituto de Astronom\'{i}a y F\'{i}sica  del Espacio (CONICET-UBA)}
\centerline{\it  Ciudad Universitaria, C.P. 1428
Buenos Aires, Argentina}

\vskip .4cm 
\centerline{\it ${}^b$ International Centre for Theoretical Sciences (ICTS-TIFR)}
\centerline{ \it Shivakote, Hesaraghatta Hobli,
Bengaluru 560089, India}

\vskip .4cm\centerline{\it  ${}^b$The Abdus Salam International Centre for Theoretical Physics}
\centerline{\it Strada Costiera 11, 34014 Trieste, Italy}

\vskip .4cm \centerline{\it ${}^c$ Michigan Center for Theoretical
Physics}
\centerline{ \it Randall Laboratory of Physics, The University of
Michigan}
\centerline{\it Ann Arbor, MI 48109-1120}

\vskip 1.5 cm
\begin{abstract}

\end{abstract}
We compute the topologically twisted index for general ${\cal N}=2$ supersymmetric field theories on $\mathbb{H}_2\times S^1$. 
We also discuss asymptotically $AdS_4$ magnetically charged  black holes with hyperbolic horizon, in four-dimensional ${\cal N}=2$ gauged supergravity. With certain assumptions, put forward by Benini, Hristov and Zaffaroni, we find precise agreement between the black hole entropy and the topologically twisted index, for $ABJM$ theories.

\end{titlepage}

\tableofcontents

\section{Introduction}

Black holes have an entropy that fits neatly in a thermodynamics framework as originally established in the works of Bekenstein and Hawking in the early 1970's. The microscopic origin, that is, the nature of the degrees of freedom that this entropy counts, has been an outstanding challenge for many decades. Any candidate to a theory of quantum gravity must provide an answer to this fundamental question. String theory, in the works of Strominger and Vafa,  has successfully passed this test for a particular type of black holes \cite{Strominger:1996sh}.  In the context of the $AdS/CFT$ correspondence, the original work of Strominger and Vafa can be interpreted as an instance of  $AdS_3/CFT_2$.  A natural question pertains higher dimensional versions of the $AdS/CFT$ correspondence.  Recent work by Benini, Hristov and Zaffaroni addresses  the microscopic counting of the entropy of certain black holes from the point of view of $AdS_4/CFT_3$ \cite{Benini:2015eyy}.

In this manuscript we explore the topologically twisted index, originally introduced by Benini and Zaffaroni in the framework of ${\cal N}=2$ supersymmetric  three-dimensional field theories in $S^2\times S^1$ \cite{Benini:2015noa}  (see also  \cite{Hosseini:2016tor,Hosseini:2016ume,Closset:2016arn, Alejandro}),  for the case of supersymmetric theories in $\mathbb{H}_2\times S^1$, where $\mathbb{H}_2$ is the hyperbolic plane.   Although we provide the ingredients for arbitrary ${\cal N}=2$ supersymmetric theories, we will particularize our results for a specific deformation of $ABJM$ theory. The holographic dual of such deformation is thought to be a hyperbolic black hole. In this work, our main motivation comes from the prospect of understanding the $D=3$ SCFT representation of the appropriate $AdS_4$ black hole microstates. With this aim we are driven to explore four dimensional ${\cal N}=2$ gauged supergravity and find black hole solutions with $\mathbb{H}_2$ horizon. Hyperbolic black holes have been discussed in the context of $AdS/CFT$ in, for example, \cite{Emparan}. 

Asymptotically $AdS_4$ black holes in $\mathcal{N}=2$ gauged supergravity, which are sourced by magnetic fluxes, have been widely studied  \cite{Cacciatori, Cacciatori0, Vandoren, Halmagyi, Halmagyi2}.  Roughly speaking, from the bulk perspective, the presence of fluxes allows to define the black hole as interpolating  from the UV $AdS_4$ to the near horizon $AdS_2\times S^2$. As a result of our study we are able to identify the role of such fluxes from the dual SCFT perspective. These flavor fluxes, together with a continuous of color fluxes, generate a one-parameter hierarchy of Landau levels on $\mathbb{H}_2$, that determines the value of the $ABJM$ index. What we are set to explore in this paper, is whether the leading behavior in the large $N$ limit of the topologically twisted index of a specific deformation of $ABJM$, evaluated on the Hilbert space composed by the aforementioned Landau levels, coincides with the Bekenstein-Hawking expression for the semiclassical entropy of the black holes in question.  We will find that indeed both results coincide.

Another important motivation for our work, is the intrinsically interesting field theory problem of localization of supersymmetric field theories in non-compact spaces. This problem naturally appears in the context of localization of supergravity theories, for an understanding of exact black hole entropy counting \cite{Sen:2008vm,Dabholkar:2010uh}. The same problem appears in holographic approaches  to Wilson loops where the world volume of the classical configuration contains an $AdS_2$ factor. For example, the excitations on a D3 brane which is dual to a Wilson loop in the totally  symmetric rank $k$  representation \cite{Faraggi:2011bb}  were identified to correspond, to an ${\cal N}=4$ vector multiplet in ${\mathbb H}_2 \times S^2$ \cite{Buchbinder:2014nia}.  Localization in non-compact spaces has recently been addressed in \cite{David:2016onq} and  \cite{Assel:2016pgi},  our work constitutes an extension to the topologically twisted case. 

The manuscript is organized as follows. In section \ref{sec:susy3d} we discuss the preliminary ingredients we need, for example, our guidance principle on the field theory side: supersymmetric localization \cite{Pestun}, the background metric, spin connection, and supersymmetric structure of the actions needed to compute $BPS$ observables in a generic three-dimensional $\mathcal{N}=2$ Chern-Simons-Matter theory on $\mathbb{H}_2\times S^1$. To complete section  \ref{sec:susy3d}, we discuss the boundary conditions to be used in the manuscript. In section \ref{sec:3} we present the space of square and delta-normalizable functions  that will be used to integrate upon, and their respective discrete and continuous spectrum. In section \ref{sec:4} we compute the one loop super-determinants. In section \ref{sec:5} we assemble our results to write down the $ABJM$ index on $\mathbb{H}_2\times S^1$, and then move on to compute its leading contribution in the large $N$ expansion, by following the procedure pioneered in \cite{Benini:2015eyy}. In section \ref{BHsec} we find what we believe to be the dual $AdS_4$ black holes and compare its Bekenstein-Hawking entropy to the leading contribution in the large $N$ expansion of the $ABJM$ index on $\mathbb{H}_2\times S^1$. In section \ref{sec:7} we conclude with a short summary of our results and comment on interesting open and related problems.  In a series of appendices we discuss more technical aspects such as, for instance, the construction of square integrable modes in appendix \ref{SQN}. 


\section{Towards the index on $\mathbb{H}_2\times S^1$ } 

\label{sec:susy3d}


In this section we summarize the building blocks that will be needed in order to compute the topologically twisted index of a generic $\mathcal{N}=2$ Chern-Simons-Matter theory on $\mathbb{H}_2\times S^1$. The zero locus will be parametrized by a continuous of color fluxes and holonomies. On $\mathbb{H}_2$, these flux $BPS$ configurations are non-normalizable but they are part of the zero locus: The localizing term $Q_\epsilon V$, which is constructed to be semi-positive definite, will vanish at them. 

First, we review the SUSY localization method to compute the partition function of 3d Chern-Simons-Matter defined over a Euclidean space $\mathcal{M}$ with off-shell supersymmetry charge $\mathcal{Q}$.   The space $\mathcal{M}$ is usually taken to be compact. The localization principle is well known  and has been elegantly summarized and interpreted in various reviews, for example,   \cite{Hosomichi:2014hja,Cremonesi:2014dva}. However, given some of the intricacies we face, for the case we discuss, we review it here, with the goal of setting up our guiding principle, notation and to highlight some of the points on which we will make a particular emphasis. 

 To close this section we elaborate on the specific set of boundary conditions that we shall use for background and fluctuations.





\subsection{SUSY localization principle}

The SUSY localization method is summarized in the following steps
\begin{itemize}
\item Select a ``middle dimensional" section ${\Gamma}$ in the space of complex fields, such as a 3d vector multiplet $\{A_\mu,\sigma,D,$ $\lambda,\bar{\lambda}\}$ of your theory. The path integral defining the SUSY partition function of a classical action $S_{cl}$, ${ Z}[{\Gamma}]$ is to be performed over ${\Gamma}$. The path $\Gamma$ must be a consistent path of integration of $S_{cl}$.
 \item
The contour $\Gamma$ intersects a set of $Q_\epsilon$-$BPS$ configurations that will be denoted as $BPS[\Gamma]$ and that is better known as: The localization locus.
\item
For each $\Gamma$ there should exist a $Q_\epsilon V$ local functional of fields whose bosonic part is semi-positive definite at $\Gamma$ and vanishes at $BPS[\Gamma]$. 
\item
Given the previous conditions, the strict limit $\tau \rightarrow \infty$ can be taken in such a way that the final result for the partition function is guaranteed to be
\begin{eqnarray}
{ Z}[{ \Gamma}]&=&\sum_{X^{(0)} \in BPS[\Gamma]} e^{-S_{cl}[X^{(0)}]}Z_{X^{(0)}}[\Gamma],\nonumber \\
 Z_{X^{(0)}}[\Gamma]&:=&\int_{\Gamma} e^{- \delta^{(2)}\left( Q_\epsilon V, ~X^{(0)} \right)  }, \label{PF}
\end{eqnarray}
\end{itemize}
where $ \delta^{(2)}\left( Q_\epsilon V,X^{(0)} \right)$ is the quadratic expansion of $Q_\epsilon V$ about $X^{(0)}$.
We have omitted the integration over the space $\mathcal{M}$ to ease the reading, but remember it is there.
Let us review the semiclassical reduction \eqref{PF}. 

 The starting point, is to notice that the partition function $Z[\Gamma]$ does not change if the initial classical action $S_{cl}$ is deformed by an arbitrary $Q_\epsilon$-exact deformation $\tau Q_\epsilon V$
\bea
\partial_\tau \bigg(Z[\Gamma]&:=&\int_{\Gamma} e^{-S_{cl}[X]-{\tau} Q_\epsilon V }\bigg)
\nonumber\\
&=&\int_{\Gamma}Q_\epsilon \left( V e^{-S_{cl}[X]-{ \tau Q_\epsilon V} }\right)=0,
\eea
provided the measure of integration in field configuration space is $Q_\epsilon$ invariant and that there are not contributions from the boundary of the latter.

Under the aforementioned conditions, we can choose a deformation term $Q_\epsilon V$ with semi-positive definite bosonic part and thereafter perform a field redefinition
\begin{eqnarray}
X\rightarrow X^{(0)}+\frac{1}{{\sqrt{\tau}}} X^{(1)}. 
\end{eqnarray}
As $Z[\Gamma]$ is independent of $\tau$ we are free to take the limit $\tau\rightarrow \infty$ and proceed as follows
\begin{eqnarray}
\int_{X} e^{-S_{cl}[X^{(0)}]- \tau QV}&=&{e^{-S_{cl}[X^{(0)}]}}\int_{X^{(1)}}  e^{-\tau QV}\nonumber \\ &\rightarrow& {e^{-S_{cl}[X^{(0)}]-\tau QV[X^{(0)}]} } \int_{X^{(1)}} e^{- \delta^{(2)}\left( Q_\epsilon V ,X^{(0)}\right)  }. \label{proc}
\end{eqnarray}

Because of the suppression factor $e^{- \tau Q V[X^{(0)}]}$ and semi-positive definiteness of the bosonic part of $Q_\epsilon V$ only classical configurations $X^{(0)}\in\Gamma$, namely $X^{(0)}\in BPS[\Gamma]$, solutions of the zero locus of $Q_\epsilon V$ ($Q_\epsilon V_{bos}=0$) contribute in this limit and \eqref{PF} is recovered.

\subsection{Background geometry and supersymmetry}

In this subsection we introduce the basic elements needed for the evaluation of the localization formula for the topologically twisted index of a generic $\mathcal{N}=2$ Chern-Simons-Matter theory. Specifically, we are interested in $U(N)_k\times U(N)_{-k}$ Chern-Simons theory coupled to matter in the bi-fundamental representation: $ABJM$ \cite{Aharony:2008ug},  living in the non-compact space $\mathcal{M}=\mathbb{H}_2\times S^1$ whose metric we will represent as
\bea
ds^{2}&=& -dt^{2}+ds_{2d}^{2},  \nonumber \\ 
 ds_{2d}^{2}&:=&- d\theta^{2}-\sinh^{2}\left( \theta\right)  d\varphi^{2},  \label{metriceta}
 \\  &&~ \varphi \sim \varphi+2\pi, ~~~ t \sim t+1.\label{metric}
 \eea
We shall use in this paper the following signature $(-,-,-)$ on the 3d boundary theory. The flat space metric is $\eta=diag(-1,-1,-1)$. 
 
In the conventions used in this section, the non trivial spin connection component is
\begin{eqnarray}
\omega^{21}_\varphi= -\cosh{\theta}.
\end{eqnarray}
 The 2d space $\mathbb{H}_2$ has infinite volume. When dealing with extensive quantities on $\mathbb{H}_2$ we will use a cut-off at large $\theta$ and drop out the dependence on such cut-off in the very end. More precisely, this recipe has been used in the context of black hole entropy in  \cite{Sen:2008vm,Banerjee:2011jp} and, in the context of holographic computations for Wilson loops it was discussed in  \cite{Faraggi:2011ge}; it amounts to  defining the volume of $\mathbb{H}_2$ as:
 \be
 \label{eq:Vol}
 vol_{\mathbb{H}_2}=-2\pi.
 \ee
 As general principle, we will consider background configurations that grow asymptotically as the volume element, or slower. As for extensive quantities constructed out of such non normalizable backgrounds, we shall apply the previous  regularization recipe  \footnote{ 
\label{boundSen} For example, to work with boundary objects - like the boundary action \eqref{sbdry}- with finite limit in the cut-off $\theta_0 \rightarrow \infty$, we follow \cite{Sen:2008yk}. The idea is to use coordinates $\left(\tilde{\q}:=\q_{0}-\q, ~\tilde\varphi=\frac{1}{2}e^{\q_{0}}\varphi\right)$, in such a way the metric
\bea
d\q^2+\sinh\q^2 d\varphi^2\nonumber,
\eea
transforms to
\bea
d\tilde{\q}^2+(e^{-\tilde{\q}}-e^{-2\q_{0}+\tilde{\q}})^2 d\tilde{\varphi}^2\nonumber,
\eea
where $\tilde{\varphi}$ is a periodic coordinate with period $\tilde{\b}=\pi e^{\q_0} $ and $0<\tilde{\q}<\q_0$.
}.

 The results of these sections allow to compute the topologically twisted index of any $\mathcal{N}=2$ Chern-Simons theory coupled to matter.  As mentioned before we are interested in the particular case of $ABJM$. The $ABJM$ theory is composed by two vector multiplets and four matter multiplets in the bi-fundamental of the gauge group. Specifically
 \begin{eqnarray}
Chern-Simons ~ \pm k&:& \left\{A_\mu, \sigma, D, \lambda_{q=1}, \bar{\lambda}_{q=1}\right\}_{\pm k},   \nonumber\\
matter&:&\left\{ \phi_q^a,\bar{\phi}_{q}^a, \psi^a_{q-1}, \bar{\psi}^a_{q-1}, F^a, \bar{F}^a\right\}, ~~~ a=1,2,3,4.  \nonumber
\end{eqnarray}
where \footnote{The supersymmetry transformation rules are defined over the complex conjugated of $(\bar{\f},\bar{\j},\bar{F})$, which are denoted as $(\bar{\f}^\dagger,\bar{\j}^{\dagger},\bar{F}^\dagger)$.}  $q$ is the charge of the corresponding field under the $R-$symmetry flux \eqref{RSymmetryb}.

We can represent $ABJM$ theories by the following standard quiver diagram:
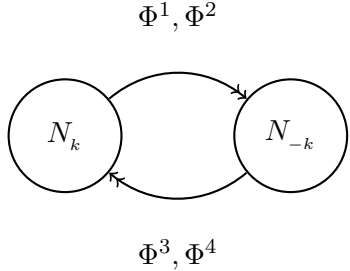
\begin{center}
\begin{tikzpicture}[every node/.style={circle,draw},thick,minimum size=1.5cm]
  \node(NL) at (0,0){$N_{_{k}}$};
  \node(NR) at (3,0){$N_{_{-k}}$};
  \draw[->>](NL) to [bend left=40] node[draw=none,minimum size=0.1cm,above]{$\F^1,\F^2$}(NR);
    \draw[->>](NR) to[bend left=40]node[draw=none,minimum size=0.1cm,below]{$\F^3,\F^4$}(NL);
\end{tikzpicture}
\end{center}
\vspace{-0.1cm}

 $ABJM$ theories have $\cn= 8$ superconformal symmetry for level $k=1,2$ and $\cn=6$ for  level $k\geq3$, the global symmetry that is manifest in the $\cn=2$ notation is $SU(2)_{1,2}\times SU(2)_{3,4}\times U(1)_T\times U(1)_R$, where each $SU(2)$ acts upon the doublets composed by the corresponding labels  \cite{Aharony:2008ug}.

  We are interested in a specific deformation of $ABJM$. Part of such deformation is a classical background for the $R$-symmetry potential
\be
V_\mu dx^\mu=\frac{1}{2} \cosh{\theta} d\varphi. \label{RSymmetryb}
\ee
The background \eqref{RSymmetryb} is non normalizable. However,  $V$ goes like the volume element of $\mathbb{H}_2$ for large $\theta$. The deformation \eqref{RSymmetryb} has non trivial consequences in the final result of the localization technic.

 The $R-$symmetry background allows for the presence of a Killing spinor $\epsilon$ with R-charge $q=1$. The Killing spinor equation (KSE) being
 \bea
  \left(\partial_\mu +\frac{1}{4} \omega^{a b}_{~~\mu} \sigma_{a b}- i V_\mu\right) \epsilon&=&0, \label{KSE}
 \eea
 with $\sigma_{a b}:=\frac{[\sigma_a, \sigma_b]}{2}$ and $\sigma_a$, $\sigma_b$ being  Pauli matrices.  As we are using negative signature it is important to keep in mind that
 \be
 \sigma^a=-\sigma_a. \label{signPa}
 \ee
 In fact, the algebrae and actions that will be defined later on, are obtained out of the results in \cite{Alejandro} by the appropriate change of signature, and \eqref{signPa}.
 
 The most general normalized solution  to the KSE  \eqref{KSE}, is proportional to
 \begin{eqnarray}
\epsilon= \left(\begin{array}{c}
 1\\0
 \end{array}\right).
 \end{eqnarray}
 Out of $\epsilon$ we can construct an off-shell supercharge $Q_\epsilon$.  Before dealing with the construction of $Q_\epsilon$, it is convenient to perform the following field redefinition 
 \bea
 \hat{A}_3:=A_3+i \sigma.
 \eea
In terms of the new variables, the offshell algebra takes the following form for the vector
 \begin{eqnarray}
Q_{\epsilon} \hat{A}_{\theta, \varphi }&=& -\frac{i}{2}\left(
 -\bar{\lambda}^\dagger \sigma_{\theta, \varphi} \epsilon\right), ~~ Q_{\epsilon} \hat{A}_t=0, \nonumber\\
Q_{\epsilon} \sigma &=& \frac{1}{2}\left(
-\bar{\lambda}^\dagger \epsilon\right),~~~~~
Q_{\epsilon} \lambda = -\frac{1}{2} \sigma^{\mu\nu} \epsilon \hat{F}_{\mu \nu}+D \epsilon
+i \sigma^3 \epsilon \hat{\mathcal{D}}_3 \sigma,\nonumber\\
Q_{\epsilon} \bar{\lambda}^\dagger&=&0,
~~~~~~~
 Q_{\epsilon} D= 
 +\frac{i}{2}  (\hat{\mathcal{D}}_\mu \bar{\lambda})^\dagger \sigma^\mu \epsilon, \label{algebra1}
 \end{eqnarray}
 and matter multiplet \footnote{Where $\epsilon^C:=C \epsilon^*$ and $C=-i \sigma_2$. Notice, that the $C$ conjugation matrix is real.}
\begin{eqnarray}
Q_{\epsilon} \phi &=&
0, ~~~~~~ Q_{\epsilon} \bar{\phi}^\dagger = - \bar{\psi}^\dagger \epsilon,\nonumber \\
Q_{\epsilon} \psi &=& 
+i \sigma^\mu \epsilon \hat{\mathcal{D}}_\mu \phi,~~~~~~
Q_{\epsilon} \bar{\psi}^\dagger= 
\bar{F}^\dagger {\epsilon^c}^\dagger, \nonumber\\
Q_{\epsilon} F&=& 
+i (\epsilon^c)^\dagger \sigma^\mu \hat{\mathcal{D}}_\mu \psi+i (\epsilon^c)^\dagger \lambda \, \phi, ~~~~~~
Q_{\epsilon} \bar{F}^\dagger= 0.
\label{algebraChiral}
\end{eqnarray}
The action of the gauge covariant derivative being
 \bea
 \hat{\mathcal{D}}_\mu:=\left\{
	\begin{array}{clcl}
		\left(\,\partial_\mu+\frac{1}{4}\sigma_{a b} \omega^{a b}_{~~\mu}-i \hat{A}_\mu \,- \, i q_{sp} V_\mu \right) & \text{ on spinors,} \\
		\left(\,\partial_\mu-i \hat{A}_\mu \, - \, i q_{sc} V_\mu \right) & \text{ on scalars. }\end{array}
\right. \label{CovDer} \eea
It can be shown, that  the Chern-Simons theory
\be
\mathcal{L}_{CS}= -\frac{i k}{4 \pi}\left( \epsilon^{\mu \nu \beta}\left(\hat{A}_\mu \partial_\nu \hat{A_\beta} -\frac{2 i}{3}\hat{A}_\mu \hat{A}_\nu \hat{A}_\beta \right)  -\bar{\lambda}^\dagger\frac{ 1-\sigma_3}{2} \lambda \right),  \label{ModifiedCS}
\ee
is annihilated by \eqref{algebra1}, up to a total derivative
 \be
-i \frac{k}{4 \pi}  \hat{\mathcal{D}}_\mu\bigg( \epsilon^{\mu \nu \beta} \left(Q_\epsilon \hat{A}_\nu \right) \hat{A}_\beta \bigg).\label{TotDerChernN}
 \ee
There can also be a mixed CS term whenever we have several Abelian factors:
\bea
\label{eq:mCS}
\cl_{mCS}=-\frac{i k_{ij}}{4\pi} \left( \e^{\m \n \b}\hat{A}^{(i)}_\m \pa_\n \hat{A}^{(j)}_\b+\bar{\l}^{(i)\dagger}(\frac{1-\s_3}{2})\l^{(j)}\right).
\eea
Where $k_{ij}$ is symmetric and $i\neq j$, in this case one similarly gets boundary pieces
\bea
\label{eq:QLmcs}
&&\cq\cl_{mCS}=\frac{-i k_{ij}}{4\pi}  \Big(\hat{\cd}_\m( \e^{\m \n \b} (Q_\e \hat{A}^{(i)}_\n) \hat{A}^{(j)}_\b \Big). \label{mCS}
\eea
The discussion of the topological current in \cite{Benini:2015noa} is valid for any Chern-Simons theory. In the case of $ABJM$, the topological $U(1)_T$ global symmetry is generated by the conserved current $J^\m_T= tr (*\hat{F}-*\hat{\tilde{{F}}})^\m$. One can couple background $U(1)_T$ gauge potentials $\hat{A}^T_\mu$, to the current $J^\m_T$. The supersymmetry completion of such term,  is a particularization of the action \eqref{eq:mCS}.  Such particularization, is given by picking $k_{ij}=k_{j i}=1$ 
and regarding just a couple of indices $(i=1,2)$. The index ``$1$'' labels a background $Q_\epsilon$-spurion vector multiplet, and the index ``$2$'' labels a $U(1)$ dynamical vector multiplet. In such a way, we obtain the corresponding mixed supersymmetric Chern-Simons action, out of \eqref{eq:mCS}. For instance, in the case of gauge group $U(N)$, there is a unique dynamical $U(1)$, and the bosonic term of the latter action is
\bea
\label{eq:TCS}
\cl^{Bos}_T=-\frac{i}{4 \pi} \e^{\m \n \b}\Big(\hat{A}^{T}_\m \pa_\n \tr[ \hat{A}_\b]+\tr[\hat{A}_\m] \pa_\n  \hat{A}^T_\b  \Big).
\eea
In the very end, we will fix the v.e.v of the spurion vector supermultiplet to specific $Q_\epsilon$-$BPS$ values
\footnote{Which is the family $\hat{A}_3^T=u^{T},~D^T= iF^{T}_{12},~\s^{T}=0$.}.
 
 At this point, we must select a ``middle dimensional''  contour of integration in field space. Let us introduce a contour $\Gamma$ consistent with the one of  \cite{Benini:2015noa}
\be
\Gamma_{vector}: ~~  Bosonic ~Fields=(Bosonic ~Fields)^* , \text{ e. g. } D=-(D)^*. \label{GammaV}
\ee
 The contour $\Gamma_{vector}$ will cross a specific family of  $Q_\epsilon$-$BPS$ configurations. 
 \bea
 BPS[\Gamma_{vector}] : \left\{\begin{array}{c}
F_{12}=-\frac{\mathfrak{m}}{2}, ~ D=-i \frac{\mathfrak{m}}{2}, ~ fermions=0.\\ \\ \hat{A}_3=u=u^*\in [0,2\pi),
 \end{array}\right. \label{BPSGamma}
 \eea
where $\mathfrak{m}$ and $u$ are Cartan valued arbitrary constants. The $u$ are the Coulomb moduli and parametrize the Coulomb branch of the theory. Expression \eqref{BPSGamma} is the most general solution - single valued at the $S^1$ factor and without fermionic zero modes- to the $BPS$ equation
\be
Q_{\epsilon} \lambda = -\frac{1}{2} \sigma^{\mu\nu} \epsilon \hat{F}_{\mu \nu}+D \epsilon+i \sigma^3 \epsilon \hat{\mathcal{D}}_3 \sigma=0,
\ee 
along the contour \eqref{GammaV} . 

As for the matter multiplet we define
\be
\Gamma_{matter}:   \bar{\phi}=\phi, ~ \bar{F}=F. \label{GammaMatter}
\ee
In our case, the zero locus of matter is 
\be
BPS[\Gamma_{matter}]:   \bar{\phi}=\phi=\bar{F}=F=fermions=0.  \label{GammaBPSMatter}
\ee

Finally, we define $Q_\epsilon$ exact terms. The $Q_\epsilon$ exact terms must be semi-positive definite along $\Gamma$, as already stressed. In the case of the vector multiplet  and the choice of $\Gamma$ \eqref{GammaV}, such a term is
\begin{eqnarray}
Q_\epsilon V^{vector}&:=&-Q_{\epsilon} \left(\left(\overset{\bullet}{Q_{\epsilon} \lambda}\right) \lambda\right), \label{locVBZ} \\
\left(\overset{\bullet}{Q_{\epsilon} \lambda}\right)&:=&\nonumber \left(Q_\epsilon \lambda\right)^* \bigg|_{\hat{A}^*\rightarrow \hat{A}, ~ \sigma^*\rightarrow \sigma, ~ D^* \rightarrow {\red-}\,D}.\\\nonumber
\end{eqnarray}
The bosonic and fermionic part of \eqref{locVBZ} are
\begin{eqnarray}
 Q_\epsilon V^{vector}_B&:=&\left(F_{1 2}+ \hat{\mathcal{D}}_3 \sigma +i D \right)^2+\left(\hat{F}_{13}\right)^2+ \left(\hat{F}_{2 3}\right)^2, \label{QVvec2BZ} \\
 Q_\epsilon V^{vector}_F &:=&- i \,\bar{\lambda}^\dagger_2\, \overleftarrow{\hat{\mathcal{D}}}_t \, \lambda_2. \label{QVvecFBZ}
\end{eqnarray}
where $\lambda_2$ is the lower component of the gaugino $\small
\lambda=\left(\begin{array}{c} \lambda_1 \\ \lambda_2 \end{array}\right)
$.

 For the matter multiplet, and given the choice of $\Gamma$ in \eqref{GammaV} and \eqref{GammaMatter}, such a term is
\begin{eqnarray}
Q_\epsilon V^{matter}&:=&
-Q_\epsilon \bigg( -i \epsilon \sigma^\mu \psi \hat{\mathcal{D}}_\mu \bar{\phi}^\dagger+ F \bar{\psi}^\dagger\epsilon^c + i \bar{\phi}^\dagger \epsilon \lambda \phi \bigg).\label{MatterAc}
 \end{eqnarray}
The bosonic and fermionic part of \eqref{MatterAc} are
\begin{eqnarray}
-Q_\epsilon V^{matter}_{B}&=& ({\hat{\mathcal{D}}}^{\mu}{\bar{\phi}})^\dagger  \, {\hat{\mathcal{D}}}_{\mu}{\phi} +  \, \bar{\phi}^\dagger\left(\hat{\mathcal{D}}_3\sigma+i D- \epsilon^{\mu \nu}_{~~\beta}v^{\beta} \left(q V_{\mu \nu}+W_{\mu \nu}\right) \right)\phi
\nonumber\\&&  +\bar{F}^\dagger F+\,\hat{\mathcal{D}}_{\mu}\left( i \, \epsilon^{\mu \nu}_{~~\beta}v^{\beta} \bar{\phi}^\dagger {\hat{\mathcal{D}}}_{\nu} \phi \right)
, \label{LBOSF0}
\\ && \nonumber \\
-Q_\epsilon V^{matter}_{F}&=&-i \, \bar{\psi}^\dagger {\sigma}^{\mu} {\hat{\mathcal{D}}}_{\mu}{\psi}- i\,  \bar{\psi}^\dagger \lambda \,\phi- i \,
\bar{\phi}^\dagger\, \bar{\lambda}^\dagger P^- \psi
+ i \hat{\mathcal{D}}_\mu\left( \bar{\psi}^\dagger P^+ \sigma^\mu \psi\right),  \label{LFERF0} \\  && \nonumber
\end{eqnarray}
where $P^{\mp}:=\frac{1\mp\sigma_3}{2} $ and $V_{\mu\nu}$, $W_{\mu \nu}$ are the field strengths of R- and flavor symmetry backgrounds, respectively. 

The term $Q_\epsilon V^{matter}_{B}$ is semi-positive definite when expanded around $BPS$[$\Gamma_{vector}$] and over $\Gamma_{matter}$. As shall be shown in due time, this last statement is implied by the requirement of square integrability over $\mathbb{H}_2$. Square integrability over $\mathbb{H}_2$, imposes bounds on the spectrum of eigenvalues of the relevant magnetic Laplacian. The aforementioned bounds imply the convergence of the Gaussian path integral $ \int_{X^{(1)}} e^{- \delta^{(2)}\left( Q_\epsilon V ,X^{(0)}\right)  }$ in \eqref{proc}.

Chern-Simons, being a gauge theory, requires gauge fixing, which we choose to be the axial condition 
\be
\hat{A}_3=const. \label{axial}
\ee
In contradistinction to 3d pure Yang-Mills theory, in 3d Chern-Simons coupled to Yang-Mills and/or matter, the constraint \eqref{axial}, fixes the gauge degeneracy completely. In the latter theory there are $3-1=2$ physical off-shell  vector degrees of freedom (DoF) meanwhile in 3d pure Yang-Mills there is $3-2=1$  massless vector offshell DoF. For a nice review on the canonical quantization of 3d Chern-Simons theory, see for instance \cite{Dunne}.

To implement the gauge fixing, we use $BRST$ method \cite{Pestun, Kapustin} and enlarge the vector multiplet, by adding the ghost fields $(c,\bar{c},\bar{b})$. We enlarge the algebra \eqref{algebra1}, by the following transformation rules
\be
Q_\epsilon c=0, ~ Q_\epsilon \bar{c}=0, ~ Q_\epsilon \bar{b}=0. \label{QExte}
\ee

Any gauge invariant functional of the physical fields is $BRST$ invariant. The $BRST$ transformations $Q_{B}$ are 
\begin{eqnarray}
Q_{B} \hat{A}_\mu= \hat{\mathcal{D}}_\mu  c, ~~ Q_{B} \bar{c}= \bar{b}, ~~ Q_{B} c =\frac{i}{2} \{c,c\} , ~~Q_{B} \lambda=i \{c, \lambda \}, \nonumber\\ 
Q_{B} \bar{\lambda}^\dagger= i \{c,\bar{\lambda}^\dagger\}, ~~ Q_{B}\sigma= i[c,\sigma], ~~ Q_{B} D=i \{c,\sigma\}, \nonumber\\
Q_{B} \phi=i[c,\sigma], ~~ Q_{B}\bar{\phi}^\dagger =i [c,\bar{\phi}^\dagger]
,~~Q_{B} \psi=i\{c,\sigma\}, ~~ Q_{B}\bar{\psi}^\dagger =i \{c,\bar{\psi}^\dagger\},
\nonumber \\ Q_{B} F= i [c,F], ~~ Q_{B} \bar{F}^\dagger= i [c,\bar{F}^\dagger],  \label{BRST}
\end{eqnarray}
from \eqref{algebra1}, \eqref{algebraChiral} and \eqref{BRST} it can be shown that
\be
\left(Q_\epsilon+Q_{B}\right)^2=\{Q_\epsilon,Q_{B}\}=0. \label{NilpBRST}
\ee 
As the $V$'s in the $Q_\epsilon V$'s localizing terms, \eqref{locVBZ} and \eqref{MatterAc}, are gauge invariant objects, then from the corresponding algebra \eqref{BRST} is easy to check that the $V$'s in \eqref{locVBZ} and \eqref{MatterAc}, are $Q_{B}$ invariant and consequently \eqref{locVBZ} and \eqref{MatterAc} are $\left(Q_\epsilon+Q_{B}\right)$-exact. 

 On top of the localizing actions  \eqref{locVBZ} and \eqref{MatterAc}, a gauge fixing term must be added. To our purposes the most convenient choice is the following  $\left(Q_\epsilon+Q_{B}\right)$-exact term 
\be
Q_{B}Tr\left(\bar{ c} \left(\hat{A}_t-const\right)\right)= \bar{ c}\hat{\mathcal{D}}_t c+  \bar{b}\left(\hat{A}_t-const\right). \label{BRSTAc}
\ee
From \eqref{QExte} and $Q_\epsilon \hat{A}_3=0$, it follows that \eqref{BRSTAc} is $\left(Q_\epsilon+Q_{B}\right)$-exact. In \eqref{BRSTAc} we wrote the gauge index trace $Tr$ only on the LHS, but the reader should keep in mind that by default we are working with gauge invariant density Lagrangians.
 
Our $BRST$ construction is conceptually  that of Pestun \cite{Pestun} and it has been previously presented in the 3d case by Kapustin, Willet and Yaakov \cite{Kapustin}.

\subsection{Boundary conditions}

In  non-compact manifolds like $\mathbb{H}_2\times S^1$  or manifolds with boundary, appropriate boundary conditions must be imposed in order to have a well defined variational -Lagrangian- problem. Once a proper classical theory has been defined, quantization is in order. Let
\bea
X^{(0)} = \{A^{(0)}_\mu, \sigma^{(0)}, D^0,\ldots\}\in BPS[\Gamma]
\eea
and
\bea
X^{(1)}=\left\{  \delta A_\mu, \delta \sigma, \delta D, \delta \lambda, \delta \bar{\lambda}, \delta c, \delta \bar{c}, \bar{b},  \delta \phi, \delta \bar{\phi},  \delta \psi, \delta \bar{\psi}, \delta F, \delta \bar{F} \right\},
\eea
be the non trivial zero locus background fields and offshell fluctuations respectively. As for the $X^{(0)}$ we define the following boundary condition
\bea
e^\mu_a A^{(0)}_\mu, ~ D^{(0)}, ~ \sigma^{(0)} &\underset{ \theta\rightarrow \infty}{\sim}& O(1). \label{BPSX0}
\eea

As for offshell fluctuations $X^{(1)}$, we define Dirichlet boundary conditions
\bea
 e^\mu_a \delta A_\mu, \delta \sigma, \delta D, \delta \lambda, \delta \bar{\lambda}, \delta c, \delta \bar{c}, \delta \bar{b}, \delta \phi, \delta \bar{\phi},  \delta \psi, \delta \bar{\psi}, \delta F, \delta \bar{F} & \underset{ \theta\rightarrow \infty}{\sim}& O(e^{-\kappa \theta}), \label{BCFluc} \\
 e^\mu_a \delta A_\mu, \delta \sigma, \delta D, \delta \lambda, \delta \bar{\lambda}, \delta c, \delta \bar{c}, \delta \bar{b},  \delta \phi, \delta \bar{\phi},  \delta \psi, \delta \bar{\psi}, \delta F, \delta \bar{F}  &\underset{ \theta\rightarrow 0}{\sim} &O(1), \label{BCFluc2}
\eea
with $\kappa\geq \frac{1}{2}$. The value of $\kappa$ defines important features of the spectrum of the associated $S^1$ quantum mechanics, if a sort of dimensional reduction is possible to perform in this case.  

The following table sketches the relation between the boundary conditions \eqref{BCFluc} and the results reported in the next section:
\begin{center}
\begin{tabular}{ |c|c|c| } 
 \hline
 $\kappa$ & Spectrum & Norm \\ 
 $\frac{1}{2}$ & Continuous: $\lambda\in [0,\infty)$ ~~~~~~~~~~~~~~& Delta-Square Integrable \\ 
 $>\frac{1}{2}$ & Discrete: $j=|s|-1,|s|-2, \ldots >-\frac{1}{2}$. & Square Integrable \\ 
 \hline
\end{tabular}
\end{center}

The total derivative part of the offshell variation of the Chern-Simons Lagrangian \eqref{ModifiedCS}, multiplied by the volume element $\sqrt{-g}$ is
\bea
-i \frac{k}{4 \pi} \hat{\mathcal{D}}_\mu\bigg(\sqrt{-g} \epsilon^{\mu \nu \beta} \left(\delta \hat{A}_\nu \right) \hat{A}_\beta \bigg),~~~\text{ with }~~~  \epsilon^{\theta \varphi t}:= \frac{1}{\sqrt{-g}}.\label{TotDerChernN2}
\eea
After integration and imposition of gauge fixing condition $\delta \hat{A}_3=0$ - see \eqref{BRSTAc}-, the total derivative \eqref{TotDerChernN2} becomes a boundary term
\be
-i \frac{k}{4 \pi} \int_{\varphi=0}^{2\pi}d\varphi \int_{t=0}^{1}dt \left(\delta \hat{A}_\varphi\right)\hat{A}_t \bigg|^{\theta=\infty}_{\theta=0}. \label{BTeRm}
\ee
Boundary conditions \eqref{BCFluc}, do not imply the vanishing of \eqref{BTeRm} at $\theta=\infty$, due to the non-compactness of $\mathbb{H}_2$ - specifically because  $e^\varphi_2\underset{\theta \rightarrow \infty }{\rightarrow} 0$-. The contribution from $\theta=0$ vanishes.

To have a well-defined variational principle, we redefine the classical action from Chern-Simons to
\be
\int \sqrt{-g} \mathcal{L}_{cl}= \int \sqrt{-g}\mathcal{L}_{CS} + S_{bdy}, ~~ S_{bdy}=+i \frac{k}{4\pi}\int_{\varphi=0}^{2\pi}d\varphi \int_{t=0}^{1}dt ~Tr \left( \hat{A}_\varphi \hat{A}_t \right)  {\text{ at } \theta=\infty}. \label{sbdry} 
\ee
Note that $S_{bdy}$ is gauge invariant, provided we restrict the derivatives of gauge transformations parameters to vanish at $\theta\rightarrow ~\infty$. In this paper, we will assume the latter condition.

 It is immediate to check that the supersymmetric transformation of $\mathcal{L}_{cl}$ is trivial by construction:  the supersymmetry variation of $S_{bdy}$ cancels the integration of the total derivative term \eqref{TotDerChernN}, as it should. 
 
 The classical action evaluated on the zero locus is
 \be
 \int \sqrt{-g} \mathcal{L}_{cl}[BPS[\Gamma]]
 =-i\frac{k}{2}  u \cdot \mathfrak{m} \label{CSOnshell}
 \ee
 Where $u \cdot \mathfrak{m}:=u_i \mathfrak{m}_i= \frac{1}{2}Tr(u~ \mathfrak{m})
$. In our conventions $h_i$ and $h_j$ are Cartan generators in the Chevalley basis, and consequently $Tr[h_i h_j]=2\delta_{i j}$. 

The contributions proportional to
$\cosh{\theta_{Max}}$ cancel out, $\theta_{Max}$ being the large cut off in $\theta$. The divergent terms coming from the integral over $\mathbb{H}_2$ and the boundary term \eqref{sbdry} cancel each other.
 Whenever we have contributions which diverge like the volume, we regulate them as we regulate the volume in \eqref{eq:Vol}, and boundary terms are regulated as  explained in footnote in page $4$. 

 It is convenient to write down the exponential of $-$\eqref{CSOnshell}
\be
x_i^{ \frac{k \mathfrak{m}_i}{2} }, \label{CSterm}
\ee
with $x_i=e^{i u_i}$. Expression \eqref{CSterm}, is the contribution to the index of a Chern-Simons term with level $k$.

The total derivative part of the variation with respect to $\bar{\phi}^\dagger$ of the bosonic localizing action of matter is
\be
+\int_{\mathcal{M}} \hat{\mathcal{D}}_{\mu}\left(\sqrt{-g} \left({{\delta \bar{\phi}}}^\dagger  \, \hat{\mathcal{D}}^{\mu}\phi+i \, \epsilon^{\mu \nu}_{~~\beta}v^{\beta} \delta\bar{\phi}^\dagger {\hat{\mathcal{D}}}_{\nu} \phi \right) \right) \label{TDMatter0}
\ee 
Under off-shell boundary conditions \eqref{BCFluc} and \eqref{BCFluc2}, the integration of \eqref{TDMatter0} gives
\bea
+i \int^{2\pi}_{0} dt \int^{2\pi}_0 d\varphi \left[ \left(\delta \bar{\phi} ^\dagger \hat{\mathcal{D}}_\varphi \phi \right) \right]_{\theta=0}. \label{OFFMat}
\eea
 
 The term \eqref{OFFMat} vanishes, when evaluated in the functional space we are going to integrate over. The explanation of the latter fact, shall be given in the beginning of subsection \ref{BosLocOp}. 
 
Notice, that the ghosts vanish at the boundary,  due to \eqref{BCFluc}. Consequently, $BRST$ gauge transformations do not affect the boundary.


Having established the localization locus, the next step is to compute one loop determinant contributions $Z_{X^{(0)}}$.  In order to do that, we need to define an appropriate functional space to integrate upon. That, will be the scope of the next section. Thereafter, we can compute $Z_{X^{(0)}}$ and use equation \eqref{PF} to evaluate our final result for the topologically twisted index.

\section{The spectrum on $\mathbb{H}_2$ with flux} \label{sec:3}

The spectrum of the Laplace operator  on $\mathbb{H}_2$ has a long history (\cite{Camporesi:1994ga} and references therein).  
Even though this section might seem just a technical remark, the result of its analysis is very  relevant in reaching our conclusions.  We therefore choose to include it in the main body of the text and provide more details in an appendix.
 
 The eigenvalue problem solved in appendix \ref{SQN}, is related to the propagation  of a scalar particle in the presence of a flux on $\mathbb{H}_2$. The hierarchy of modes to be reported in this section, can be interpreted as a series of Landau levels on $\mathbb{H}_2$, that emerge due to the presence of a flux $s$ \cite{Comtet, Antoine, Lisovyy}. These alternative viewpoints deserve further attention and we are keen to pay them so in forthcoming works. 

In this section, we will present the outcome of the analysis that shall be reported in appendix \ref{SQN}. We encourage the reader looking for a detailed understanding, to go through that appendix. 

The Laplacian in the presence of a flux $s$, coming from a potential
\be
A=s \cosh{\theta} d\varphi,  \label{gaugeP}
\ee
is given by
\be
\square_{s}:=-\partial_{\theta}^{2}-\coth^{2}\theta\partial_{\theta}+\frac
{1}{\sinh^{2}\theta}(j_{3}-s\cosh\theta)^{2},
\ee
with $j_{3}=-i\partial_{\varphi}.$ 


The equation that defines the functional space upon which we will compute
determinants is%
\be
\left(  \square_{s}+\Delta\right)  f_{\Delta,j_{3}}=0. \label{DefEq}
\ee
The boundary conditions that will define our functional space are
\be
\eqref{BCFluc} \text{  and  } \eqref{BCFluc2}  \text{ with  }  \kappa > \frac{1}{2}. \label{BC2}
\ee

\subsection{The discrete spectrum}

First, we parametrize the eigenvalue as:
\be
\Delta=j(j+1)-s^{2}.
\ee
{The quantization conditions}
\be
j_3-s,~  j-|s|\in \mathbb{Z}, \label{QCPrimary}
\ee  
{together with equation \eqref{DefEq}, define a finite (resp. infinite) dimensional space of square integrable functions on $\mathbb{H}_2$, that we denote as}
\be
\Xi_{j}^{(1)}%
(s):=\left\{  f_{\Delta,j_{3}}^{(1)}\right\}  _{j_{3}},
\ee
{respectively
}
\be
\Xi_{j}^{(2)}(s):=\left\{  f_{\Delta,j_{3}}^{(2)}\right\}  _{j_{3}}.%
\ee
The explicit form of the eigenfunctions $f_{\Delta,j_{3}}^{(1)}$ and $f_{\Delta,j_{3}}^{(2)}$, is defined in appendix \ref{f1app} and \ref{f2app}, respectively, for the case $s>\frac{1}{2}$. The case $s<-\frac{1}{2}$ can be worked out analogously.

{The range of $j_3$ is given by the relations}%
\be
\begin{tabular}
[c]{lll}%
$s\geq j_{3}\geq\max(|j|,|j+1|)$ & if & $s>+\frac{1}{2}$,\\ \\
$-j_{3}\geq-s\geq\max(|j|,|j+1|)$ & if & $s<-\frac{1}{2}$, \label{rest1}%
\end{tabular}
\ee
{respectively}%
\be
\begin{tabular}
[c]{lll}%
$j_{3}\geq s\geq\max(|j|,|j+1|)$ & \ \ if \  & $s>+\frac{1}{2}$,\\ \\
$-s\geq-j_{3}\geq\max(|j|,|j+1|)$ & \ \ if & $s<-\frac{1}{2}$. \label{rest2}%
\end{tabular}
\ee
There are additional constraints to the value of $j$. Indeed, the eigenfunctions
$f^{(1,2)}_{\Delta,j_{3}}$ and henceforth the spectrum $\Delta=j(j+1)-s^{2}$, { are invariant under the transformation}
\[
j\rightarrow-\left(  j+1\right).
\]
Thenceforth we must restrict $j$ to be either%

\be%
\begin{array}
[c]{ccc}%
j>-\frac{1}{2} & \text{ \ \ or } & j<-\frac{1}{2}
\end{array}\label{regj}
.
\ee
In order to have square integrable functions
\be
j\neq-\frac{1}{2}. \label{regj2}
\ee
 For the choice $j>-\frac{1}{2}$, restrictions \eqref{rest1} and \eqref{rest2}, imply an upper bound for $j$

\be
j=|s|-1, |s|-2,... > -\frac{1}{2}.
\ee
Interestingly, for
\be
0\leq|s| \leq \frac{1}{2},
\ee
there are no square integrable modes. A particular conclusion of this last statement, is the known fact that in $\mathbb{H}_2$ there are no square integrable scalar modes. In the presence of flux  $s: |s|> \frac{1}{2}$, square integrable modes emerge. 
 
 In appendix \ref{AppA2}, it is proven that in the case $|s|=1$, our square integrable eigenmodes match those well known discrete modes, of the vector Laplace-Beltrami operator in $\mathbb{H}_2$, with helicity $s=\pm1$. This last statement, suggests to explore the possibility that our spectrum encodes the full tower of higher spin square integrable eigenmodes of the Laplace-Beltrami operator on $\mathbb{H}_2$.  We hope to come back to this point in the future.

{The relevant scalar product is}%

\be
\mathit{<f,g>:=}\int_{0}^{\infty}\mathit{d\theta}\int_{0}^{2\pi}%
\mathit{d\varphi}\sinh\left(  \theta\right)  \mathit{f}^{\ast}\mathit{(\theta
,\varphi)g(\theta,\varphi)}. \label{ScalarP}%
\ee

{As already mentioned, and proven in appendix \ref{SQN}, square integrability of}$\ f_{\Delta,j_{3}%
}^{(1)}${ (resp. }$\ f_{\Delta,j_{3}}^{(2)}${) is interrelated to the
specific bounds on }$j_{3}$ and $j$  that were previously written. 

{ Different states $f_{\Delta,j_{3}}^{(1,2)}e^{ij_{3}\varphi}%
$ in $\Xi_{j}^{(1,2)}(s)$ are orthogonal with
respect to the scalar product \eqref{ScalarP}.  \text{Spaces }$\Xi_{j}^{(1,2)}(s)$ with
different label $ j>-\frac{1}{2}$ $(\ ${or} $j<-\frac{1}{2})$ are orthogonal.
This is because $\square_{s}$ is Hermitian in $\Xi_{j}(s)$
and spaces with different label $j>- \frac{1}{2}$ $(${or} $j<-\frac{1}{2})$, have
different eigenvalues $\Delta$  under $\square_{s}$.}

Summarizing, the space of square integrable modes for a given $s$ is
\be
\Xi(s)= \bigoplus^{|s|-1}_{j>-\frac{1}{2}} \left(\Xi^{(1)}_j(s) \oplus \Xi^{(2)}_j(s)\right).\label{FuncSpaceTot0}
\ee

 In the next section, we will refer to the following spaces
 \be
 \Xi_j(s):=%
{}
\left(  \Xi_{j}^{(1)}(s)~%
{\displaystyle\oplus}
~\Xi_{j}^{(2)}(s)\right). \label{FuncSpacej}
 \ee
The spaces \eqref{FuncSpacej} are subspaces of \eqref{FuncSpaceTot0}.

\subsection{The continuous spectrum}

The continuous spectrum is a direct generalization of the spectrum reported by Higuchi and Camporesi \cite{Camporesi:1994ga} to the case where there is  a constant flux on $\mathbb{H}_2$. The corresponding eigenmodes  solve the defining equation 
\be
\left(
\square_{s}+\Delta_{(\lambda,s)}\right)  f_{\Delta_{(\lambda, s)},j_{3}}=0,
\ee 
with%
\[
\Delta_{(\lambda,s)}:=-\lambda^{2}-s^{2}-\frac{1}{4},\text{ \ }\lambda\in \mathbb{R},\text{
}\lambda\geq0,
\]
and boundary conditions 
\bea
f_{\Delta_{(\lambda,s)},j_{3}}(x) &\underset{x\rightarrow -\infty}{\sim}&   c_{1(\lambda
,j_{3},s)}\frac{x^{+i\lambda}}{x^{\frac{1}{2}}}+c_{2(\lambda,j_{3},s)}%
\frac{x^{-i\lambda}}{x^{\frac{1}{2}}},  \label{UVCond} \\ f_{\Delta_{(\lambda,s)},j_{3}}(x) &\underset{x\rightarrow 0}{\sim}&O(1). \label{IRCond}
\eea
Conditions \eqref{UVCond} and \eqref{IRCond} are given in coordinates $x$, but they are equivalent to the particularization $\kappa=\frac{1}{2}$  of \eqref{BCFluc}.

The final solution to the boundary problem just presented, is obtained by imposing \eqref{IRCond} on the most general solution \eqref{MostGSol}. The result is
\bea
f^{(1)}_{\Delta_{(\l,s)}}(x) &\text{ ~~ if~~ } &   j_3 \geq s,  \\
f^{(2)}_{\Delta_{(\l,s)}}(x) &\text{ ~~ if~~ } &   j_3 < s.
\eea

The norm of $f_{\Delta_\lambda j_3}$ under the scalar product \eqref{ScalarP} is infinite. By choosing appropriately the remaining 
integration constant one can set 
\be
\langle f_{\Delta_{(\lambda,s)}, j_3}, f_{\Delta_{(\lambda^\prime,s)}, j^\prime_3}\rangle =\delta(\l-\l')\del_{j_3 j_3'}. \label{Norma}
\ee

\paragraph{Comments:}

Our thermal cycle is not the $S^1$ inside the $\mathbb{H}_2$, but the trivially fibered one. The latter fact, is related to an important conceptual difference between the physical framework of our approach and the one of, for instance, \cite{Almheiri}. In physical terms, our $\mathbb{H}_2$ modes are not probing the near horizon limit of a black hole in the presence of electric flux \cite{Sachdev, Maldacena, Polchinski, Maldacena2, Engelsoy:2016xyb,Jensen:2016pah, Maldacena3}, but the boundary dynamics of a magnetically charged hyperbolic $AdS_4$ black hole. 

That said, if we interpret our $\varphi$-cycle as the thermal one, our hierarchy of square integrable modes is certainly probing Euclidean $AdS_2$, in the presence of an electric flux deformation. That is closely related to the problem addressed in \cite{Almheiri}. To have a self-consistent approach, coming from supersymmetric localization, to the problem studied in \cite{Almheiri}, one should try to localize an appropriate off-shell supercharge on the quantum gravity side. In that spirit, in \cite{Sameer2}, $AdS_2\times S^2$ was shown to be the unique ungauged $BPS$ localizing solution, to 4d $\mathcal{N}=2$ Super-Conformal gravity in a convenient gauge fixing.  It is plausible, that other electric or magnetic $AdS_2\times \Sigma$ localizing solutions can be found, by relaxing some of the conditions used in \cite{Sameer2}, as suggested by the results in \cite{Halmagyi2}. If that is the case, it would be quite interesting to explore what can supersymmetric localization say about the problems addressed in \cite{Almheiri}.

\section{One-loop determinants}  \label{sec:4}

Having clarified the structure of the spectrum for the flux Laplacian on $\mathbb{H}_2$ (see appendix \ref{SQN}), we have all the ingredients to address the computation of one-loop determinants.

\subsection{Bosonic localizing operator} \label{BosLocOp}
\bigskip

For square integrable modes, the total derivatives of the quadratic expansion in the localizing
terms, integrate to zero. In the case of the  matter multiplet, the total derivate term is 
\bea
+\int_{\mathcal{M}} \hat{\mathcal{D}}_{\mu}\left(\sqrt{-g} \left({{ \bar{\phi}}}^\dagger  \, {\hat{\mathcal{D}}}^{\mu}{\phi}+i \, \epsilon^{\mu \nu}_{~~\beta}v^{\beta} \bar{\phi}^\dagger {\hat{\mathcal{D}}}_{\nu} \phi\right) \right) \label{TDMatter}\\
=+i \int d t d \varphi  \left[ \bar{\phi}^\dagger \hat{\mathcal{D}}_\varphi \phi\right]_{\text{ At } \theta=0}, \label{TOTD2}
\eea
where the result in the second line follows from the asymptotic behavior \eqref{BCFluc} and \eqref{BCFluc2}. This boundary term vanishes because, as proven in appendix \ref{SQN}, the only modes of $\phi$ that do not vanish at the contractible cycle $\theta=0$, have the following angular dependence
\bea
e^{i ~s ~\varphi},\ \text{   with  }~~~ s&:=&-\frac{\rho(\mathfrak{m})-q_{R}}{2}, \label{fluxqR} 
\eea
and they are annihilated by
\be
\hat{\mathcal{D}}_\varphi:=\partial_\varphi -i s.
\ee
 Due to a careful choice of boundary conditions, total derivatives are irrelevant for the current
discussion, as they do not contribute to the 1-loop determinant.
%

After integration by parts, the quadratic expansion of the bosonic part of the Lagrangian density of the matter localizing term \eqref{LBOSF0}, takes the form
\bea
\bar{\phi}^\dagger O_{B} \phi &  
:= 
\bar{\phi}^\dagger \bigg(  (\rho(u)+i\partial_{t})^{2}+\left( \square_{s}-s\bigg)  \right) \phi. \label{OB}
\eea

Notice that the operator $O_{B}$ is 
positive definite on \textquotedblleft representations\textquotedblright%
\ $\Xi_{j}(s)$\ labeled by $j$ running at step $1$ down from $|s|-1$ but
larger than $-\frac{1}{2}$.  For \textquotedblleft representations\textquotedblright\ labeled by $j:-\frac{1}{2} <
j\leq|s|-1$ (or $-|s|\leq j<-\frac{1}{2}$), the operator $\left(  \square_{s}-s\right)  $
has eigenvalues that obey the following inequality$\ -j(j+1)+s(s-1)\geq0$ and
hence is semi-positive definite. Consequently, $O_B$ is positive definite if $(\rho(u)+i\partial_{t})^{2}>0$. This last condition is guaranteed, provided we avoid points in the Coulomb branch such that $\rho(u) \in \mathbb{Z}$.

 Having in mind the particular case
\[
j,~j_3,~s \in \mathbb{Z},
\]
at some stages, we will denote the aforementioned set of
$j^{\prime}s$ as follows%
\[
j:0\leq j\leq|s|-1.
\]
The union of the aforementioned $\Xi_{j}(s)$, is the maximal space of square integrable modes \eqref{FuncSpaceTot0}.

For latter convenience, let us define%
\bea
\square_{s}^{\pm}&:=&\square_{s}\pm s.
\eea

 Should we select $\phi$ in the vector space spanned by $\Xi_{j}(s)$,
thence for $s>\frac{1}{2}$%
\begin{align}
\det_{\Xi_{j}^{(1)}(s)}(O_{B})  &  =%
{\displaystyle\prod\limits_{j_{3}=j+1}^{s}}
{\displaystyle\prod\limits_{
k}}
\left(  (\rho(u)+k)^{2}-j(j+1)+s(s-1)\right)  , \label{EqRefe1}\\
\det_{\Xi_{j}^{(2)}(s)}(O_{B})  &  =%
{\displaystyle\prod\limits_{j_{3}=s+1}^{\infty}}
{\displaystyle\prod\limits_{
k}}
\left(  (\rho(u)+k)^{2}-j(j+1)+s(s-1)\right) \label{EqRefe2} ,
\end{align}
where $k=i \partial_t$. The result for $s<-\frac{1}{2}$ is obtained analogously.

As $\phi$ is a complex scalar, the functional integration%

\[\int[D\phi^\dagger D\phi] ~
Exp\left[-%
{\displaystyle\int\limits_{\mathbb{H}_2\times S^{1}}}
\phi^{\dagger}\cdot(O_{B})\cdot\phi%
\right],
\]
is proportional to $\frac{1}{\det(O_{B})}$, where%

\[
\det_{\Xi(s)}(O_{B})=\left\{
\begin{tabular}
[c]{lll}%
$%
{\displaystyle\prod\limits_{j=0}^{s-1}}
{\displaystyle\prod\limits_{j_{3}=j+1}^{\infty}}
{\displaystyle\prod\limits_{\rho^{\ast},k}}
\left(  (\rho(u)+k)^{2}-j(j+1)+s(s-1)\right)  $ & \ \ \ if & $s>\frac{1}{2}$\\
$%
{\displaystyle\prod\limits_{j=0}^{-s-1}}
{\displaystyle\prod\limits_{j_{3}=-\infty}^{-j-1}}
{\displaystyle\prod\limits_{\rho^{\ast},k}}
\left(  (\rho(u)+k)^{2}-j(j+1)+s(s-1)\right)  $ & \ \ if & $s<-\frac{1}{2}$%
\end{tabular}
\ \right.  .
\]

\subsection{Fermionic localizing operator }

To compute the fermionic determinant,  we used the square of the kinetic operator that appears in the quadratic expansion of the fermionic part \eqref{LFERF0} of the localizing term,  we also specify the space of functions on which each component acts%
\[
\left(
\begin{array}
[c]{cc}%
O_{B} & 0\\
0 & (u+i\partial_{t})^{2}+\square_{s-1}^{+}%
\end{array}
\right)  \binom{\psi^{+}\in Span(\Xi(s))}{\psi^{-}\in Span(\Xi(s-1))}.
\]

While reproducing the computations that will be reported in section \ref{supdet}, it will be convenient  to use the following identity%
\[
\square_{s-1}^{+}\text{ }f^{(1,2)}_{\Delta(s-1),\text{ }j_{3}}=\left(
-j(j+1)+s(s-1)\right)  \text{ }f^{(1,2)}_{\Delta(s-1),\text{ }j_{3}}.%
\]
\subsection{$\zeta$-function regularization: $s> \frac{1}{2}$}
\bigskip
We use $\zeta$-function regularization to compute the determinants of $O_{B}$
upon the functional space%
\[
\Xi_{j}(s):=\Xi_{j}^{(1)}(s)%
~{\displaystyle\oplus}~
\Xi_{j}^{(2)}(s),\ \ j=s-1\ \ (or\ \ -s).
\]
which is the space of zero modes of \ 
\be
\square_{s}^{-} \nonumber
\ee
 (the space of eigenstates of $O_{B}$ with eigenvalue $(\rho(u)+k)^{2}$). We stress  that in the case $s>\frac{1}{2}$ and after
cohomological cancellations, these zero modes are the only ones that contribute
to the one loop super-determinant.

In order to compute the heat kernel $K(0,0)$ associated to the eigenspaces in
question, we need to analize the relevant case $j_{3}=s$. In the
latter case and after particularizing to \ $j=\ s-1$\ or $-s$, the square
integrable modes $f^{(1)}$ and $f^{(2)}$ drastically simplify to%
\[
f^{(1,2)}_{\Delta(s), j_3=s}=\chi \text{ }\left(  x-1\right)  ^{-s}.
\]%
The constant $\chi$ is determined from the normalization condition
\bea
|\chi|^{2}2\pi vol_{S_{1}}\int_{-\infty}^{0}dx(2)\text{ } |  x-1|  ^{-2s}  &
=\frac{|\chi|^{2} 2\pi vol_{S^{1}}}{s-\frac{1}{2}}=1.\label{normal}
\eea
The $2$ in the LHS of \eqref{normal} is the line element in coordinates $x$.
From the value of $|\chi|^2$, we obtain the heat kernel at origin
\be
K(t;0,0)=\frac
{1}{2\pi vol_{S^{1}}}(s-\frac{1}{2})e^{-t\text{ }(\rho(u)+k)^{2}}.\label{heatK}
\ee
From \eqref{heatK} we obtain the zeta function 
\bea
\zeta(z)  &  =\frac{vol_{\mathbb{H}_2}vol_{S^{1}}}{2\pi vol_{S^{1}}}\frac{(s-\frac{1}{2}%
)}{\text{ }\left(\left(\rho(u)+k\right)^2\right)^{z}}=-\frac{(s-\frac{1}{2})}{\text{ }\left(\left(\rho(u)+k\right)^2\right)^{z}%
},
\eea
with
\be
 vol_{\mathbb{H}_2}=-2\pi, ~~ vol_{S^1}=1.
\ee


After using $\zeta$-function method, we obtain the desired determinant%

\bea
\det\limits_{\Xi_{j=-s}(s)}(O_{B})=\det\limits_{\Xi_{j=s-1}(s)}(O_{B}%
)=e^{-\zeta^{\prime}(0)}&=&\left((\rho(u)+k)^2\right)^{-s+1/2} \\
&=&|(\rho(u)+k)|^{-2s+1}. \label{absPar}
\eea
In appendix \ref{reg} we obtain the same result, by using an alternative procedure. Notice that in \eqref{absPar} we could have also written 
\bea
&=&\bigg(-|(\rho(u)+k)|\bigg)^{-2s+1}.
\eea
It is possible that such a change in the election of sign, changes the value of the partition function. From now on, we will ignore this second choice, except for specific steps where having it in mind will be useful.



\subsection{$\zeta$-function regularization: $s<\frac{1}{2}$}


In the case $s<\frac{1}{2}$, the contribution to the super-determinant is coming from
the following set of eigenfunctions
\be
\Xi_{j}(s-1):=\Xi_{j}^{(1)}(s-1)%
{\displaystyle\oplus}
~\Xi_{j}^{(2)}(s-1),\ \ j=-s\ \ (\text{or}\ \ s-1),
\ee
which is the space of zero modes of 
\be
\square_{s-1}^{+}.
\ee
In the case $s<\frac{1}{2}$, and after cohomological cancellations, these zero modes are the only modes that contribute
to the one loop super-determinant.

Following the very same steps described in the previous section, we focus on the
solutions obtained for $\ j_{3}=s-1$ and $\ \ j=-s\ \ (\text{or}\ \ s-1).$ In this
case, the zero mode solutions reduce to%
\[
f_{\Delta(s-1),\text{ }j_{3}=s-1}^{(1,2)}=\chi \left(  x-1\right)  ^{s-1},
\]
from where it is straigthforward to compute the $\zeta$-function.%

The constant $\chi$ is determined from the normalization condition
\bea
|\chi|^{2}2\pi vol_{S_{1}}\int_{-\infty}^{0}dx\text{ }(2)  |x-1|  ^{2(s-1)}  &
=\frac{|\chi|^{2}2 \pi vol_{S^{1}}}{-s+\frac{1}{2}}=1.
\eea
From the value of $|\chi|^2$, we obtain the heat kernel at origin
\be
K(t;0,0)=\frac
{1}{2 \pi vol_{S^{1}}}(-s+\frac{1}{2})e^{-t\text{ }(\rho(u)+k)^{2}}. \label{HeatK2}
\ee
From \eqref{HeatK2}, we obtain the zeta function 
\bea
\zeta(z)  &  =\frac{vol_{\mathbb{H}_2}vol_{S^{1}}}{2 \pi vol_{S^{1}}}\frac{(-s+\frac{1}{2}%
)}{\text{ }\left(\left(\rho(u)+k\right)^2\right)^{z}}=-\frac{(-s+\frac{1}{2})}{\text{ }\left(\left(\rho(u)+k\right)^2\right)^{z}}.
\eea


Finally, we obtain the desired determinant%
\bea
\det_{\Xi_{j=-s}(s-1)}\left(  (u+i\partial_{t})^{2}+\square_{s-1}^{+}\right)
&=&\det_{\Xi_{j=s-1}(s-1)}\left(  (u+i\partial_{t})^{2}+\square_{s-1}%
^{+}\right) \\ & =&e^{-\zeta^{\prime}(0)}=\left((\rho(u)+k)^2\right)^{s-1/2} \\ &=&|\left(\rho(u)+k\right)|^{2s-1}.%
\eea

\subsection{Super-determinant} \label{supdet}
In the computation presented in this section, we will assume $j,j_{3},s\in \mathbb{Z}$, the other cases can be worked out in complete analogy. The super-determinant in the case $s>\frac{1}{2}$ is%

\begin{align*}
\frac{\sqrt{\det\limits_{\Xi(s)}(O_{B})\det\limits_{\Xi(s-1)}((\rho(u)+k)^{2}+\square_{s-1}^{+})}}{\det\limits_{\Xi%
(s)}(O_{B})}  &  =\sqrt{\frac{\det\limits_{\Xi(s-1)}((\rho
(u)+k)^{2}+\square_{s-1}^{+})}{\det\limits_{\Xi(s)}(O_{B})}}\\
&  =\sqrt{\frac{%
{\displaystyle\prod\limits_{j=0}^{s-2}}
{\displaystyle\prod\limits_{j_{3}=j+1}^{\infty}}
\left(  (\rho(u)+k)^{2}-j(j+1)+s(s-1)\right)  }{%
{\displaystyle\prod\limits_{j=0}^{s-2}}
{\displaystyle\prod\limits_{j_{3}=j+1}^{\infty}}
\left(  (\rho(u)+k)^{2}-j(j+1)+s(s-1)\right)  }}\\
&  \times\sqrt{\frac{1}{\det\limits_{\Xi_{j=s-1}(s)}(O_{B})}},\\
&  =\sqrt{\frac{1}{\det\limits_{\Xi_{j=s-1}(s)}(O_{B})}}=|(\rho(u)+k)|^{s-\frac
{1}{2}}.
\end{align*}

Notice that in the RHS of the second line, we have a quotient of two identical infinite products. This cancellation, occurs due to the supersymmetric pairing of eigenmodes: cohomological cancellations.%


Let us comment about the particular case $s=1.$ In that case, there are not cohomological cancellations. The reason is, that when $s=1$, only the
space of zero modes $j=s-1$ (or $j=-s$), for the scalar $\phi$, and the
\textquotedblleft chiral\textquotedblright\ spinor $\psi^{+}$ exist. In the
case $s=1$, there is not \textquotedblleft anti-chiral\textquotedblright%
\ square integrable mode $\psi^{-}$ on $\mathbb{H}_2$, because for such spinors the
effective flux is $s-1=0.$

Next, let us compute the super-determinant in the case $s<\frac{1}{2}$%
\begin{align*}
\frac{\sqrt{\det\limits_{\Xi(s)}(O_{B})\det\limits_{\Xi(s-1)}%
((\rho(u)+k)^{2}+\square_{s-1}^{+})}}{\det\limits_{\Xi(s)}(O_{B})}  &
=\sqrt{\frac{\det\limits_{\Xi(s-1)}((\rho(u)+k)^{2}+\square_{s-1}^{+})}%
{\det\limits_{\Xi(s)}(O_{B})}}\\
&  =\sqrt{\frac{%
{\displaystyle\prod\limits_{j=0}^{-s-1}}
{\displaystyle\prod\limits_{j_{3}=-\infty}^{-j-1}}
\left(  (\rho(u)+k)^{2}-j(j+1)+s(s-1)\right)  }{%
{\displaystyle\prod\limits_{j=0}^{-s-1}}
{\displaystyle\prod\limits_{j_{3}=-\infty}^{-j-1}}
\left(  (\rho(u)+k)^{2}-j(j+1)+s(s-1)\right)  }}\\
&  \times\sqrt{\det_{\Xi_{j=-s}(s-1)}\left(  (u+i\partial_{t})^{2}%
+\square_{s-1}^{+}\right)  },\\
&  =\sqrt{\det_{\Xi_{j=-s}(s-1)}\left(  (u+i\partial_{t})^{2}+\square
_{s-1}^{+}\right)  }=|(\rho(u)+k)|^{s-\frac{1}{2}}.
\end{align*}

The final expression coincides with the one of the case $s>\frac{1}{2}.$ However, in the case
of $s<\frac{1}{2}$ the unpaired modes are the \textquotedblleft
anti-chiral\textquotedblright\ square integrable modes $\psi^{-}$, labeled by a
radial number $j=-s$, and perceiving a flux $s-1$ on $\mathbb{H}_2.$

Let us treat separately the case $s=0$. In that case,
there are not square integrable ``chiral'' modes $\psi^{+}$, neither scalar
ones $\phi.$ However, there exist \textquotedblleft
anti-chiral\textquotedblright\ square integrable modes $\psi^{-}$, perceiving
an effective magnetic flux of $-1$ and labeled by radial number $j=0.$ In this
case the regularized super-determinant is given by%
\[
\sqrt{%
{\displaystyle\prod\limits_{j=0}^{0}}
\det\limits_{\Xi(-1)}((\rho(u)+k)^{2}+\square_{-1}^{+})}=|(\rho(u)+k)|^{-\frac
{1}{2}}.
\]

Collecting partial results, not only for the case $j,j_3,s \in \mathbb{Z}$, but for the most general spectrum  $j,j_{3}$ and $s=-\frac{\rho(\mathfrak{m})-q_R}{2}$ obeying ``discreteness" conditions \eqref{QCPrimary},
is straigthforward to obtain the final result for the one loop super-determinant part of the index on $\mathbb{H}_2$%

\[
Z_{1-loop}^{matter}(\mathbb{H}_2,\mathfrak{m},q_{R})=%
{\displaystyle\prod\limits_{\rho:\text{ }s\neq\frac{1}{2}}}\left[
 |C_{reg} \sin\left( \frac{\rho(u)}{2}\right)|\right]  ^{\frac{-\rho(\mathfrak{m})+q_{R}-1}{2}},
\]
where $C_{reg}=-2 i$. 

The restriction to  $s \neq \frac{1}{2},$ is a necessary
condition to have square integrable modes. However, $|(\rho(u)+k)|^{s-\frac{1}{2}}=1$ and consequently  $s\neq \frac{1}{2}$ can very well be ignored and the result for $Z_{1-loop}^{matter}(\mathbb{H}_2,\mathfrak{m},q_{R})$ will be the same.

Interestingly, under GNO conditions 
\[%
 s\in \mathbb{Z} \text{ ~~ or ~~ } s\in \mathbb{Z}+\frac{1}{2},%
\]
the one loop result for the index on $\mathbb{H}_2\times S^1$
\[
Z_{1-loop}^{matter}(\mathbb{H}_2,\mathfrak{m},q_{R})=%
{\displaystyle\prod\limits_{\rho}}
\left[ |C_{reg} \sin\left(  \frac{\rho(u)}{2}\right)|\right]  ^{\frac{-\rho(\mathfrak{m})+q_{R}-1}{2}},
\]
coincides with the square root of the analog result on $S^{2}\times S^{1}$, under the identification of $s_{\mathbb{H}_2}$ with
$ s_{\mathbb{S}^{2}}$, for each mode ($\rho,k$). 
 
 Namely, under GNO quantization conditions
\bea
Z_{1-loop}^{matter}(\mathbb{H}_2,\mathfrak{m},q_{R})&=&%
{\displaystyle\prod\limits_{\rho}}
\left[ |C_{reg} \sin\left( \frac{\rho(u)}{2}\right)|  \right]^{\frac{-\rho(\mathfrak{m})+q_{R}-1}{2}%
} \nonumber\\&=&\sqrt{Z_{1-loop}^{matter}(S^{2},\mathfrak{m},q_{R})}. \label{Statement}%
\eea
Let us remark that we are not forced to impose GNO conditions on $\mathbb{H}_2$. Consequently, we shall not impose GNO quantization conditions 

The one loop determinant of a matter multiplet in the adjoint representation of the gauge group, with R-charge $q_{R}=2$ (which coincides with
the vector multiplet super-determinant, see appendix \ref{VecBZ}) in the presence of flux is%
\bea
Z_{1-loop}^{vector}(\mathbb{H}_2,\mathfrak{m})&=&%
{\displaystyle\prod\limits_{\alpha}}
\left[| C_{reg}\sin\left(  \frac{\alpha(u)}{2}\right)| \right] ^{\frac{-\alpha(\mathfrak{m})+1}{2}}\nonumber \\%
&\sim& \left[{\displaystyle\prod\limits_{\alpha>0}} \sin\left(  \frac{\alpha(u)}{2}\right)^2 \right]^{1/2} \label{VectorInd0}
\\
&=&\sqrt{Z_{1-loop}^{vector}(S^{2},\mathfrak{m})}. \label{VectorInd}
\eea


Notice that the result for the vector multiplet \eqref{VectorInd0}  -- which is independent on $\mathfrak{m}$--  , matches with the result of \cite{David:2016onq} in their
flat space limit $L\rightarrow\infty$ and under the transformation $u \rightarrow i u$. Notice that their transformation of $A_\mu$ under $\epsilon$, $\delta_\epsilon A_\mu$ matches with ours, if only if $A_\mu$ is substituted by $i A_\mu$.

We tried to obtain the cohomological cancellations from $\zeta$-regularisation
of the non zero modes. However, the heat kernel method for spinors of
\cite{Camporesi:1994ga}, definitely, does not respect supersymmetry
(in the case of the discrete spectrum when the absolute value of the total flux felt by the mode must be larger than $\frac{1}{2}$) and breaks the cancellations between
fermions and bosons, unless the normalization associated to the heat kernel of
antichiral modes $\psi^{-}$ is modified in a non elegant way.

\subsection{What about the continuous spectrum?}

So far, we have focused on the contribution of the discrete spectrum to the index. It is time to find out what is the contribution of the continuous spectrum.
The eigenvalues of the relevant differential operators when acting upon the  eigenfunctions of the continuous spectrum are
\bea
O_{B} f_{\Delta_{\left(\lambda,s\right)},j_3}&=&\left(  (\rho(u)+k)^{2}+\lambda^2+\frac{1}{4}+s(s-1)\right) f_{\Delta_{\left(\lambda,s\right)},j_3} \nonumber\\
\left( (\rho(u)+k)^{2}+\square_{s-1}^{+}f_{\Delta_{\left(\lambda,s-1\right)},j_3}\right)&=&\left(  (\rho(u)+k)^{2}+\lambda^2+\frac{1}{4}+s(s-1)\right) f_{\Delta_{\left(\lambda,s-1\right)},j_3}. \nonumber\\ \label{EVc}
\eea

The super-determinant to compute is
\be
\frac{\sqrt{\det(O_{B})\det%
((\rho(u)+k)^{2}+\square_{s-1}^{+})}}{\det(O_{B})}.  \label{Det}
\ee

The determinants in \eqref{Det} are computed by the method of Heat Kernel. Once the $f_{\Delta_{(\lambda,s)}, j_3}$ are normalized
as in \eqref{Norma} the Heat Kernel of an operator $\co$ with eigenvalues $E[\l,s]$ when acting upon $f_{\D_{(\l,s)},j_3}$ is defined as
\bea
\label{eq:spefunc}
K^{(k,\a)}[p,p',\t]=\int_0^{\infty} d\l \sum_{j_3 \in \mathbb{Z}} f^*_{\D_{(\l,s)},j_3}(p) f_{\D_{(\l,s)},j_3}(p') e^{-E[\l,s] \t},
\eea
where $p=\{\theta,\varphi,t\}$ and $p'$ labels the set of coordinates of a given point.  

We do not need the full heat kernel, since for the $\z$-function  all we need is its value at the origin $p=p'=0$.
As in the case of square integrable modes, all the eigenmodes $f_{\D_{(\l,s)},j_3}$ vanish at the origin, except for those with $j_3=s$. After some work, the spectral function $\mu^{(s)}(\lambda)$ is found to be
\bea
\frac{1}{vol_{\mathbb{H}_2} vol_{S^1}}\mu^{(s)}(\l)&:=&\sum_{j_3 \in \mathbb{Z}} f^*_{\D_{(\l,s)},j_3}(0) f_{\D_{(\l,s)},j_3}(0)=f^*_{\D_{(\l,s)},s}(0) f_{\D_{(\l,s)},s}(0) \nonumber\\
&=&\frac{1}{(2\pi)^2}\frac{\lambda  \sinh (2 \pi  \lambda )}{\cosh (2 \pi  \lambda )+\cos (2 \pi  s)}. \label{spectral}
\eea
Having the spectral function in \eqref{spectral} we are ready to compute the $\z$-functions by using the following definition
\be
\z^{\co}(z;s)= \int_0^{\infty}d\l \frac{\m^{(s)}(\l)}{(E_{\co}[\l,s])^z} \label{zeta}.
\ee
From the definition of $\z$-function \eqref{zeta} we compute the relevant determinants
\be
\det \co=e^{-\partial_z \z(0,s)}. \label{det}
\ee

Notice that, the spectral function obeys the following property
\be
\m^{(s)}=\m^{(s-1)}, \label{prop}.
\ee
and the eigenvalues of the operators $O_B$ and $((\rho(u)+k)^{2}+\square_{s-1}^{+})$ upon the respective eigenfunctions $f_{\Delta_{\left(\lambda,s\right)},j_3}$ and $f_{\Delta_{\left(\lambda,s-1\right)},j_3}$ are the same, see equations \eqref{EVc}. From the latter facts and after the definitions  \eqref{zeta} and \eqref{det} it follows the triviality of \eqref{Det}
\be
\frac{\sqrt{\det(O_{B})\det%
((\rho(u)+k)^{2}+\square_{s-1}^{+})}}{\det(O_{B})}=1.  \label{DetUNO}
\ee
In conclusion, the continuous spectrum provides a trivial contribution to the topologically twisted index.

\subsection{GNO condition?} \label{GNO}

The gauge potential representative \eqref{gaugeP} is singular at the contractible cycle $\theta=0$. There are ways to solve this issue. One of them is to impose the holonomy of the gauge potential $\theta=0$ to be in the centre of the group, we shall not resort to this way.   A second way is to simply perform a non trivial gauge transformation with parameter
\be
\L(\varphi)= -s \varphi.
\ee
The new potential
\be
A= s\left(\cosh{\theta} -1 \right)d \varphi, \label{newGaugeP} 
\ee
 is regular at $\theta=0$ and has the same behavior at the boundary $\theta \rightarrow \infty$ as \eqref{gaugeP}. In fact \eqref{newGaugeP} is the analytic continuation of the section on the north chart of the magnetic monopole bundle on $S^2$ to the single chart that covers $\mathbb{H}_2$.
 
  To appreciate the consequence of the non triviality of \eqref{gaugeP} at $\theta=0$, let us comment on its effect on matter. Out of the hierarchy of eigenmodes, 
the only modes that do not vanish at the contractible cycle, are those with $j_3=s$, see equation \eqref{fluxqR}. When cycling around the contractible cycle, these modes exhibit an Aharonov-Bohm phase of
\be
2 \pi s, \label{BAPH}
\ee
 due to the non triviality of \eqref{gaugeP} at $\theta=0$. Should we impose the scalars to be periodic at $\theta=0$, the phase \eqref{BAPH} must be an integer multiplet of $2\pi$. In consequence, periodicity of scalars is not implying GNO quantization conditions. 
  The GNO conditions \cite{Goddard} are
 \begin{eqnarray}
\alpha\left(\mathfrak{m}\right)\in \mathbb{Z}, \label{GN}
\end{eqnarray}
where $\alpha$ is any element of the root lattice of the gauge group $\mathcal{G}$. Condition \eqref{GN} states that
$\mathfrak{m}$ is in the co-root lattice of $\mathcal{G}$.

 If we particularize \eqref{fluxqR} to $\rho=\alpha$ and $q_R=0$ then \eqref{GN} implies
 \be
 s \in \mathbb{Z} \text{ or } \mathbb{Z}+\frac{1}{2}. \label{eqr}
 \ee
As already explained, \eqref{eqr} is consistent with square integrability. However, if $s\in \mathbb{Z}+\frac{1}{2}$, the Aharonov-Bohm phase is an odd multiple of $\pi$ and the scalar modes are multivalued at $\theta=0$. Notice that with the smooth representative \eqref{newGaugeP} such issue is no longer present. In this representation the only modes that do not vanish at $\theta=0$ are those with $j_3=0$. This is because by performing a gauge transformation to a regular gauge potential the $j_3$ gets substituted by $j_3-s$. In fact, in this smooth representation, the Aharonov-Bohm phase at any $\varphi$-cycle becomes an integer multiple of $2\pi$ in virtue of our quantization conditions \eqref{QCPrimary}.

Conditions \eqref{QCPrimary} are less restrictive than GNO conditions as they include not just \eqref{eqr}, but a continuous family of flux configurations. Consequently, we must not restrict our zero locus $BPS[\Gamma]$ by the GNO quantization conditions.  

Notice that in the case of $S^2$, the monopole bundle consists of two charts. In that case, the GNO condition comes from imposing single-valuedness of the structure group transformation that relates the sections at north and south \cite{Nakahara}. In the case of $\mathbb{H}_2$, there is not such a feature.


 We must say, however, that in our case, relaxing GNO conditions has consequences on global gauge invariance  ---see for instance section 2.1 of \cite{Witten2}, to appreciate a related discussion--- . Let us analyze the case of $ABJM$. In that case, indeed, under a large gauge transformation $u\rightarrow u+2\pi$ and $\tilde{u}\rightarrow \tilde{u}+2 \pi$, the Chern-Simons terms $x^{i k \mathfrak{m}_i/2}$ and $\tilde{x}^{-i k \tilde{\mathfrak{m}}_i/2}$, change by a phase of $e^{ \pi i k \mathfrak{m}_i}$ and $e^{- \pi i k \tilde{\mathfrak{m}}_i}$, respectively. Those phases can be absorbed by a couple of topological $U(1)_T$ holonomies \footnote{These terms arise from a couple of mixed Chern-Simons terms of the form \eqref{eq:TCS}. Specifically, from the coupling of the $U(1)_{L,R}$ dynamical vector multiplets and $Q_\epsilon$-spurion vector multiplets: $U(1)_L-{spurion_L}$,  $U(1)_R- spurion_{R}$ . To obtain \eqref{topHol} we have considered the following non trivial v.e.v's ${{\hat{A}_{L 3}^T}}=\pi p$ and ${\hat{A}^T_{R 3}}=-\pi \tilde{p}$ for the $L$ and $R$ spurion multiplets. Allover our discussion, we will fix the spurion fluxes to zero $\mathfrak{t}_L=\tilde{\mathfrak{t}}_R=0$.
 } of the form
\begin{eqnarray}
\prod^{N}_{i=1}  \xi _{p}(\mathfrak{m}_{i}/2) &:=&e^{\pi i\, p \, \sum^{N}_{i=1}\mathfrak{m}_{i}}, \\
\prod^{N}_{i=1}\widetilde{\xi }_{\widetilde{p}}(\widetilde{\mathfrak{m}}_{i}/2) &:=&e^{-\pi i\, \widetilde{p} \sum^N_{i=1}\, %
\widetilde{\mathfrak{m}}_{i}}, \label{topHol}
\end{eqnarray}
after the change of labels
\bea
p  &\rightarrow& p-k, \nonumber\\ 
\tilde{p} &\rightarrow& \tilde{p}-k. \label{changeP}
\eea
Transformation \eqref{changeP}, is a symmetry of the measure $\sum_{p, 
\tilde{p} \in \mathbb{Z}}$, if $k\in \mathbb{Z}$, and consequently, symmetry under large gauge transformations is restored, if we perform an average over $p$ and $\tilde{p}$.  The one loop contributions do not spoil the previous procedure, because the determinants in the case of $ABJM$ are invariant under $u\rightarrow u+2\pi$. 


\paragraph{Comment} In this section, it was proven that if we discard the discrete modes, and consider only the continuous spectrum, the index is trivial. In contradistinction,  the index becomes non trivial when evaluated on square integrable eigenfunctions. The index is somehow encoding information about the tower of normalizable modes.
 Indeed, its one loop contribution is determined by the zeta regularized number of zero modes \cite{Benini:2015eyy} of the operators $\square_{s}^{-}$  --- if $s>\frac{1}{2}$ ---  and $\square_{s-1}^{+}
$ --- if $s<-\frac{1}{2}$---.

\section{The $ABJM$ index on $\mathbb{H}_2\times S^1$} \label{sec:5}

It is time to analyze the $ABJM$ index on $\mathbb{H}_2\times S^1$. In this section, we borrow notation and strategy from section 2.1 of \cite{Benini:2015eyy}. The final scope is to obtain the leading large-$N$ behavior of the index, in terms of flavor fluxes and holonomies. In order to do that, we will show that the corresponding large-$N$ Bethe Ansatz equations (BAE), are equivalent to the ones defined in \cite{Benini:2015eyy}. In fact, the leading large-$N$ solution presented in \cite{Benini:2015eyy} will be a solution to our BAE, and consequently, can be used to evaluate the leading large-$N$ result for the $ABJM$ index on $\mathbb{H}_2\times S^1$. 

Let us start by writing down the localization formula for the $ABJM$ index. The Chern-Simons plus boundary term contribution is
\begin{equation*}
\underset{i=1}{\overset{N}{\prod }}x_{i}^{\frac{1}{2}k \mathfrak{m}_{i}}\widetilde{x}%
_{i}^{-\frac{1}{2}k \widetilde{\mathfrak{m}}_{i}}.
\end{equation*}

After collecting classical and 1-loop contributions, we can write down the expression of the $ABJM$ index on $\mathbb{H}_2\times S^1$
\begin{eqnarray}
Z_{\mathbb{H}_2\times S^{1}} &:= & c \sum\limits_{p,\widetilde{p}\in \mathbb{Z}}\left( 
\frac{1}{N!}\right) ^{2} \int_{|x|=|\tilde{x}|=1} \prod\limits_{i=1}^{N}\frac{%
dx_{i}}{2\pi ix_{i}}\frac{d\widetilde{x}_{i}}{2\pi i\widetilde{x}_{i}} \int\limits^{M_{+}}_{-M_- } \int\limits^{\tilde{M}_{+}}_{-\tilde{M}_- }  \prod\limits_{i=1}^{N} d\mathfrak{m}_i d\tilde{\mathfrak{m}}_i 
\nonumber\\&&\times
\prod\limits_{i=1}^{N}{x_{i}^{k \mathfrak{m}_{i}/2}\widetilde{x}_{i}^{-k%
\widetilde{\mathfrak{m}}_{i}/2} \xi _{p}(\mathfrak{m}_{i}/2) \widetilde{\xi }_{\widetilde{p}}(\widetilde{\mathfrak{m}}_{i}/2)  }\underset{i\neq j}{\overset{N}\prod }\sqrt{\left( 1-\frac{x_{i}}{x_{j}}\right) \left( 1-%
\frac{\widetilde{x}_{i}}{\widetilde{x}_{j}}\right)} 
 \times \nonumber \\
&& \prod\limits_{i,j=1}^{N} \prod\limits_{a=1,2} \left( \pm\left| {\frac{\sqrt{\frac{%
x_{i}}{\widetilde{x}_{j}}y_{a}}}{1-\frac{x_{i}}{\widetilde{x}_{j}}y_{a}}}\right|%
\right) ^{\frac{\mathfrak{m}_{i}-\widetilde{\mathfrak{m}}_{j}-\mathfrak{n}_{a}+1}{2}}  \prod\limits_{b=3,4}\left(\pm\left| \frac{%
\sqrt{\frac{\widetilde{x}_{j}}{x_{i}}y_{b}}}{1-\frac{\widetilde{x}_{j}}{x_{i}%
}y_{b}} \right| \right) ^{\frac{\widetilde{\mathfrak{m}}_{j}-\mathfrak{m}_{i}-\mathfrak{n}_{b}+1}{2}}, \nonumber \\ 
\nonumber\\
\label{ZH2}
\end{eqnarray}
where $c:=\frac{1}{\sum_{p, \tilde{p} \in \mathbb{Z}}1}$ and $\mathfrak{n}_a$ are flavor fluxes. Notice that we have written back the sign degeneracy, mentioned below equation \eqref{absPar}. If we suppose
\be
N \in 2 \mathbb{N}, \label{Even}
\ee
the former signs and the absolute values between parenthesis, become spurious in the contour of integration to be defined below, and consequently we drop them from now on. As we are interested in the large-$N$ limit, \eqref{Even} is enough to our purposes.  

The integration over fluxes and eigenvalues is dictated by the localization principle: they are the zero locus $BPS[\Gamma]$ associated to our contour of field-integration $\Gamma$ and their values are not fixed by boundary conditions. The color holonomies $x_i=e^{i u_i}$, $\tilde{x}_j=e^{i \tilde{u_j}}$ are integrated along $S^1$ as follows from our reality conditions on $\Gamma$:  $Im[u_i]=0$ and $2\pi-$periodicity of the integrand dependence on $u$ and $\tilde{u}$.

 The general idea is to pick up certain residues in the large-$N$ limit. In order to find out the position of the relevant simple poles,  we will need to compute the large-$N$  solution to the very same Bethe ansatz equation of \cite{Benini:2015eyy}. In \cite{Benini:2015eyy} it was suggested that in the large-$N$ limit, the set of Bethe ansatz eigenvalues, which is the set of simple poles enclosed by our contour, condense to a support included in the region \footnote{Even though the evidence presented in \cite{Benini:2015eyy} is quite convincing, it would be nice to have a proof of the absence of extra eigenvalues outside the region $\mathbb{X}(\Delta_a)$. We believe this is a point that deserves further understanding, but we will leave that analysis for future work.   }
 \bea
\mathbb{X}(\Delta_a): =\left\{\left(\tilde{u}_j-u_i\right)\in \mathbb{C}: \left\{\begin{tabular}{ccc}0<$-\mathbb{R}e [\tilde{u}_j -u_i]+\Delta_a<2\pi$ &\text{ ~~ if~~ } &   $a=1,2$ \\
$0<~~\mathbb{R}e[\tilde{u}_j -u_i]+\Delta_a<2\pi $&\text{ ~~ if~~ } & $  a=3,4$ \end{tabular}\right\} \right\}. \label{X}
\eea
The region $\mathbb{X}(\Delta_a)$ is the union of the regions covered by angles of the $(i,j)$-complex planes with coordinate $\tilde{x}_{ji}:=e^{i (\tilde{u}_j -u_i)}$. Each $(i,j)$-angle is defined as
\be
\max\left( -\Delta_3,-\Delta_4, \max\left(\Delta_1,\Delta_2\right)-2\pi \right)<\arg\left(\tilde{x}_{ji}\right)<\min\left(\Delta_1,\Delta_2 ,2\pi -\max{\left(\Delta_3,\Delta_4\right)}\right).
\ee
 To compute our residues we have two possibilities. Either we deform the $(ij)$-$S^1$: $|\tilde{x}_{ji}|=1$, to the perimeter of the inner region of the $(i,j)$-angle including the origin, or we deform it to the perimeter of the outer region, and close the contour at infinity $\tilde{x}_{ji}=\infty$ \footnote{By inner (resp. outer) region we mean the region of the $(i,j)$-angle that is in (resp. out) of the $(i,j)$-$S_1$.}. Both choices are equivalent, provided we include the respective ``boundary contribution'' at $\tilde{x}_{ji}=0$ or $\tilde{x}_{ji}=\infty$, depending on the case. The ``inner'' choice corresponds to selecting poles with $\mathbb{I}m(\tilde{u}_j-u_i)>0$. The ``outer'' choice corresponds to poles with $\mathbb{I}m(\tilde{u}_j-u_i)<0$. Notice that for the large $N$ solution of our interest, an outer $(i,j)$-pole implies the presence of an inner $(j,i)$-pole and $vice~versa $ (see equation (2.39) of \cite{Benini:2015eyy}). In fact in the limit $N\rightarrow \infty$ all of such poles will condense either at $0$ or $\infty$, except for the case $i=j$. In the particular case $i=j$, the poles condense to an arc of $(i,i)$-$S^1$. For each $(i,j)$, we will choose to deform the $(i,j)$-$S^1$ contour in the way that encloses the poles at the ``bulk'' (these, are the poles whose positions are the eigenvalues in equation (2.39) of \cite{Benini:2015eyy}). 
  
The projection of the domain $\mathbb{X}(\Delta_a)$ upon our integration contour, which is obtained by demanding $\mathbb{I}m[u]=\mathbb{I}m[\tilde{u}]=0$ on the former, is a parametrization of the maximally connected region without Coulomb branch singularities \footnote{In fact in integrating $x_i$ over $S_1$ we should avoid colliding with such singularities, either by slightly deforming the contour, or by turning on an infinitesimal mass regulator in the matter one loop determinant.}.
Actually, the intersection of the boundary of the complex region $\mathbb{X}(\Delta_a)$, $\partial \mathbb{X}(\Delta_a)$ with our contour of integration $S^1$, is a parametrization of the domain of such singularities. By a singularity of the Coulomb branch we mean a point $(u_i,\tilde{u}_j)$ such that the quantity that defines the one loop contributions
\begin{eqnarray}
\prod_{a=1,2}\left( \frac{\sqrt{\frac{
x_{i}}{\widetilde{x}_{j}}y_{a}}}{1-\frac{x_{i}}{\widetilde{x}_{j}}y_{a}}\right)^{-1} \prod_{b=3,4} \left( \frac{%
\sqrt{\frac{\widetilde{x}_{j}}{x_{i}}y_{b}}}{1-\frac{\widetilde{x}_{j}}{x_{i}%
}y_{b}} \right)&=&  \frac{\sin\frac{\left(-\tilde{u}_j+ {u}_i+\Delta_1\right)}{2}\sin\frac{\left(-\tilde{u}_j+u_i+\Delta_2\right)}{2}}{\sin\frac{\left(\tilde{u}_j-u_i+\Delta_3\right)}{2}\sin\frac{\left(\tilde{u}_j-u_i+\Delta_4\right)}{2}},
\end{eqnarray}
becomes $0$ or $\infty$.

Notice that, as we are not imposing the GNO conditions, we have to integrate over the values of fluxes $\mathfrak{m}_i$ and $\tilde{\mathfrak{m}}_i$ along the Cartan directions. As we have already stated, the fluxes $\mathfrak{m}$ and $\tilde{\mathfrak{m}}$ are non-normalizable modes, even though they are in $BPS[\Gamma]$. In that respect, our approach is reminiscent of the one advocated in \cite{Rey}. In \cite{Rey}, integration over non-normalizable modes belonging to the zero locus of the relevant supercharge, was suggested for the localization on $\mathbb{H}_2\times S^1$. Although the localization  performed there, was on the branched sphere, the integration over the Coulomb branch parameter completes the nice picture suggested by (4.15) of \cite{Rey} \footnote{A second possibility we will not explore in this work, is to fix the values of non normalizable modes in $Q_\epsilon[\Gamma]$ to specific values. However, in order to match the final result to the supergravity dual, an extremization procedure should be engineered for those values. In some sense, integration over these non normalizable color - not flavor- modes is such sort of extremization. In this second, more open, line of thought, perhaps one could relax our reality condition on $u_i$ and simply define the latter extremization as integration over the complex and more abstract Jeffrey-Kirwan (JK) contour. It would be quite remarkable  if with such an alternative approach, one obtains the same result for the index. In that case, the approach followed in this manuscript, would provide a less abstract viewpoint of the JK contour. We suspect this is indeed the case, but as it is not the final goal of our study, we shall not check so in this manuscript. }.

We use a couple of very large cut-offs $ M_{\pm}, ~ \tilde{M}_{\pm}>0$, because the volume of the moduli space of fluxes $(\mathfrak{m}, \tilde{\mathfrak{m}})$ is infinite. After computing the integral over the holonomies for fixed values of $M_{\pm}$ and $\tilde{M}_{\pm}$, we are free to send one and only one, of either $M_+$(resp. $\tilde{M}_+$) or $M_-$(resp. $\tilde{M}_-$), to infinity. The other one, remains as a regulator that we will redefine as $M$ (resp. $\tilde{M}$). Thereafter, we pick up the residues of the analytical continuation of the regulated integrand, the final result will be independent on $M$ and $\tilde{M}$, and we are free to take $M, ~\tilde{M}\rightarrow \infty$ on such residues 
 \footnote{The consequences of using this regularization procedure are somehow reminiscent of the consequences of applying the Jeffrey-Kirwan recipe in \cite{Benini:2015eyy}.  }.

Next, we shall show independence of the regulated expression on the cutoffs $M$ and $\tilde{M}$, but before proceeding, let us remind a couple of conditions that we have implicitly used so far. The topological twisting condition is
\begin{eqnarray}
\sum_a \mathfrak{n}_a=2. \label{TTcondition}
\end{eqnarray}
To understand how \eqref{TTcondition} is the topological twisting condition, please, refer to section 2.1 of \cite{Benini:2015eyy}.
From conservation of flavor symmetry, it follows that
\begin{eqnarray}
\prod_{a} y_a=1. \label{globalS}
\end{eqnarray}
For clarity, it is convenient to re-organize the RHS of \eqref{ZH2} as follows
\begin{eqnarray}
&&c \sum\limits_{p,\widetilde{p} \in \mathbb{Z}}\left( \frac{1}{N!}\right) \int \underset{i=1}{\overset{N}{\prod }}%
\frac{dx_{i}}{2\pi ix_{i}}\frac{d\widetilde{x}_{i}}{2\pi i\widetilde{x}_{i}}%
\int\limits_{-M_-}^{M_+} \int\limits_{-\tilde{M}_-}^{\tilde{M}_+}\underset{i=1}{\overset{N}{%
\prod }}d\mathfrak{m}_{i}d\widetilde{\mathfrak{m}}_{i}\left( {\underset{i\neq j}{\overset{N}{\prod }}\sqrt{\left( 1-\frac{x_{i}}{%
x_{j}}\right) \left( 1-\frac{\widetilde{x}_{i}}{\widetilde{x}_{j}}\right)} A}%
\right)\nonumber  \\
&&\;\times \left( \underset{i=1}{\overset{N}{\prod }}\exp[\Upsilon _{i}(x,%
\widetilde{x})\mathfrak{m}_{i}]\right) \left( \underset{j=1}{\overset{N}{\prod }}\exp[%
\widetilde{\Upsilon _{j}}(x,\widetilde{x})\widetilde{\mathfrak{m}}_{j}]\right), 
\end{eqnarray}
where
\begin{eqnarray}
\Upsilon _{i}(x,\widetilde{x}) &=&\log \left( x_{i}^{k} e^{2 \pi i p} \prod%
\limits_{j=1}^{N} \frac{\sqrt{\frac{x_{i}}{\widetilde{x}_{j}}y_{1}\frac{x_{i}%
}{\widetilde{x}_{j}}y_{2}}}{\left( 1-\frac{x_{i}}{\widetilde{x}_{j}}%
y_{1}\right) \left( 1-\frac{x_{i}}{\widetilde{x}_{j}}y_{2}\right) }\frac{%
\left( 1-\frac{\widetilde{x}_{j}}{x_{i}}y_{3}\right) \left( 1-\frac{%
\widetilde{x}_{j}}{x_{i}}y_{4}\right) }{\sqrt{\frac{\widetilde{x}_{j}}{x_{i}}%
y_{3}\frac{\widetilde{x}_{j}}{x_{i}}y_{4}}} \right)^{1/2}\nonumber  \\
&=&\log \left( {x_{i}^{k}e^{2 \pi i p}  \overset{N}{\underset{j=1}{\prod }}\frac{%
\left( 1-y_{3}\frac{\widetilde{x}_{j}}{x_{i}}\right) \left( 1-y_{4}\frac{%
\widetilde{x}_{j}}{x_{i}}\right) }{\left( 1-y_{1}^{-1}\frac{\widetilde{x}_{j}%
}{x_{i}}\right) \left( 1-y_{2}^{-1}\frac{\widetilde{x}_{j}}{x_{i}}\right) }}%
\right)^{1/2} \nonumber \\
&:=&\frac{1}{2}\log \left( e^{i B_{i} }\right)  \label{eBi}
\end{eqnarray}
and
\begin{equation*}
\widetilde{\Upsilon }_{j}(x,\widetilde{x})=\frac{1}{2} \log \left( e^{i \widetilde{B%
}_{j}}\right),
\end{equation*}
with
\bea
e^{i \widetilde{B}_{j}}&:=&\widetilde{x}_{j}^{k}e^{2 \pi i \tilde{p}}\overset{N}{\underset{i=1}{\prod }}%
\left. \frac{\left( 1-y_{3}\frac{\widetilde{x}_{j}}{x_{i}}\right) \left( 1-y_{4}%
\frac{\widetilde{x}_{j}}{x_{i}}\right) }{\left( 1-y_{1}^{-1}\frac{\widetilde{%
x}_{j}}{x_{i}}\right) \left( 1-y_{2}^{-1}\frac{\widetilde{x}_{j}}{x_{i}}%
\right) }\right. , \label{expB} \\
A&:=&\prod\limits_{i,j=1}^{N}\left(\prod\limits_{a=1,2}\left( \left. {\frac{\sqrt{\frac{%
x_{i}}{\widetilde{x}_{j}}y_{a}}}{1-\frac{x_{i}}{\widetilde{x}_{j}}y_{a}}}%
\right.\right) ^{\frac{-\mathfrak{n}_{a}+1}{2}}\prod\limits_{b=3,4}\left(\left.{ \frac{%
\sqrt{\frac{\widetilde{x}_{j}}{x_{i}}y_{b}}}{1-\frac{\widetilde{x}_{j}}{x_{i}%
}y_{b}} }\right.\right) ^{\frac{-\mathfrak{n}_{b}+1}{2}}\right).  \label{A}
\eea
Definitions \eqref{eBi}, \eqref{expB} and \eqref{A} are the ones in equations (2.21) and (2.20) of \cite{Benini:2015noa}. 

 After evaluating the integral over fluxes, we obtain
 \begin{eqnarray}
Z_{\mathbb{H}^2\times {S}^{1}}(\mathfrak{n},y,M)&=&c \sum\limits_{p,\widetilde{p} \in \mathbb{Z}}\left( \frac{1}{N!}\right)^2 \int_{|x|=|\tilde{x}|=1} \underset{i=1}{%
\overset{N}{\prod }}\frac{dx_{i}}{2\pi ix_{i}}\frac{d\widetilde{x}_{i}}{%
2\pi i\widetilde{x}_{i}}\left( {\underset{i\neq j}{\overset{N}{\prod }}%
\sqrt{\left( 1-\frac{x_{i}}{x_{j}}\right) \left( 1-\frac{\widetilde{x}_{i}}{%
\widetilde{x}_{j}}\right)} A}\right)\nonumber \\ &\times&\underset{i=1}{\overset{N}{ \prod 
}}\frac{s_{\Upsilon_i} \exp \bigg[ s_{\Upsilon_i} \Upsilon _{i}(x,\widetilde{x}) M \bigg]
}{\frac{1}{2}\log \left( e^{i B_{i}}\right) }\times \underset{j=1}{\overset{N}{\prod }}\frac{s_{\tilde{\Upsilon}_j}\exp\bigg[s_{\tilde{\Upsilon}_j}  \widetilde{%
\Upsilon _{j}}(x,\widetilde{x}) \tilde{M}\bigg]}{\frac{1}{2} \log \left( e^{i\widetilde{B}_{j}}\right) }. \label{RemInt}
\end{eqnarray}
where
\be
s_{{\Upsilon}_i}:=\text{ sign }\mathbb{R}e{\Upsilon}_i(x,\tilde{x}), ~~~s_{\tilde{\Upsilon}_j}:=\text{ sign } \mathbb{R}e {\tilde{\Upsilon}_j(x,\tilde{x})}. \nonumber
\ee
In \eqref{RemInt} we have already taken $M_{- \,\text{or}\, +}\rightarrow \infty$ and $\tilde{M}_{- \,\text{or}\, +}\rightarrow \infty$.

 The next step is to evaluate the residues of the analytical continuation of the regulated integrand in \eqref{RemInt}, at the simple poles enclosed by our contour.  Such poles are located at positions $x_*$ and $\tilde{x}_*$, defined by the eigenvalues of the BAE
\be
 e^{i%
 B_{i}(x_{\ast },\widetilde{x}_{\ast })}=1, ~~ e^{i%
\widetilde{B}_{i}(x_{\ast },\widetilde{x}_{\ast })}=1. \label{BAEq}
\ee
 Solutions of \eqref{BAEq} are not onto our integration contour \footnote{This is because the imaginary part of both eigenvalues $u$ and $\tilde{u}$ is different from $0$ (see equation (2.39) of \cite{Benini:2015eyy}).}, but are enclosed by it.
In this way we solve the remaining integrals in \eqref{RemInt}.
 Notice that we are naively focusing on the contribution coming from simple poles generated by the BAE \eqref{BAEq}. Next, we shall see how the solution to \eqref{BAEq} is independent on the topological holonomies $p$ and $\tilde{p}$. The integrand of \eqref{RemInt} is also independent on $p$ and $\tilde{p}$. Consequently, the average over $p$ and $\tilde{p}$ will be trivial.

The final result for the index is
\begin{equation}
Z_{\mathbb{H}_2 \times S^{1}}(\mathfrak{n},y)=\prod\limits_{a=1}^{4}y_{a}^{-\frac{
N^{2}n_{a}}{4}}\sum\limits_{I \in BAE}\frac{ 2^{2N}}{\det \mathbb{B}}\left({\frac{{\overset{N}{\underset{i=1}{\prod }}x_{*i}^{N}\widetilde{x}_{*i}^{N}%
\underset{%
i\neq j}{\overset{N}{\prod }}}\left( 1-\frac{x_{*i}}{x_{*j}}\right) \left( 1-\frac{%
\widetilde{x}_{*i}}{\widetilde{x}_{*j}}\right) }{\underset{i\neq j}{\overset{N}%
{\prod }}\prod\limits_{a=1}^{2}\left( \widetilde{x}_{*j}-y_{a}x_{*i}\right)
^{1-n_{a}}\prod\limits_{a=3}^{4}\left( x_{*i}-y_{a}\widetilde{x}_{*j}\right)
^{1-n_{a}}}}\right)^{1/2}, \label{Zh2S1BH}
\end{equation}
where
\begin{equation}
\det \mathbb{B}:=\frac{\partial \left( e^{i B_{j}},e^{i\widetilde{B}_{j}}\right) }{%
\partial \left( \log x_{l},\log \widetilde{x}_{l}\right) }=%
\begin{pmatrix}
x_{l}\frac{\partial e^{i {B_{j}}}}{\partial x_{l}} & \widetilde{x}_{l}%
\frac{\partial e^{i{B_{j}}}}{\partial \widetilde{x}_{l}} \\ 
x_{l}\frac{\partial e^{i{\widetilde{B}_{j}}}}{\partial x_{l}} & 
\widetilde{x}_{l}\frac{\partial e^{i{\widetilde{B}_{j}}}}{\partial 
\widetilde{x}_{l}}%
\end{pmatrix}%
_{2N\times 2N}.
\end{equation}
Notice that the cut-off dependence disappeared in \eqref{Zh2S1BH} due to the BAE. 



 We are interested in the large-$N$ limit of $Z_{\mathbb{H}_2\times S^1}.$ In that limit,
there is a unique BAE solution thence the summation over the label $%
I$ becomes spurious. 

It is convenient to define
\begin{equation}
D(z):=\frac{\left( 1-z y_{3}\right) \left( 1-z y_{4}\right) }{\left(
1-z y_{1}^{-1}\right) \left( 1-z y_{2}^{-1}\right) }.\label{D}
\end{equation}

In terms of $D(z)$, the LHS of the BAE are
\begin{equation*}
e^{i {B_{i}(x_{\ast },\widetilde{x}_{\ast })}}=x_{i}^{k} e^{2 \pi p} \overset{N}{%
\underset{j=1}{\prod }}\left. D\left( \frac{\widetilde{x}_{j}}{x_{i}}\right)\right.
,\;\;e^{i {\widetilde{B}_{j}(x_{\ast },\widetilde{x}_{\ast })}}=%
\widetilde{x}_{j}^{k}e^{2\pi \tilde{p}}\overset{N}{\underset{i=1}{\prod }}\left. D\left( 
\frac{\widetilde{x}_{j}}{x_{i}}\right)\right. .
\end{equation*}

In terms of the quantity
\begin{equation*}
G_{ij}:=\left. \frac{\partial \log D}{\partial \log z}\right\vert _{z=%
\widetilde{x}_{j}/x_{i}}, 
\end{equation*}
the matrix $\mathbb{B}$ takes the form
\begin{equation}
\left. \mathbb{B}\right\vert _{BAE}=%
\begin{pmatrix}
\delta _{jl}[k-\overset{N}{\underset{m=1}{\sum }}G_{jm}] & G_{jl} \\ 
-G_{lj} & \delta _{jl}[k+\overset{N}{\underset{m=1}{\sum }}G_{mj}]%
\end{pmatrix}.%
\label{B}\end{equation}
The next step is to write down the BAE in ``angular'' coordinates $u_i$, $\tilde{u}_i$, and $\Delta_a$, which are defined from
\begin{equation*}
x_{i}=e^{iu_{i}},\;\widetilde{x}_{j}=e^{i\widetilde{u}_{j}},\;y_{a}=e^{i%
\Delta _{a}}.
\end{equation*}
In these coordinates, the constraint $\prod\limits_{a}y_{a}=1$ looks like $\sum\limits_{a}\Delta _{a}=0 ~(%
\text{mod}\;2\pi ).$ 

In ``angular'' coordinates, the BAE \eqref{BAEq} are
\begin{eqnarray}
0 &=&ku_{i}+i\sum_{j=1}^{N}\left[
\sum_{a=3,4}Li_{1}\left( e^{i\left( \widetilde{u}_{j}-u_{i}+\Delta
_{a}\right) }\right) -\sum_{a=1,2}Li_{1}\left( e^{i\left( \widetilde{u}%
_{j}-u_{i}-\Delta _{a}\right) }\right) \right] -2\pi \left(n_{i} -p \right), \nonumber \\
0 &=&k\widetilde{%
u}_{j}+i\sum_{j=1}^{N}\left[ \sum_{a=3,4}Li_{1}\left( e^{i\left( \widetilde{u%
}_{j}-u_{i}+\Delta _{a}\right) }\right) -\sum_{a=1,2}Li_{1}\left( e^{i\left( 
\widetilde{u}_{j}-u_{i}-\Delta _{a}\right) }\right) \right] -2\pi \left( \widetilde{%
n}_{j}-\tilde{p} \right),\nonumber \\ \label{BAEangular}
\end{eqnarray}
with $ n_{i}, ~ \tilde{n}_j \in \mathbb{Z}$.  Equations \eqref{BAEangular} are the same BAE given in (2.32) of \cite{Benini:2015eyy}.

To fix the values of $n_i$ and $\tilde{n}_j$, we use the identity
\begin{equation*}
Li_{1}\left( e^{iu}\right) -Li_{1}\left( e^{-iu}\right) =-iu+i\pi 
\end{equation*}
together with the assumption of absence of ``long range interaction'' \cite{Benini:2015eyy}. The latter condition, implies
\begin{eqnarray}
  2\pi n_{i}  &=&2 \pi p +\left( \sum_{a}\Delta _{a}-4\pi \right)
\sum_{j}\Theta(\mathbb{I}{m}\left( u_{i}-\widetilde{u}_{j}\right) ), \label{LRange1} \\
 2\pi \widetilde{n}_{i}  &=&2 \pi \tilde{p}+\left( \sum_{a}\Delta
_{a}-4\pi \right) \sum_{i} \Theta(\mathbb{I}{m}\left( u_{i}-\widetilde{u}%
_{j}\right) ). \label{LRange2}
\end{eqnarray}
We can solve the constraints \eqref{LRange1} and \eqref{LRange2}, for $n_i$ and $\tilde{n}_j$, with the choice
\begin{equation}
\sum_{a}\Delta _{a}=2\pi . \label{DeltaConstP}
\end{equation}
It is key to observe that \eqref{LRange1} and \eqref{LRange2}, imply that solutions to the BAE are independent of the topological holonomies $p$, $\tilde{p}$. Additionally, \eqref{B} is also independent of $p$ and $\tilde{p}$ and we conclude that the average on $p$ and $\tilde{p}$ can be substituted by one. 

The solution of \eqref{BAEq}, obeying \eqref{DeltaConstP}, is precisely the one used in \cite{Benini:2015eyy}, to obtain their quite nice result, (2.89), in the large-$N$ limit. In the next section, for completeness, we evaluate the aforementioned solution. We have tried other choices, such as $\sum_a \Delta_a=0$. However, as pointed out in \cite{Benini:2015eyy}, there are always some issues with the potential solutions. 

\subsection{Large-$N$ behavior of the index} 

From now on, we take the limit $N\rightarrow \infty $, assume the Chern-Simons level $k=1$,  introduce the density of eigenvalues $\rho (t)=\frac{1}{N}%
\sum_{i=1}^{N}\delta (u_{i}-t)$ and the quantity $\delta v(t)$, as precisely done in
section 2.3 of \cite{Benini:2015eyy}. In this continuous limit, the BAE arise from the
variations of the auxiliary Lagrangian
\begin{eqnarray}
\frac{\mathcal{V}}{iN^{\frac{3}{2}}} &=&\int dt\left[ t\rho (t)\delta v(t)+\rho
(t)^{2}\left( \sum\limits_{a=3,4}g_{+}(\delta v(t)+\Delta
_{a})-\sum\limits_{a=1,2}g_{-}(\delta v(t)-\Delta _{a})\right) \right]  \nonumber\\
&&-\mu \left[ \int dt\rho (t)-1\right] -\frac{i}{N^{1/2}}\int dt\rho (t)%
\left[ \sum\limits_{a=3,4}Li_{2}\left( e^{i\left( \delta v(t)+\Delta
_{a}\right) }\right) -\sum\limits_{a=1,2}Li_{2}\left( e^{i\left( \delta
v(t)-\Delta _{a}\right) }\right) \right] , \label{BAEs} \nonumber\\
\end{eqnarray}
where
\bea
g_{\pm}(u):=\frac{u^3}{6} \mp \frac{\pi}{2} u^2+\frac{\pi^2}{3}u.
\eea
It is easy to follow the steps in \cite{Benini:2015eyy}. Indeed
for $\sum_{a}\Delta _{a}=2\pi $ and under the assumptions 
\bea
\mu >0,\;\exists
\tilde{t}:\delta v(\tilde{t})=0,\;\Delta _{1}<\Delta _{2} < \Delta _{3}<\Delta _{4}, \label{BAEsConditions}
\eea
together with \eqref{X}
\bea
0<-\delta v(t)+\Delta_a<2\pi &\text{ ~~ if~~ } &   a=1,2, \nonumber \\
0<~~\delta v(t)+\Delta_a<2\pi &\text{ ~~ if~~ } &   a=3,4,
\eea
the large-$N$ relevant part of the  solution to the continuous limit of the BAE, coming from
Lagrangian \eqref{BAEs} is

\begin{equation*}
\rho (t):=\left\{ 
\begin{array}{ccc}
-\frac{\mu +\text{$\Delta_3 $} t}{(\text{$\Delta_1 $}+\text{$\Delta_3 $}) (\text{$\Delta_2
   $}+\text{$\Delta_3 $}) (\text{$\Delta_3$}-\text{$\Delta_4 $})}& $if$ & t_{0}<t<t_{1} \\ 
\frac{2 \pi  \mu +t (\text{$\Delta_3 $} \text{$\Delta_4 $}-\text{$\Delta_1 $}
   \text{$\Delta_2 $})}{(\text{$\Delta_1 $}+\text{$\Delta_3 $}) (\text{$\Delta_1
   $}+\text{$\Delta_4 $}) (\text{$\Delta_2 $}+\text{$\Delta_3 $}) (\text{$\Delta_2
   $}+\text{$\Delta_4 $})}& $if$ & t_{1}<t<t_{2} \\ 
\frac{\text{$\Delta_1 $} t-\mu }{(\text{$\Delta_1 $}-\text{$\Delta_2 $}) (\text{$\Delta
   $1}+\text{$\Delta_3 $}) (\text{$\Delta_1 $}+\text{$\Delta_4 $})}& $if$ & t_{2}<t<t_{3}%
\end{array}%
\right. .
\end{equation*}

\begin{equation*}
\delta v(t):=\left\{ 
\begin{array}{ccc}
-\Delta_3 + e^{-N^{1/2} Y_3(t)} & $if$ & t_{0}<t<t_{1} \\ 
\frac{\mu  (\text{$\Delta_1$} \text{$\Delta_2$}-\text{$\Delta_3$} \text{$\Delta_4 $})+t
   (\text{$\Delta_1 $} \text{$\Delta_2 $} \text{$\Delta_3 $}+\text{$\Delta_1 $}
   \text{$\Delta_2 $} \text{$\Delta_4 $}+\text{$\Delta_1 $} \text{$\Delta_3 $}
   \text{$\Delta_4 $}+\text{$\Delta_2 $} \text{$\Delta_3 $} \text{$\Delta_4 $})}{2 \pi  \mu
   +\text{$\Delta_1 $} \text{$\Delta_2 $} (-t)+\text{$\Delta_3 $} \text{$\Delta_4 $}\; t}& $if$ & t_{1}<t<t_{2} \\ 
\Delta_1-e^{-N^{1/2} Y_1(t)}& $if$ & t_{2}<t<t_{3}%
\end{array}%
\right. .
\end{equation*}

\bigskip 
\begin{eqnarray}
Y_{1}(t) &=&\left\{ 
\begin{array}{ccc}
\frac{(\text{$\Delta_1 $}+\text{$\Delta_4 $}) (\mu +\text{$\Delta_3 $} t)}{(\text{$\Delta_2
   $}+\text{$\Delta_3 $}) (\text{$\Delta_3 $}-\text{$\Delta_4 $})}+t& $if$ & t_{0}<t<t_{1} \\ 
0& $if$ & t_{1}<t<t_{2} \\ 
\frac{\mu -\text{$\Delta_2 $} t}{\text{$\Delta_1 $}-\text{$\Delta_2 $}}& $if$ & t_{2}<t<t_{3}%
\end{array}%
\right.  \nonumber \\
Y_{2}(t)&=&\left\{ 
\begin{array}{ccc}
\frac{(\text{$\Delta_2 $}+\text{$\Delta_4 $}) (\mu +\text{$\Delta_3$} t)}{(\text{$\Delta_1
   $}+\text{$\Delta_3 $}) (\text{$\Delta_3 $}-\text{$\Delta_4 $})}+t& $if$ & t_{0}<t<t_{1} \\ 
0& $if$ & t_{1}<t<t_{2} \\ 
t-\frac{(\text{$\Delta_2 $}+\text{$\Delta_3 $}) (\text{$\Delta_2 $}+\text{$\Delta_4 $})
   (\text{$\Delta_1 $} t-\mu )}{(\text{$\Delta_1 $}-\text{$\Delta_2 $}) (\text{$\Delta_1
   $}+\text{$\Delta_3 $}) (\text{$\Delta_1 $}+\text{$\Delta_4 $})}& $if$ & t_{2}<t<t_{3}%
\end{array}%
\right. \nonumber  \\
Y_{3}(t) &=&\left\{ 
\begin{array}{ccc}
\frac{\mu +\text{$\Delta_4 $} t}{\text{$\Delta_3 $}-\text{$\Delta_4 $}}& $if$ & t_{0}<t<t_{1} \\ 
0& $if$ & t_{1}<t<t_{2} \\ 
-\frac{(\text{$\Delta_2 $}+\text{$\Delta_3 $}) (\text{$\Delta_1 $} t-\mu )}{(\text{$\Delta
   $1}-\text{$\Delta_2 $}) (\text{$\Delta_1 $}+\text{$\Delta_4 $})}-t& $if$ & t_{2}<t<t_{3}%
\end{array}%
\right. \nonumber  \\
Y_{4}(t) &=&\left\{ 
\begin{array}{ccc}
\frac{(\text{$\Delta_1 $}+\text{$\Delta_4 $}) (\text{$\Delta_2 $}+\text{$\Delta_4 $}) (\mu
   +\text{$\Delta_3 $} t)}{(\text{$\Delta_1 $}+\text{$\Delta_3 $}) (\text{$\Delta_2
   $}+\text{$\Delta_3 $}) (\text{$\Delta_3 $}-\text{$\Delta_4 $})}-t& $if$ & t_{0}<t<t_{1} \\ 
0& $if$ & t_{1}<t<t_{2} \\ 
-\frac{(\text{$\Delta_2 $}+\text{$\Delta_4$}) (\text{$\Delta_1$} t-\mu )}{(\text{$\Delta_1
   $}-\text{$\Delta_2 $}) (\text{$\Delta_1 $}+\text{$\Delta_3$})}-t& $if$ & t_{2}<t<t_{3}%
\end{array}%
\right. ,\label{BAEsol}
\end{eqnarray}
with
\begin{equation*}
t_{0}=-\frac{\mu }{\text{$\Delta_3 $}},\;t_{1}=-\frac{\mu }{\text{$\Delta_4 $}}\;,t_{2}=\frac{\mu }{\text{$\Delta_2 $}}, \; t_{3}= \frac{\mu }{\text{$\Delta_1 $}}\;.
\end{equation*}
From \eqref{BAEsConditions} it follows the ordering of transition times

\begin{equation*}
t_{0}<t_{1}<t_{2}<t_{3},\;\rho >0.
\end{equation*}

From the normalization condition $\int\limits_{t_{0}}^{t_{3}}dt\rho (t)=1$ it follows that
\begin{equation*}
\mu =\sqrt{2\Delta _{1}\Delta _{2}\Delta _{3}\Delta _{4}}.
\end{equation*}

To obtain the leading free energy in the limit $N \rightarrow \infty $,
one evaluates \eqref{Zh2S1BH} at the BAE solution \eqref{BAEsol}. The final result can be easily inferred, given the fact that our BAE
solution is the same one found in \cite{Benini:2015eyy}, for the case $k=1$,  $%
\sum_{a}\Delta _{a}=2\pi .$ In the latter case, and from the fact that the
summand in \eqref{Zh2S1BH} is the square root 
of the one in eq
(2.24) of \cite{Benini:2015eyy}, it results that
\begin{equation*}
\text{ $\mathbb{R}$e} \log \left. Z_{\mathbb{H}_{2}\times {S}^{1}}\right\vert _{\text{Large }N\text{ BAE
solution}}=\frac{1}{2}\text{ $\mathbb{R}$e} \log \left. Z_{{S}^{2}\times{S}^{1}}\right\vert _{\text{%
Large }N\text{ BAE solution}}+\text{ sub.  terms}
\end{equation*}
 Finally, one can arrive to the result
\begin{eqnarray}
\text{ $\mathbb{R}$e} \log Z_{\mathbb{H}_{2}\times S^{1}}^{k=1} &=&-F_{\mathbb{H}_{2}\times
S^{1}}^{k=1}(\mathfrak{n},\Delta )\nonumber\\&=&-\left( \frac{1}{2}\right) \times \frac{N^{\frac{3}{2%
}}}{3}\sqrt{2 \Delta _{1}\Delta _{2}\Delta _{3}\Delta _{4}}\sum%
\limits_{a=1}^{4}\frac{n_{a}}{\Delta _{a}}+\text{sub. terms}.\;  \label{SCFT}\\
\;\sum_{a}\Delta _{a} &=&2\pi .
\end{eqnarray}

After extremizing with respect to $\Delta_1$, $\Delta_2$ and $\Delta_3$, we obtain the following relation between fluxes and holonomies  
\begin{eqnarray}
\mathfrak{n}_1&=&\frac{\text{$\Delta_1 $} (\text{$\Delta_1 $}-\pi )}{\text{$\Delta_1 $}^2+\text{$\Delta_1 $}
   (\text{$\Delta_2$}+\text{$\Delta_3$})-2 \pi  (\text{$\Delta_1$}+\text{$\Delta_2$}+\text{$\Delta_3$})+\text{$\Delta_2 $}^2+\text{$\Delta_2 $} \text{$\Delta_3
   $}+\text{$\Delta_3 $}^2+\pi ^2}, \nonumber\\
\mathfrak{n}_2 &=&\frac{\text{$\Delta_2 $} (\text{$\Delta_2 $}-\pi )}{\text{$\Delta_1 $}^2+\text{$\Delta_1 $}
   (\text{$\Delta_2 $}+\text{$\Delta_3 $})-2 \pi  (\text{$\Delta_1 $}+\text{$\Delta_2
   $}+\text{$\Delta_3 $})+\text{$\Delta_2 $}^2+\text{$\Delta_2 $} \text{$\Delta_3$}+\text{$\Delta_3 $}^2+\pi ^2}, \nonumber\\
\mathfrak{n}_3&=& \frac{\text{$\Delta_3 $} (\text{$\Delta_3 $}-\pi )}{\text{$\Delta_1 $}^2+\text{$\Delta_1 $}
   (\text{$\Delta_2 $}+\text{$\Delta_3 $})-2 \pi  (\text{$\Delta_1 $}+\text{$\Delta_2
   $}+\text{$\Delta_3 $})+\text{$\Delta_2 $}^2+\text{$\Delta_2 $} \text{$\Delta_3
   $}+\text{$\Delta_3 $}^2+\pi ^2} , \nonumber\\ \label{InversionSol}
\end{eqnarray}
that will prove to be useful later on, when comparing with the conjectured $AdS/CFT$ dual quantity.

\subsection{Comments on the index} 

The index computed in this work, which follows closely \cite{Benini:2015noa}, has the canonical interpretation of a Witten index counting ground states according to $ Z(\mathfrak{n}_a, \Delta_a)={\rm Tr} (-1)^F e^{-\beta H} e^{iJ_a \Delta_a}$. From the 3d perspective, we are simply counting operators with the corresponding relation among  quantum numbers. Now, assuming that the deformed $ABJM$ theory flows to an effective quantum mechanics in the IR, then the index computes the degeneracy of ground states in  the quantum mechanics. Since the index is an invariant of the flow, we connect directly the 3d and 1d perspective. 

On the gravity side, we have, similarly, the possibility of viewing the counting from the 4d or 2d perspectives. The better formulated one, at the moment, turns out to be the 2d perspective, which Sen has developed in the framework of $AdS_2/CFT_1$ \cite{Sen:2008yk,Sen:2008vm}. In this context, the ground state degeneracy in the quantum mechanics is computed by an $AdS_2$ partition function  with specific boundary conditions, which leads precisely to the quantum black hole entropy. There are, of course, some open issues with the application of Sen's proposal, in the context of asymptotically $AdS$ black holes, but it certainly provides a solid starting point.

\section{The hyperbolic $AdS_4$ black hole}\label{BHsec}

In this section we construct what we believe are the holographic dual to the $ABJM$ configuration discussed thus far. Namely, we construct magnetically charged, asymptotically $AdS_4$ black holes with non-compact $\mathbb{H}_2$ horizon that are embedded in M-theory. Our construction follows similar spherical solutions in ${\cal N}=2$ gauged supergravity, see, for example, \cite{Cacciatori, Cacciatori0, Vandoren, Halmagyi, Halmagyi2}. We will comment on some similarities and differences with the solutions with spherical horizon in subsection \ref{subsec:spherical}. We shall focus on the case of $n_V=3$ vector multiplets. In this way the $n_V+1$ vector fields-- counting also the one in the graviphoton vector multiplet -- are set to be identified as dual to the global charges of $ABJM$. 

\subsection{A brief summary of 4d $\mathcal{N}=2$ gauged SUGRA with $n_V=3$}

For completeness, let us briefly introduce the concepts that we shall use. The central object is the pre-potential
\be
\mathcal{F}= \mathcal{F}(X^\D),
\ee
which is a holomorphic function of the holomorphic sections $X^\D(z^i)$, $\D=1,2,3 \text{ and }4$. The symplectic sections are functions of the physical scalars $z^i$ with $i=1,2,3$. 

Another important object is the K\"ahler potential 
\be
\mathcal{K}=- \log{i \left(\bar{X}^\Lambda \mathcal{F}_{\Lambda}-X^\Lambda \bar{\mathcal{F}}_{\Lambda} \right)},  ~~ \mathcal{F}_{\Lambda}:=\frac{\partial \mathcal{F}}{\partial
X^{\Lambda }}.
\ee
 We will also need to use the period matrix
\begin{equation*}
N_{\Lambda \Sigma }:=\overline{\mathcal{F}}_{\Lambda \Sigma }+2i\frac{\mathbb{I}{m}\left(
\mathcal{F}_{\Lambda \Gamma }\right) X^{\Gamma }\mathbb{I}{m}\left( \mathcal{F}_{\Sigma \Delta
}\right) X^{\Delta }}{X^{\Gamma }\mathbb{I}{m}\left( \mathcal{F}_{\Gamma \Delta }\right)
X^{\Delta }},\;\;\;\mathcal{F}_{\Gamma \Delta }:=\frac{\partial \mathcal{F}_{\Gamma }}{\partial
X^{\Delta }}.
\end{equation*}
and the following auxiliary variables
\begin{equation*}
\left( L^{\Lambda },M_{\Lambda }\right) :=e^{K/2}\left( X^{\Lambda
},\mathcal{F}_{\Lambda }\right) ,\;\;\left( f_{i}^{\Lambda },h_{\Lambda ,i}\right)
:=e^{K/2}\left( D_{i}X^{\Lambda },D_{i}\mathcal{F}_{\Lambda }\right),
\end{equation*}
where the covariant derivative $D_i$ is defined as $D_{i}:=\partial _{z^{i}}+K_{i}$.%

In our case we will be interested in real holomorphic sections
\be
\bar{X}^\D=X^\D, ~~ \bar{z}^i=z^i.
\ee

To construct black holes, we shall set the fermions, and fermionic variations, to zero. The supersymmetry variation of the gravitino is
\begin{equation}
\delta \psi _{\mu A}:=\nabla _{\mu }\varepsilon _{A}+2iF_{\mu \nu
}^{-\Lambda }I_{\Lambda \Sigma }L^{\Lambda }\gamma^{\nu }\epsilon
_{AB}\varepsilon ^{B}-\frac{g}{2}\sigma _{AB}^{3}\xi _{\Lambda }L^{\Lambda
}\gamma _{\mu }\varepsilon ^{B}, \label{gravitino}
\end{equation}
where the covariant derivative of the Killing spinor $\varepsilon_A$ is
\begin{equation*}
\nabla _{\mu }\varepsilon _{A}=\left( \partial _{\mu }-\frac{1}{4}\omega
_{\mu }^{ab}\gamma _{ab}\right) \varepsilon _{A}+\frac{1}{4}\left(
K_{i}\partial z^{i}-K_{\overline{i}}\partial \overline{z}^{\overline{i}%
}\right) \varepsilon _{A}+\frac{i}{2}g\xi _{\Lambda }A_{\mu }^{\Lambda
}\sigma _{A}^{3\;B}\varepsilon _{B}.
\end{equation*}
The supersymmetry variation of the gaugino 
\begin{equation}
\delta \lambda ^{iA}=i\partial _{\mu }z^{i}\gamma ^{\mu }\varepsilon
^{A}-g^{i\;\overline{j}}\overline{f}_{j}^{\Lambda }I_{\Lambda \Sigma }F_{\mu
\nu }^{\Sigma \;-}\gamma ^{\mu \nu }\epsilon ^{AB}\varepsilon _{B}+igg^{i%
\overline{j}}\overline{f}_{\overline{j}}^{\Lambda }\xi _{\Lambda }\sigma
^{3,AB}\varepsilon _{B}, \label{gaugino}
\end{equation}
will be used too. In equations \eqref{gravitino} and \eqref{gaugino} we are discarding higher order terms in fermions. These terms are not relevant to our discussion.

To reproduce our results it will be useful to have the following definitions \cite{Bertolini}
\begin{eqnarray*}
F_{\Lambda }^{-} &:=&\frac{1}{2}\left( F_{\Lambda }-i\ast F_{\Lambda
}\right) , \\
{\ast F_{\Lambda}}_{ \mu \nu} &:=&\frac{1}{2}\epsilon _{\mu \nu \alpha \beta }F^{\alpha
\beta },
\end{eqnarray*}%
\begin{equation*}
\frac{1+\gamma _{5}}{2}\varepsilon _{A}=\frac{1-\gamma _{5}}{2}\varepsilon
^{A}=0.
\end{equation*}

\subsection{Hyperbolic black holes}\label{hypbhsec}

To avoid confusion, the index $\Lambda=\{1,2,3,4\}$ is equivalent to the index $a=\{1,2,3,4\}$ that will be introduced in the next subsection.

We are interested in the STU model. Thence, we fix the Fayet-Iliopoulus parameters in an isotropic manner 
\begin{eqnarray}
\xi_0=\xi_1=\xi_2=\xi_3=\xi_V.
\end{eqnarray}
The relevant pre-potential will be
\begin{equation*}
\mathcal{F}(X)=-2 i \sqrt{X_1 X_2 X_3 X_4 }.
\end{equation*}
We consider real sections, with the following parametrization
\be\small
X^{\Lambda}=\bar{X}^\Lambda=\left\{-\frac{z^1}{z^1+z^2+z^3+3},-\frac{z^2}{z^1+z^2+z^3+3},-\frac{z^3}{z^1+z^2+z^3+3}
   ,-\frac{1}{z^1+z^2+z^3+3}\right\},
\ee
and propose the following static, spherically symmetric ansatz, for the metric and sections
\begin{eqnarray}
ds^{2}&=&-U^{-2}(r)dr^{2}-h^{2}(r)d\theta ^{2}-h^{2}(r)\sinh ^{2}\theta \;d\varphi
^{2}+U^{2}(r)dt^{2},  \label{BH}\\
X^{\Lambda}&=&\alpha+ \frac{\beta_\L}{r}.
\end{eqnarray}
The non trivial components of the spin connection are \footnote{In this section we used different conventions than in section \ref{sec:susy3d}. We have used standard conventions on four-dimensional $\mathcal{N}=2$ gauged supergravity, which are the ones given in \cite{Bertolini} (see also \cite{Vandoren}). For instance, the definition of spin connection is minus the one used in section \ref{sec:susy3d}. Consequently, in the covariant derivatives there is a relative minus sign in front of the term proportional to the spin-connection between this section and section 2. }
\begin{eqnarray*}
\omega _{t}^{14} &=&-U(r)U^{\prime }(r),\;\omega _{\theta
}^{12}=-U(r)h^{\prime }(r), \\
\omega _{\varphi }^{13} &=&-U(r)h^{\prime }(r)\sinh \theta ,\;\omega _{\varphi
}^{23}=-\cosh \theta .
\end{eqnarray*}
In this section
\begin{eqnarray*}
\epsilon _{4123}&=&1, \\  \eta _{ab}&=&(-1,-1,-1,1), \\  \gamma _{5}&=&i\gamma ^{4}\gamma^{1}\gamma ^{2}\gamma ^{3}.
\end{eqnarray*}
We use the following Ans\"{a}tze for the functions $U(r)$ and $h(r)$ 
\begin{eqnarray}
U(r) &:=&e^{K/2}(g\ r-\frac{c}{2g\ r}),\label{U} \\
h(r) &:=&d\ e^{-K/2}r, \label{h}
\end{eqnarray}
where $g$, $c$ and $d$ are constants. 

The corresponding black holes, are sourced by magnetic fluxes $p_\L$:
\begin{equation}
A_{\Lambda \varphi }=-p_{\Lambda }\cosh \theta ,\;F_{\Lambda \theta \varphi }=-%
\frac{p_{\Lambda }}{2}\sinh \theta .
\end{equation}
The non trivial components of the anti-selfdual field strength are
\begin{equation}
\;\;F_{\Lambda _{\theta \varphi }}^{-}=-F_{\Lambda _{\theta \varphi }}^{-}=-\frac{%
p_{\Lambda }}{4}\sinh \theta ,\;\;F_{\Lambda _{rt}}^{-}=-F_{\Lambda
_{tr}}^{-}=i\frac{p_{\Lambda }}{4h^{2}(r)}\cdot 
\end{equation}
The chiral and anti-chiral Killing spinors $\epsilon_A$ and $\epsilon^B$, have to obey the following relation - these conditions are obtained from the vanishing of the gravitino supersymmetric transformation-
\begin{equation}
\varepsilon _{A}=\epsilon _{AB}\gamma ^{4}\varepsilon ^{B},\;\varepsilon
_{A}=\pm \sigma _{AB}^{3}\gamma ^{1}\varepsilon ^{B}. \label{KSEq}
\end{equation}
The most general solution to \eqref{KSEq} is
\begin{equation}
\varepsilon _{A}=%
\begin{pmatrix}
\pm \varkappa (r) & \mp i\;\overline{\varkappa }(r) \\ 
\pm i\;\varkappa (r) &\pm \overline{\varkappa }(r) \\ 
-i\;\varkappa (r) & \overline{\varkappa }(r) \\ 
-\varkappa (r) & -i\;\overline{\varkappa }(r)%
\end{pmatrix}%
,\;\varepsilon ^{B}=%
\begin{pmatrix}
\overline{\varkappa }(r) & i\varkappa (r) \\ 
-i\overline{\varkappa }(r) & \varkappa (r) \\ 
\pm i\overline{\varkappa }(r) &\pm \varkappa (r) \\ 
\mp \overline{\varkappa }(r) & \pm i\varkappa (r)%
\end{pmatrix}.%
\end{equation}
Solving the $BPS$ conditions, leads to relations 
\begin{eqnarray}
\alpha  &=&\mp\frac{1}{4\xi _{V}}, \\
c &=&\frac{1+8~ d^{2}g^{2}\xi _{V}^{2}\left( \beta _{4}^{2}+\beta
_{1}^{2}+\beta _{2}^{2}+\beta _{3}^{2}\right) }{d^{2}}, \\
0 &=&\beta _{4}+\beta _{1}+\beta _{2}+\beta _{3},\label{beta4} \\
1 &=&\pm g\xi _{V}\left( p_{4}+p_{1}+p_{2}+p_{3}\right). 
\end{eqnarray}
Notice that the constant $c$ is positive.
The relation between fluxes and the parameters $\beta_a$ is also obtained from the $BPS$ conditions
\begin{eqnarray}
p_{1} &=&\frac{\pm1+16 d^{2}g^{2}\xi _{V}^{2}\;\left( -\beta _{1}^{2}+\beta
_{2}^{2}+\beta _{3}^{2}+\beta _{1}\beta _{2}+\beta _{2}\beta _{3}+\beta
_{1}\beta _{3}\right) }{4g\xi _{V}},\nonumber \\
p_{2} &=&\frac{\pm 1+16 d^{2}g^{2}\xi _{V}^{2}\;\left( +\beta _{1}^{2}-\beta
_{2}^{2}+\beta _{3}^{2}+\beta _{1}\beta _{2}+\beta _{2}\beta _{3}+\beta
_{1}\beta _{3}\right) }{4g\xi _{V}}, \nonumber\\
p_{3} &=&\frac{\pm1+16 d^{2}g^{2}\xi _{V}^{2}\;\left( +\beta _{1}^{2}+\beta
_{2}^{2}-\beta _{3}^{2}+\beta _{1}\beta _{2}+\beta _{2}\beta _{3}+\beta
_{1}\beta _{3}\right) }{4g\xi _{V}}. \label{paba}
\end{eqnarray}
The warping of the Killing spinor is also fixed by the $BPS$ conditions
\begin{equation*}
\varkappa =\varkappa _{0}\sqrt{U(r)},\;\overline{\varkappa }=\overline{%
\varkappa }_{0}\sqrt{U(r)}.
\end{equation*}
We have, finally, completely solved the $BPS$ conditions and constructed our hyperbolic $AdS_4$ black holes.

\subsection{Spherical black holes}\label{subsec:spherical}

A prevalent intution in the context of supergravity states that changing the horizon from spherical to hyperbolic, leads  from black holes to naked singularities and {\it vice versa}  \cite{Chamseddine:1999xk,Klemm:2000nj}.  In this section we explore the details of this intuition in the context of the magnetically charged black holes. 

Let us first solve the $BPS$ equations for the spherical black hole ansatz
\begin{eqnarray}
ds^{2}&=&-U^{-2}(r)dr^{2}-h^{2}(r)d\theta ^{2}-h^{2}(r)\sin^{2}\theta \;d\varphi
^{2}+U^{2}(r)dt^{2},  \label{BHS2}\\
X^{\Lambda}&=&\alpha+ \frac{\beta_\L}{r}. \label{BHS22}
\end{eqnarray}
with $U(r)$ and $h(r)$ defined in \eqref{U} and \eqref{h}.

The non vanishing components of the spin connection are
\begin{eqnarray*}
\omega _{t}^{14} &=&-U(r)U^{\prime }(r),\;\omega _{\theta
}^{12}=-U(r)h^{\prime }(r), \\
\omega _{\varphi }^{13} &=&-U(r)h^{\prime }(r)\sin \theta ,\;\omega _{\varphi
}^{23}=-\cos \theta .
\end{eqnarray*}

For technical convenience, let us parametrize the gauge potential as follows \footnote{The definition of field strength used in this section differs from the one used in section \ref{sec:susy3d}.}
\begin{equation}
A_{\Lambda \varphi }=-p_{\Lambda }\cos \theta ,\;F_{\Lambda \theta \varphi }=%
\frac{p_{\Lambda }}{2}\sin \theta .
\end{equation}
The non trivial components of the antiselfdual potential are
\begin{equation}
\;\;F_{\Lambda _{\theta \varphi }}^{-}=-F_{\Lambda _{\theta \varphi }}^{-}=\frac{%
p_{\Lambda }}{4}\sin \theta ,\;\;F_{\Lambda _{rt}}^{-}=-F_{\Lambda
_{tr}}^{-}=i\frac{p_{\Lambda }}{4h^{2}(r)}.
\end{equation}

After solving the $BPS$ equations, we arrive to
\begin{eqnarray}
\alpha  &=&\mp \frac{1}{4\xi _{V}}, \\
c &=&\frac{-1+8~ d^{2}g^{2}\xi _{V}^{2}\left( \beta _{4}^{2}+\beta
_{1}^{2}+\beta _{2}^{2}+\beta _{3}^{2}\right) }{d^{2}},  \label{cS2}\\
0 &=&\beta _{4}+\beta _{1}+\beta _{2}+\beta _{3}, \\
-1 &=& \pm g\xi _{V}\left( p_{4}+p_{1}+p_{2}+p_{3}\right). 
\end{eqnarray}
and 
\begin{eqnarray}
p_{1} &=&\frac{\mp 1+16 d^{2}g^{2}\xi _{V}^{2}\;\left( -\beta _{1}^{2}+\beta
_{2}^{2}+\beta _{3}^{2}+\beta _{1}\beta _{2}+\beta _{2}\beta _{3}+\beta
_{1}\beta _{3}\right) }{4g\xi _{V}},\nonumber \\
p_{2} &=&\frac{\mp 1+16 d^{2}g^{2}\xi _{V}^{2}\;\left( +\beta _{1}^{2}-\beta
_{2}^{2}+\beta _{3}^{2}+\beta _{1}\beta _{2}+\beta _{2}\beta _{3}+\beta
_{1}\beta _{3}\right) }{4g\xi _{V}}, \nonumber\\
p_{3} &=&\frac{\mp 1+16 d^{2}g^{2}\xi _{V}^{2}\;\left( +\beta _{1}^{2}+\beta
_{2}^{2}-\beta _{3}^{2}+\beta _{1}\beta _{2}+\beta _{2}\beta _{3}+\beta
_{1}\beta _{3}\right) }{4g\xi _{V}}. \label{pabaS2}
\end{eqnarray}
The warping of Killing spinor is also fixed by the $BPS$ conditions
\begin{equation*}
\varkappa =\varkappa _{0}\sqrt{U(r)},\;\overline{\varkappa }=\overline{%
\varkappa }_{0}\sqrt{U(r)}.
\end{equation*}
In contradistinction to the hyperbolic solution, in the spherical case, $c$ can be negative, see equation \eqref{cS2}. 

The position of the curvature singularity is
\be
r_s= \left \{ \begin{tabular}{ccc} 
$ -\frac{1}{\alpha} \max \left\{ \beta_1, \beta_2 ,\beta_3 ,\beta_4 \right\}>0  $& if & $ \alpha<0$
 \\ \nonumber \\$ -\frac{1}{\alpha} \min \left\{ \beta_1, \beta_2 ,\beta_3 ,\beta_4 \right\}>0$  & if &  $\alpha>0$  \end{tabular} \right. .
\ee
If $\beta_a=0$, the position of the curvature singularity is $r_s=0$. In the case $\beta_a=0$, one encounters a hyperbolic black hole since $c>0$ and there is a horizon. However for the spherical solutions, $c<0$, and we encounter a naked singularity.  It is straightforward to check that, when $\beta_a=0$, the change 
\be
\left(r, ~\theta,~ t, ~p_a,~ c \right) \rightarrow \left(i r, ~ i \theta, ~i t, ~-p_a, ~-c\right)
\ee
transforms the hyperbolic black hole $BPS$ solutions of the previous subsection, into the spherical $BPS$ solution of this subsection, particularized to $\beta_a=0$. The latter has a naked singularity. Actually, the exchange $p_a \rightarrow -p_a$ can be cancelled by an exchange of Killing spinor --- only because $\beta_a=0$ ---. By an exchange of Killing spinor, we mean to change the choice of sign in the constraint \eqref{KSEq}. Such a change, has physical meaning, as it leads to a configuration that is $BPS$ with respect to a different supercharge.

We  emphasize that the intuition emanating from \cite{Chamseddine:1999xk,Klemm:2000nj}, is restricted to the case of $\beta_a=0$, that is, the case of constant sections; it is, therefore, relaxed in the case $\beta_a\neq 0$.

\subsection{The Bekenstein-Hawking entropy: $\mathbb{H}_2$ vs $S^2$ }
In this subsection, we compare the entropy of hyperbolic  and spherical black holes. We first consider the case
of isotropic fluxes. In the end, we will find that the entropy density of the hyperbolic solution coincides with the entropy density of the spherical one which were discussed in \cite{Benini:2015eyy}. 



From now on, we particularize our hyperbolic solutions to the following case
 \be
 \alpha =-\frac{1}{4},\ \xi _{V}=1,\ g=\frac{1}{\sqrt{2}}. \label{PartBH}
 \ee

A sufficient -- not necessary -- condition for the existence of hyperbolic $AdS_4$ black holes is
\begin{eqnarray}
\beta_1,\beta_2, \beta_3>0, \; r_h=\sqrt{c}> r_s=  4 \max{\left(\text{$\beta_1$},\text{$\beta _2$},\text{$\beta_3$}\right)},
\end{eqnarray}
where the domain of the radial coordinate $r$ is $r > r_s$. The constant $r_s$ is the radial position of the singularity, which is covered by the horizon at $r_h=\sqrt{c}$.

A particular solution to these conditions is
\begin{equation*}
\beta _{1}=\ \beta _{2}=\ \beta _{3}=\beta >0.
\end{equation*}
In that case, the fluxes are
\begin{equation*}
p_{1}=p_{2}=p_{3}=p=\frac{1+32 d^{2}\ \beta ^{2}}{2\sqrt{2}}>0.
\end{equation*}

The regularized area density of the horizon is
\begin{eqnarray}
\frac{A_{\mathbb{H}_2}(\beta)}{vol_{\mathbb{H}_2}}&=& \frac{1}{2}\sqrt{1+512 d^{3}\beta ^{3}(-6 d \beta +\sqrt{%
1+48 d^{2} \beta ^{2}})} \\
&=& \frac{ \sqrt{-3 \left(1-2 \sqrt{2} p\right)^2+2 \left(2 \sqrt{2} p-1\right)^{3/2}
   \sqrt{6 \sqrt{2} p-1}+1}}{2} \nonumber\\ \label{eqq1}\\
\frac{A_{\mathbb{H}_2}(p)}{vol_{\mathbb{H}_2}}&\underset{p\rightarrow +\infty }{\sim }& \sqrt{\left(4\sqrt{3}-%
6\right)} \, p,   ~~~~~  p>0.
\end{eqnarray}

\bigskip 

 Next, we compare the entropy density \eqref{eqq1} to the one of spherical black holes used in \cite{Benini:2015eyy}. The isotropic solution presented in section 4.1 of \cite{Benini:2015eyy} is
\begin{equation*}
\mathfrak{n}_{1}=\mathfrak{n}_{2}=\mathfrak{n}_{3}=\sqrt{2}p^\prime ,\;\;\mathfrak{n}_{4}=2-3\sqrt{2}p^\prime, \; \; p^\prime<0. 
\end{equation*}
From the following quantities \cite{Benini:2015eyy}
\begin{eqnarray*}
F_{2}(p) &:&=-\left( 12p^{\prime\, 2}-6\sqrt{2}p^\prime +1\right) ,\;\; \\
\Theta (p) &:&=192 p^{\prime\, 4}-160\sqrt{2}p^{\prime\, 3}+96p^{\prime \, 2}-12%
\sqrt{2}p^\prime +1, \label{largefhyp}
\end{eqnarray*}
one arrives to the following expression for the area density, of the spherical horizon in terms of the flux $p^\prime$
\begin{eqnarray}
\frac{A_{S^2}(p^\prime)}{vol_{S^2}}&:=&  \frac{\sqrt{F_{2}+\sqrt{\Theta }}}{\sqrt{2}}
\\ &=&\frac{\sqrt{\sqrt{4 p^\prime \left(8 p^\prime \left(6 p^{\prime\, 2}-5 \sqrt{2} p^\prime+3\right)-3
   \sqrt{2}\right)+1}+6 \left(\sqrt{2}-2 p^\prime\right) p^\prime-1}}{\sqrt{2}} \nonumber \\ \label{eqq2}\\
 \frac{A_{S^2}(p^\prime)}{vol_{S^2}}  &\underset{p^\prime \rightarrow -\infty }{\sim }&-\sqrt{\left(4\sqrt{3}-%
6\right)} \, p^\prime,   \;~~~~   \;   \;   p^\prime<0. \label{largefsph} 
\end{eqnarray}

We can do the comparison with the result obtained from scratch, with our spherical solutions, however, in order to match our results with the ones in \cite{Benini:2015eyy}, we report the comparison by using theirs. Notice that the large flux limits \eqref{largefhyp} and \eqref{largefsph}, do coincide. In fact, it can be checked that for any value of $u$ the entropy density as a function of flux $p$ and $p^\prime$ coincide, as
\begin{eqnarray}
\frac{A_{\mathbb{H}_2}(|u|)}{vol_{\mathbb{H}_2}}=\frac{A_{S^2}(|u|)}{vol_{S^2}}. \label{coincH2}
\end{eqnarray}
Equation \eqref{coincH2} can be checked by comparing \eqref{eqq1} and \eqref{eqq2}, order by order in Taylor expansions in $u$ about $0$ and $\infty$, or by simply working out the expressions.

\subsection{Matching results}

In this final section, we compare  the $AdS/CFT$ dual results. On one side, we have the result for the $ABJM$ index on $\mathbb{H}_2\times S^1$ \eqref{SCFT}. On the other side, we have the entropy of the hyperbolic magnetic $AdS_4$ black holes \eqref{BH}. The first thing to do, is to compute the Bekenstein-Hawking entropy of \eqref{BH}. Thereafter, we check the relation between the classical entropy and the value of the holomorphic  sections $X_a$ -- or as were denoted in the previous subsection $X^\L$-- at the horizon $r_h$. The aforementioned relation is identical, up to a relabelling of variables, to the relation between the logarithm of the $ABJM$ index and the holonomies $\Delta_a$ \eqref{SCFT}. Finally, we prove that under the appropriate relabeling of variables and extremization of the logarithm of the $ABJM$ index  \eqref{SCFT} with respect to the holonomies $\Delta_a$, the bulk and SCFT results coincide, as was the case in \cite{Benini:2015eyy}.

 For generic $p_1$, $p_2$ and $p_3$, we have checked that the classical entropy
\begin{eqnarray}
S_{BH}&=&\frac{A_{\mathbb{H}_2}}{4 G_{4d}}  \nonumber \\ &=&-\frac{\pi}{4 G_{4 D}}  \sqrt{(\Psi -4 d \text{$\beta_1 $}) (\Psi -4 d \text{$\beta_2 $}) (\Psi -4 d
   \text{$\beta_3 $}) (4 d (\text{$\beta_1 $}+\text{$\beta_2 $}+\text{$\beta_3 $})+\Psi )}, \nonumber\\\end{eqnarray}
where
\begin{eqnarray}
\Psi(\beta_1,\beta_2,\beta_3):=\sqrt{8 d^2 \left(\text{$\beta_1 $}^2+\text{$\beta_1 $} \text{$\beta_2 $}+\text{$\beta_1 $} \text{$\beta_3$}+\text{$\beta_2 $} \text{$\beta_3 $}+\text{$\beta_2 $}^2+\text{$\beta_3
   $}^2\right)+1},
\end{eqnarray}
coincides with the expression
\begin{eqnarray}
+\frac{2 \pi}{4G_{4d}} \sqrt{X_1(r_h) X_2(r_h) X_3(r_h) X_4(r_h)} \sum_{a=1}^4 \frac{ \sqrt{2} p_a }{X_a(r_h)}, \label{formulaHor}
\end{eqnarray}
where
\begin{eqnarray}
X_a(r_h)=\frac{d ~\beta _a}{\sqrt{4~ d^2 \left(\text{$\beta_1 $}^2+\text{$\beta_2 $}^2+\text{$\beta_3
   $}^2+\text{$\beta_4 $}^2\right)+1}}-\frac{1}{4}. \label{Simplct}
\end{eqnarray}
The value of $\beta_4$ is determined by \eqref{beta4}. The $p_a$'s as function of $\beta_a$'s have been given in equation \eqref{paba} which follows from the $BPS$ equations. 
 
We will prove next, that equation \eqref{formulaHor}  -- that comes from the analysis in the bulk--  is equal to the extremal value of the SCFT topologically twisted index \eqref{SCFT}, under the specific dictionary
\bea
\sqrt{2} p_a &\leftrightarrow& \mathfrak{n}_a,  \label{fluxID}\\
-2\pi X_a(r_h)&\leftrightarrow& \bar{\Delta}_a, ~~ a=1,2,3,4,
\eea
where the  $\bar{\Delta}_a$ are the solutions to the variables $\Delta_a$ that come out of the inversion of equation \eqref{InversionSol}.

There are many ways to prove that \eqref{formulaHor} is equivalent to the extreme value of \eqref{SCFT}, the simplest one is to evaluate \eqref{InversionSol} on 
\begin{eqnarray}
\Delta_a=\Delta_a(\beta_1,\beta_2,\beta_3)=-2\pi X_a(r_h),
\end{eqnarray}
to obtain
\begin{eqnarray}
\mathfrak{n}_1=\frac{1}{2}+4 d^2 \left(-\text{$\beta_1 $}^2+\text{$\beta_2 $}^2+\text{$\beta_3$}^2+\text{$\beta_1 $} \text{$\beta_2 $}+\text{$\beta_1 $}
   \text{$\beta_3 $}+\text{$\beta_2 $} \text{$\beta_3$}\right), \nonumber\\
   \mathfrak{n}_2=\frac{1}{2}+4 d^2 \left(+\text{$\beta_1 $}^2-\text{$\beta_2 $}^2+\text{$\beta_3$}^2+\text{$\beta_1 $} \text{$\beta_2 $}+\text{$\beta_1 $}
   \text{$\beta_3 $}+\text{$\beta_2 $} \text{$\beta_3 $}\right), \nonumber\\
   \mathfrak{n}_3=\frac{1}{2}+4 d^2 \left(+\text{$\beta_1 $}^2+\text{$\beta_2 $}^2-\text{$\beta_3
   $}^2+\text{$\beta_1 $} \text{$\beta_2 $}+\text{$\beta_1 $}
   \text{$\beta_3 $}+\text{$\beta_2 $} \text{$\beta_3 $}\right). \label{fluxSCFT}
\end{eqnarray}
Notice that
\begin{eqnarray}
\text{Equation }\eqref{fluxSCFT}= \sqrt{2} \text{ Equation } \eqref{paba},\label{RelationExtrema}
\end{eqnarray} 
under \eqref{PartBH} and identification \eqref{fluxID}. 

The relation \eqref{RelationExtrema} implies that the positions $\bar{\Delta}_a$ of the saddle points of \eqref{SCFT} coincide with the values of the  sections $X_a$ at the horizon \eqref{Simplct}, under  identification \eqref{fluxID}. As the logarithm of the $ABJM$ index \eqref{SCFT} and the Bekenstein-Hawking entropy \eqref{formulaHor} are the same under identification \eqref{fluxID} and 
\be
\frac{1}{G_{4d}}= \frac{2 \sqrt{2}}{3} N^{3/2},
\ee
we have thence proven that under the aforementioned identifications, boundary ``degeneracy of states'' and bulk black hole entropy coincide.

As a final comment, we notice that the identifications \eqref{fluxID} are not directly obtained from the $AdS/CFT$ dictionary. The $AdS/CFT$ dictionary is naturally formulated in the UV, the UV value of the holomorphic sections $X_{a}$ is $-\frac{1}{4}$. To obtain agreement, the use of extremization principle of the result of the SCFT side  is crucial \cite{Benini:2015eyy}. We believe there is a proper way to clarify some of these {\it ad hoc} issues but we leave the discussion  for future work. 



\section{Conclusions} \label{sec:7}

In this manuscript we have first studied topologically twisted localization of ${\cal N}=2$ supersymmetric field theories in $\mathbb{H}_2\times S^1$. Our work differs in various important points from  recent work on localization on this space by  \cite{David:2016onq}. In particular, we have crucially considered topologically twisted theories and extended the type of theories under consideration beyond vector multiplets, to include, for example, matter multiplets.

At a technical level, we have also discussed explicitly subtle aspects of the eigenvalue problem corresponding to the Laplacian in the presence of a background magnetic field and we expect that such results could have wide application in the general context of localization. Quite interestingly, we have found a hierarchy of normalizable modes and its corresponding discrete spectrum. A particular sub-family of the aforementioned hierarchy,  corresponds to the vector zero modes of the Laplace-Beltrami operator on $\mathbb{H}_2$, that were introduced in \cite{Camporesi:1994ga} and figure prominently in, for example, \cite{Banerjee:2010qc} and more recently in \cite{David:2016onq}. The full hierarchy of normalizable modes exists due to the presence of magnetic fluxes $s$ over a specific threshold:  $|s|>\frac{1}{2}$. We strongly suspect, that the discrete spectrum is encoding the full hierarchy of higher spin normalizable modes of the Laplace-Beltrami operator on $\mathbb{H}_2$.  If this is indeed the case, it would be very interesting to pursue a study of the potential traces of  2d higher spin symmetry, on the set of black holes microstates \cite{Benincasa}.  As a first step toward formulating  such a problem, it would be useful to start by identifying the square integrable modes in the language of \cite{Benincasa}. 

We have also studied  ${\cal N}=2$ gauged supergravity and found magnetically charged supersymmetric solutions with hyperbolic horizon. We have shown that under assumptions similar to those advanced in \cite{Benini:2015eyy} the entropy of these solutions coincides with the real part of the logarithm of the topologically twisted index of the dual field theory. In conclusion, we have provided evidence in favor of identifying the set of square integrable modes in the presence of a constant flux on $\mathbb{H}_2\left(\times S^1\right)$, precisely speaking a very restricted set of zero modes out of the maximal set, as the boundary microstates responsible for the Bekenstein-Hawking entropy of the $AdS_4$ hyperbolic black holes presented in section \ref{BHsec}. One important further test for this identification, would  be to  compute quantum corrections to the Bekenstein-Hawking entropy, on both sides of the duality.  





On the 3d SCFT side we have made crucial use of the extremization approach advocated in \cite{Benini:2015eyy}. The result of this approach is consistent with the constraints from the $BPS$ equations on the gravity side and has been argued to be equivalent to the attractor mechanism. Under these conditions we  have found precise agreement between the leading large-$N$ results on the two sides.  However, it would be important to elucidate the role of extremization intrinsically in the field theory  but also from the gravity perspective. This is particularly important because in some cases the attractor mechanism has been shown to apply away from the strictly supersymmetric context. 


Another natural generalization of this work, following \cite{Benini:2016rke}, is to extend the analysis to dyonic black holes. More generally, it would be interesting to consider mapping the full space of deformations on both sides of the correspondence and, in particular, its modifications on the free energy and the entropy.  Another interesting direction concerns potential factorizations of the index on $S^2\times S^1$, introduced in  \cite{Benini:2015noa}, in terms of blocks  given by the partition functions in $\mathbb{H}_2\times S^1$. A similar factorization principle has been uncovered in various theories and in different dimensions, see, for example, \cite{Alejandro,Pasquetti:2011fj,Gava:2016oep,Hori:2013ika,Nieri:2015yia}.  In this manuscript we have found a particular relation but it should be pointed out that we have set  all fermionic zero modes to zero and have integrated over a particular set of modes.  Clearly, to achieve a {\it bona fide} factorization formula we will need to consider more general  boundary conditions and contemplate retracing some of the steps suggested in  \cite{Rey}.  Indeed, such an approach with general boundary conditions has been implemented for GLSM's  in  \cite{Hori:2013ika}. 

Finally, it would be interesting to discuss the microstate counting of magnetically charged strings in asymptotically $AdS_5$ spacetimes. Such magnetically charged solutions have a long history in supergravity dating back to explorations in \cite{Chamseddine:1999xk}. It is logical to expect that the microscopic explanation should be found within 4d topologically twisted field theories on $S^2\times T^2$ or possibly $\mathbb{H}_2\times T^2$. Indeed, as a natural starting point along these lines,  the topologically twisted index introduced by Benini and Zaffaroni in \cite{Benini:2015noa} for supersymmetric field theories on $S^2\times S^1$, was briefly discussed for 4d theories in $S^2\times T^2$ in their original work, and was also addressed in \cite{Honda:2015yha}. It has recently been shown that, in the high temperature limit, the index produces a central charge that matches the supergravity answer \cite{Hosseini:2016cyf} therefore providing  a strong argument in favor of the identification. We hope to report on some of these interesting directions soon.

\section*{Acknowledgments}
We are first and foremost grateful to the Abdus Salam Centre for Theoretical Physics where this work originated and was largely conducted; L. PZ., in particular, acknowledges sabbatical support. The work of ACB is supported by CONICET.

We would like to thank Junya Yagi for collaboration in the early stages of this project. We  are especially thankful to E. Gava and K. Narain for discussions and clarifications at various  stages of this work. We are thankful to G. Bonelli and A. Tanzini for comments on a preliminary presentation by V. G-R. of parts of this work in summer 2016. We are also  grateful to  R.K. Gupta, K. Intriligator, U. Kol, J.  Liu, S. J. Rey, A. Sen and V. Rathee for discussions. ACB is very grateful to Carmen N\'u\~{n}ez for her support and encouragement.

 \appendix

 \section{Comments on the discrete spectrum} \label{SQN}

In this appendix we report details on the construction of the square integrable modes defined in section \ref{sec:3}.

 The general solution to the defining equation \eqref{DefEq} is%
\bea
f  & =&\chi_{1}\left(  \frac{\left(  \cosh\theta-1\right)  ^{s}}{\sinh^{j_{3}%
+s}\theta}{}_2F_{1}(a_{1},b_{1},c_{1};-\sinh^{2}\frac{\theta}{2})\right) \nonumber\\
& & ~~~~~~~ +\chi_{2}\left(  \frac{\left(  \cosh\theta-1\right)  ^{j_{3}}}{\sinh^{j_{3}%
+s}\theta}{}_2 F_{1}(a_{2},b_{2},c_{2};-\sinh^{2}\frac{\theta}{2})\right),\label{genSOlLat}
\eea
where $\chi_{1,2}$ are arbitrary integration constants. It is very useful to work in the following coordinates
\bea
x  &  :=-\sinh^{2}\frac{\theta}{2},\text{ \ \ \ \ }-\infty<x\leq0.
\eea
In  $x$ coordinate, \eqref{genSOlLat} takes the form
\bea
f  &=&\chi_{1}  \left(  x^{s}\left(  x\text{ }\left(  x-1\right)  \right)
^{-\frac{j_{3}+s}{2}}\text{ }{}_2 F_{1}(a_{1},b_{1},c_{1};x)\right) \nonumber\\
&  &~~~~~~~+\chi_{2}\left(  x^{j_{3}}\left(  x\text{ }\left(  x-1\right)  \right)
^{-\frac{j_{3}+s}{2}}\text{ }{}_2 F_{1}(a_{2},b_{2},c_{2};x)\right)  , \label{MostGSol}
\eea
%
where the parameters $a_{1,2}$, $b_{1,2}$ and $c_{1,2}$ are
\begin{align}
a_{1}  &  =\frac{1}{2}-j_{3}-\sqrt{\frac{1}{4}+\Delta+s^{2}},\text{ \nonumber\ \ }%
b_{1}=\frac{1}{2}-j_{3}+\sqrt{\frac{1}{4}+\Delta+s^{2},} \nonumber\\
c_{1}  &  =1-j_{3}+s, \nonumber\\
a_{2}  &  =\frac{1}{2}-s-\sqrt{\frac{1}{4}+\Delta+s^{2}},\text{  \nonumber \ \ }%
b_{2}=\frac{1}{2}-s+\sqrt{\frac{1}{4}+\Delta+s^{2}},\nonumber\\
c_{2}  &  =1-s+j_{3}. \label{abc}
\end{align}
Let us particularize $\Delta$ to 
\be
\Delta= j\left(j+1\right)-s^2, ~~~ \text{ with }  j \in \mathbb{R}. \label{DELTA}
\ee
The choice \eqref{DELTA} completes perfect square inside the square roots in \eqref{abc}.

The asymptotic behavior of the linear solutions proportional to $\chi_1$ and $\chi_2$  --- from now on $f_1$ and $f_2$ ---,
is 
\bea
f_{1}(x) &\underset{x\rightarrow -\infty}{\sim}& \chi_{1}^{-}~  x^{-1-j}\left(  1+O(\frac{1}{x})\right)  +\chi_{1}^{+}~%
x^{j}\left(  1+O(\frac{1}{x})\right),  \label{UVCondDisc0} \\ f_{1}(x) &\underset{x\rightarrow 0}{\sim}&x^{\frac{s-j_{3}}{2}}\left( O(0)\right), \label{IRCondDisc0}
\eea
and
\bea
f_{2}(x) &\underset{x\rightarrow -\infty}{\sim}& \chi_{2}^{-}~x^{-1-j}\left(  1+O(\frac{1}{x})\right)  +\chi_{2}^{+}%
~x^{j}\left(  1+O(\frac{1}{x})\right),  \label{UVCondDisc} \\ f_{2}(x) &\underset{x\rightarrow 0}{\sim}&x^{\frac{j_{3}-s}{2}}\left( O(0)\right). \label{IRCondDisc}
\eea

Regularity at the contractible cycle $x=0$, conditions to pick up
\bea
f_1 &\text{ ~~if~~  } &   j_3 \leq s,    \label{condf1} \\
f_2 & \text{~~if~~} &   j_3 > s.         \label{condf2}
\eea

We demand $\mathcal{C}^{\infty}$-differentiability at the contractible cycle $x=0$. In the vicinity of $x=0$, $f_{1}$\ and
$f_{2}$ go like $x^{\frac{s-j_{3}}{2}}$ and $x^{\frac{j_{3}-s}{2}}$,
respectively.  Thus  $\mathcal{C}^\infty$-differentiability at the contractible cycle $x=0$
implies
\begin{eqnarray}
j_3 -s \in \mathbb{Z}, \label{QC1}
\end{eqnarray}
and the appropriate choice among \eqref{condf1} and \eqref{condf2}. Notice that from \eqref{QC1}, it follows that: $j_{3}\in \mathbb{Z}$  implies $s\in \mathbb{Z}.$. However, we are not forced to impose integrality of $j_3$, $j$ or $s$.


For the time being, let us assume $ s\geq0$. In due time, we extend the analysis to the case of generic $s$.

It is important to stress that square integrability condition is equivalent to impose the conditions
\bea
\chi^+_1=0&\text{ if }&   j_3 \leq s  \text{ and }  j> -\frac{1}{2} \nonumber\\
\chi^-_1=0 &\text{ if }& j_3 \leq s  \text{ and }  j< -\frac{1}{2}  \nonumber\\
\chi^+_2=0&\text{ if }&   j_3>s  \text{ and } j > -\frac{1}{2} \nonumber\\
\chi^-_2=0 &\text{ if }& j_3 \leq s  \text{ and }  j< -\frac{1}{2} . 
\label{SquareInt}
\eea

\subsection{The quantization conditions: $f_{1}$ } \label{f1app}

 Let us find out the quantization conditions that guarantee \eqref{SquareInt}. Our starting point in this subsection is
 \be
 j_3\leq s  \text{ ~~ and~~ }  s\geq 0 . \label{Cc1}
 \ee
 For pedagogical reasons, let us assume for the time being
 \begin{eqnarray}
 j,~j_3,~ s \in \mathbb{Z} \text{ or } \mathbb{Z}+\frac{1}{2}. \label{QuantuC}
 \end{eqnarray}
We shall see in due time that assumption \eqref{QuantuC} is not necessary. As for $j$, let us not assume nothing else at this point. On the track, we will comment on the restrictions that arise for $j$.
  
 Condition \eqref{Cc1} selects the solution $f_{1}$%
\begin{align}
f_{\Delta(s),\text{ }j_{3}}^{(1)} 
&  :=\chi_{1}\left(  x^{s}\left(  x\text{ }\left(  x-1\right)  \right)
^{-\frac{j_{3}+s}{2}}{}_2 F_{1}(-j_{3}-j,1-j_{3}+j,1-j_{3}+s;x)\right). \label{f1}
\end{align}
Notice that this solution is invariant under the transformation%
\[
j\rightarrow-\left(  j+1\right)
\]
and in consequence we have to restrict $j$ to be either
\be
j>- \frac{1}{2}  \text{  ~~    or    ~~  }   j<- \frac{1}{2}, \label{j12}
\ee
as preferred. 

 Notice that $j=-\frac{1}{2}$ is left invariant by the transformation above. For $j=-\frac{1}{2}$ both independent solutions have
the same asymptotic behavior $x^{j}$ and $x^{-j-1}$, and they are not
square integrable. 

In order to have square integrability it is necessary to have%
\[
j\neq -\frac{1}{2}.%
\]
We exclude the particular case  $j=-\frac{1}{2}$. 
 

Before writing down the quantization conditions, let us comment on the strategy. 
It turns out that the quantization conditions are the conditions for which the hypergeometric factor
in \eqref{f1} truncates to a specific polynomial. The sum of the degree of such polynomial with the degree of the leading power of the prefactor in \eqref{f1} in the limit $x\rightarrow-\infty$ must equate to
\bea
-1-j  & ~\text{ if }~ & j>-\frac{1}{2}, \\
j & ~\text{ if }~ & j<-\frac{1}{2}.
\eea
The conditions to obtain the previously mentioned goal are
\bea
1-j_{3}+j\leq0~&\text{    if   }&~j> -\frac{1}{2}, \\
-j_{3}-j \leq0~&\text{ if  }&~j<-\frac{1}{2}.
\eea
Together with \eqref{Cc1} these conditions are compactly written in the following one
\be
\max(|j|,|j+1|)\leq j_{3}\leq s. \label{CondComp}
\ee


It is straightforward to check that, if we assume \eqref{QuantuC} together with \eqref{CondComp} the desired truncation holds:
\begin{align*}
{}_2 F_{1}(-j_{3}-j,1-j_{3}+j,1-j_{3}+s;x)  &  =\\
&  =1+%
{\displaystyle\sum\limits_{n=0}^{\infty}}
\frac{(a)_{n}(b)_{n}}{(c)_{n}}\frac{x^{n+1}}{(n+1)!}\\
&  =1+\left\{
\begin{tabular}
[c]{lll}%
$0$ & \ if & $d^{(1)}=0$\\
$%
{\displaystyle\sum\limits_{n=0}^{d^{(1)}-1}}
\frac{(a)_{n+1}(b)_{n+1}}{(c)_{n+1}}\frac{x^{n+1}}{(n+1)!}$ & \ if &
$d^{(1)}>0$%
\end{tabular}
\right.  ,\\
(a)_{n+1}  &  :=%
{\displaystyle\prod\limits_{i=0}^{n}}
(a+i),
\end{align*}
The degree of the polynomial $d^{(1)}$ being
\[
d^{(1)}:=\left\{
\begin{tabular}
[c]{lll}%
$j_{3}-j-1$ & \ \ if \  & $-\frac{1}{2}< j<j_{3}$\\
$j_{3}+j$ & \ \ if & $-j_{3}\leq j<-\frac{1}{2}$%
\end{tabular}
\right.  .
\]
At this point is easy to check  that indeed the aforementioned asymptotic
behavior of \eqref{f1} about $x=0$ and $x=-\infty$ holds.

\subsection{The quantization conditions: $f_{2}$}  \label{f2app}

In this subsection we analyze the case
\be
j_3 > s  \text{ ~~ and~~ }  s\geq 0. \label{CCf2}
\ee
We assume again \eqref{QuantuC}. Condition \eqref{CCf2} selects the solution $f_2$
\begin{align}
f_{\Delta(s),\text{ }j_{3}}^{(2)} 
&  :=\chi_{2}\left(  x^{j_{3}}\left(  x\text{ }\left(  x-1\right)  \right)
^{-\frac{j_{3}+s}{2}} {}_2 F_{1}(-s-j,1-s+j,1-s+j_{3};x)\right)
\end{align}
As the previous case, this solution is invariant under the transformation%
\[
j\rightarrow-\left(  j+1\right)
\]
and in consequence at some stage we shall be forced to assume \eqref{j12}. Let us, however, not assume
the latter restriction on $j$ yet. Let us just assume \eqref{QuantuC}.
 
 The quantization conditions are%
\bea
1-s+j\leq0~&\text{    if   }&~j>-\frac{1}{2}, \\
-s-j \leq0~&\text{ if  }&~j<-\frac{1}{2}.
\eea
Together with \eqref{Cc1}, these conditions are compacted in the following one
\be
\max(|j|,|j+1|)\leq s < j_3. \label{CondComp2}
\ee
It is straightforward to check that, if we assume \eqref{QuantuC} together with \eqref{CondComp2} the desired truncation holds:
\begin{align*}
{}_2F_{1}(-s-j,1-s+j,1-s+j_{3};x)  &  =\\
&  =1+\left\{
\begin{tabular}
[c]{lll}%
$0$ & \ if \  & $d^{(2)}=0$\\
$%
{\displaystyle\sum\limits_{n=0}^{d^{(2)}-1}}
\frac{(a)_{n+1}(b)_{n+1}}{(c)_{n+1}}\frac{x^{n+1}}{(n+1)!}$ & \ if &
$d^{(2)}>0$%
\end{tabular}
\right.  .
\end{align*}
The degree of the polynomial $d^{(2)}$ being
\[
d^{(2)}:=\left\{
\begin{tabular}
[c]{lll}%
$s-j-1$ & \ \ if \  & $-\frac{1}{2}\leq j<s$\\
$s+j$ & \ \ if & $-s\leq j<-\frac{1}{2}$%
\end{tabular}
\ \right.  ,
\]
and $f_{\Delta(s),\text{ }j_{3}}^{(2)}$ is square integrable. Notice that, there are not
square integrable modes for $s=0$, as well known.



\subsection{The case of negative flux $s<0$}

So far, we have focused on the case of positive magnetic flux $s>0$, or being more specific on the case $s>\frac{1}{2}$. However, there are square integrable
modes when $s<0$ too --as parity preservation dictates--. To find those, it is convenient to use the identity%
\[
_2F_{1}(a,b,c;x)=(1-x)^{c-a-b} {}_2F_{1}(c-a,c-b,c;x)
\]
upon the previously written solutions $f_{\Delta(s),\text{ }j_{3}}^{(1)}$ and $f_{\Delta
(s),\text{ }j_{3}}^{(2)}$, to obtain%
\begin{align*}
f_{\Delta(s),\text{ }j_{3}}^{(1)}  &  =\chi_{1}~\left(  x^{-j_{3}}\left(  x\text{
}\left(  x-1\right)  \right)  ^{\frac{j_{3}+s}{2}}{}_2F_{1}(s-j,1+s+j,1+s-j_{3}%
;x)\right)  .\\
f_{\Delta(s),\text{ }j_{3}}^{(2)}  &  =\chi_{2}~\left(  x^{-s}\left(  x\text{
}\left(  x-1\right)  \right)  ^{\frac{j_{3}+s}{2}}{}_2F_{1}(j_{3}-j,1+j_{3}%
+j,1+j_{3}-s;x)\right)  .
\end{align*}
Again, these eigenfunctions are invariant under the change \eqref{j12} and in consequence
\be
j>-\frac{1}{2} \text{ ~~  or ~~} j<-\frac{1}{2}.
\ee

The hypergeometric factors written above, truncate to polynomials ---and in consequence
$f_{\Delta(s),\text{ }j_{3}}^{(1)}$ and $f_{\Delta(s),\text{ }j_{3}}^{(2)}%
$\ square integrable--- provided the following quantization conditions hold
\begin{align*}
\max(|j|,|j+1|) &  \leq-s\leq-j_{3},\\
\max(|j|,|j+1|) &  \leq-j_{3}\leq-s,
\end{align*}
--- and for the time being \eqref{QuantuC} ---, for $f_{\Delta(s),\text{ }j_{3}}^{(1)}$ and $f_{\Delta(s),\text{
}j_{3}}^{(2)}$, respectively. The explicit form of these square integrable
modes can be obtained by repeating the analysis\ done for the case $s>\frac{1}{2}$, and they exist if and only if
\be
s< -\frac{1}{2}.
\ee

\subsection{Generalized conditions}

So far we have been assuming 
\be
j_{3}, ~j, ~s \in \mathbb{Z} \text{ or } \mathbb{Z}+\frac{1}{2}.
\ee
However, the aforementioned GNO conditions -- see subsection \ref{GNO}-- can be relaxed.

 As already stated, to achieve regularity at the contractible cycle
$x=0$, the following necessary condition
\[
j_{3}-s\in \mathbb{Z},
\]
must hold. To have discrete spectrum there are necessary conditions too:
\[
\begin{tabular}
[c]{lll}%
$f^{(2)}\rightarrow-s+j\in \mathbb{Z},~f^{(1)}\rightarrow-j_{3}+j\in \mathbb{Z}$ & ~~ if &
$\text{\ }s>+\frac{1}{2}$, \\ \\
$~f^{(1)}\rightarrow s+j\in \mathbb{Z},~~~f^{(2)}\rightarrow j_{3}+j\in \mathbb{Z}$ & ~~ if &
$~s<-\frac{1}{2}$.
\end{tabular}
\]
Notice that the conditions in the right (resp. left) side, follow from a linear
combination of the condition of regularity at the contractible cycle, and the respective conditions
in the left (resp. right) side. Hence, we can write down the more compact and equivalent statement
\be
j_3-s \in \mathbb{Z} \text{  and  } j-|s| \in \mathbb{Z}.  \label{QCondFinal}
\ee

In the table below, we write down the explicit form of the spectrum. For simplicity of presentation but without lack of generality, let us take
$ j>-\frac{1}{2}$. In that case, the relevant spectrum is%

\[
\frame{%
\begin{tabular}
[c]{|l|l|l|}\hline
$\forall s$ such that & $s>\frac{1}{2}$ & $s<-\frac{1}{2}$\\\hline
$j_{3}$ & $j+1,j+2,...,j+k,...,\infty$ & $-\infty,...,-k-j,...,-2-j,-1-j$%
\\\hline
$j$ & $s-1,s-2,...,s-k,...> -\frac{1}{2}$ & $-s-1,-s-2,...,-s-k,...>-\frac{1}{2}$\\\hline
\end{tabular}
}.
\]
A particular case is when $j,j_{3},s\in \mathbb{Z}+\frac{1}{2}$. In that case, the table
above reduces to%

\[
\frame{%
\begin{tabular}
[c]{|l|l|l|}\hline
$\forall s$ such that & $s >\frac{1}{2}$ & $s<-\frac{1}{2}$\\\hline
$j_{3}$ & $j+1,j+2,...,j+k,...,\infty$ & $-\infty,...,-k-j,...,-2-j,-1-j$%
\\\hline
$j$ & $s-1,s-2,...,s-k,...,\frac{1}{2}.$ & $-s-1,-s-2,...,-s-k,...,\frac{1}%
{2}.$\\\hline
\end{tabular}
}.
\]
The corresponding eigenfunctions can be recovered from the summary that shall
be presented next, and the results in previous sections.

\subsection{Collecting the eigenfunctions}

The maximal functional space of square integrable modes is%

\be
\Xi(s):=%
{\displaystyle\bigoplus\limits_{ -\frac{1}{2}<j<|s|}}
\left(  \Xi_{j}^{(1)}(s)%
{\displaystyle\oplus}
\Xi_{j}^{(2)}(s)\right)  , \label{FuncSpaceTot}
\ee
where the subspace $\Xi_{j}^{(1)}(s)$ is defined as
\[
\Xi_{j}^{(1)}(s):=\left\{f_{\Delta(s),j_{3}}^{(1)}\right\}_{j_{3}}\text{, with }%
\Delta:=j(j+1)-s^{2},
\]
together with conditions \eqref{QCondFinal} and


\be
\begin{tabular}
[c]{lll}%
$\max(|j|,|j+1|)\leq j_{3}\leq s,$ & if & $\text{\ }s>\frac{1}{2}$,\\\\
$\max(|j|,|j+1|)\leq-s\leq-j_{3}$ & if & $s<-\frac{1}{2}$.%
\end{tabular}  \label{QCondFinalS}
\ee

The subspace $\Xi_{j}^{(2)}(s)$ is defined as%
\[
\Xi_{j}^{(2)}(s):=\left\{f_{\Delta(s),j_{3}}^{(2)}\right\}_{j_{3}}~~\text{, with }%
\Delta:=j(j+1)-s^{2},\text{ }%
\]
together with conditions \eqref{QCondFinal} and
\[
\begin{tabular}
[c]{lll}%
$\max(|j|,|j+1|)\leq s\leq j_{3}$ & \ if & $\text{\ }s > \frac{1}{2}$,\\ \\
$\max(|j|,|j+1|)\leq-j_{3}\leq-s$ & \ if & $s <-\frac{1}{2}$.%
\end{tabular}
\]

\bigskip
Of special interest will be the following limiting spaces%
\bigskip
\[
\Xi_{s-1\text{ (or }-s\text{)}}^{(1,2)}(s):=\{f_{\Delta,j_{3}}^{(1,2)}%
\}\text{, with }j:=\text{ }s-1\text{ (or \ }-s\text{) and }\Delta=-s.
\]

\bigskip
These spaces are the ones that contribute to the super-determinant that concerns
us, when $s> \frac{1}{2}$ and cohomological cancellations are performed.

It will be useful to keep in mind that for every $j,$ in the direct sum space
$\Xi_{j}(s):=\Xi_{j}^{(1)}(s)\oplus\Xi_{j}^{(2)}(s)$ the angular number
$j_{3}$ will range at step $1,$ departing from the lower (resp. upper) bound
given below%

\[
\begin{tabular}
[c]{lll}%
$\max(|j|,|j+1|)$ $<j_{3}<\infty$ & \ if & $s> ~ \frac{1}{2}$,\\\\
$-\infty$ $<j_{3}<-\max(|j|,|j+1|)$ & \ if & $s < -\frac{1}{2}$.%
\end{tabular}
\]

\textbf{Comment:} Notice that upon square integrable \textquotedblleft
representations\textquotedblright\ $\Xi_{j}(s),$\ which are labeled by $j$
running at step $1$ down from $|s|-1$ and greater than $-\frac{1}{2}$, namely%
\[
j:-\frac{1}{2}< j\leq|s|-1,
\]
the bosonic operator $O_{B}$ \eqref{OB}, is positive definite if $\left(\rho(u)+k \right)^2>0$.
Indeed, that operator needs to be positive definite in order to have
convergence of the functional integral of the exponential of the quadratic
expansion of the bosonic localizing term.

 \subsection{Normalizable modes from asymptotics}

One can also find the discrete spectrum by looking at  the asymptotic expansion of the general solutions \eqref{f1} (one can repeat the procedure for the other solution). We choose to focus on $ f_{1}$ and for values $s>\frac{1}{2}$,  from regularity  and  smoothness at $x=0$ it follows that $\ j_{3}\leq s$ and the difference $s-j_3\in \mathbb{Z}_+$.

As before, we define $\Delta:=j(j+1)-s^{2}.$

At $x=-\infty$  $f_{1}$  is
\be
f_{1}\sim \chi_{1}^{-}~x^{-1-j}\left(  1+O(\frac{1}{x})\right)  +\chi_{1}^{+}~%
x^{j}\left(  1+O(\frac{1}{x})\right).
\ee




The coefficients above are:

\begin{align*}
\chi_{1}^{-}  &  \sim\frac{\Gamma\lbrack1+s-j_{3}]\Gamma\lbrack-1-2j]}{\Gamma\lbrack-j_{3}-j
]\Gamma\lbrack s-j]},\\
\chi_{1}^{+}  &  \sim\frac{\Gamma\lbrack1+s-j_{3}]\Gamma\lbrack1+2j]}{\Gamma\lbrack1-j_{3}+j
]\Gamma\lbrack s+j]}.
\end{align*}
Suppose $j>-\frac{1}{2}$, to cancel out the $x^{j}$ behavior of $f_{1}$ while preserving the $x^{-1-j}$, we need to have that $\chi_1^{+} $ vanishes, this is achieved when either of the arguments of the $\G$'s in the denominator is 0 or a negative integer. Then
\bea
1-j_{3}+j=-n
\quad \text{or}\quad s+j=-n\quad \text{with}\quad n\in \mathbb{Z}_+ ,
\eea
the second choice is out of order given our assumptions ($s>\frac{1}{2},\,j>-\frac{1}{2}$)
therefore $1-j_{3}+j=-n$ when replacing this value in $\chi_1^{-}$ one has to be careful, since the arguments of $\G$'s in the denominator might be also a negative integer, 
\bea
\chi_{1}^{-}  &  \sim\frac{\Gamma\lbrack1+s-j_{3}]\Gamma\lbrack1+2n-2j_3]}{\Gamma\lbrack-2j_3+n+1
]\Gamma\lbrack s-j]} & \sim\frac{\Gamma\lbrack1+2n-2j_3]}{\Gamma\lbrack-2j_3+n+1
]},
\eea
We have not replaced the value of $j$ in terms of $n$ and $j_3$, these $\G$ functions, given our assumptions will be a number different from zero and finite and will not play a role. Looking at the denominator in the last expression one naively conclude that there are values of $j_3$ and  $n$ for which the argument is negative integer (since $j_3>n+1$), and therefore $\chi_1^-$ is also  zero, but it is not the case since, for each of this values, the argument of $\G$ in the numerator is also a negative integer, and the divergences cancel.  One can recast the ratio above as:
\bea
\frac{\Gamma\lbrack1+2n-2j_3]}{\Gamma\lbrack-2j_3+n+1]}=\frac{\G[-m+n]}{\G[-m]}  \quad\text{where}\quad m>n, 
\eea  
using :
\bea
\Gamma [\epsilon -m]=\frac{(-1)^{m-1} \Gamma [-\epsilon ] \Gamma [\epsilon +1]}{\Gamma [m-\epsilon +1]} \quad \text{where}\quad \e \quad\text{is very small.}
\eea
Applying this relation on both numerator an denominator and taking $\e\rightarrow0$
\bea
\frac{\G[\e-m+n]}{\G[\e-m]}\rightarrow(-1)^n \frac{\G[m+1]}{\G[m-n+1]},
\eea  
therefore  for $j>-\frac{1}{2}$ we  conclude:
\begin{align*}
\chi_{1}^{-}  &  =O(1),\\
\chi_{1}^{+}  &  =0,
\end{align*}
We can then proceed analogously for $j<-\frac{1}{2}$ to get:
\begin{align*}
\chi_{1}^{-}  &  =0,\\
\chi_{1}^{+}  &  =O(1),
\end{align*}
 We then have:
\begin{align*}
\text{for}\quad s>\frac{1}{2}\quad & \left \{ 
\begin{tabular}[c]{lll}%
$j<j_{3}\leq s$ & \ if & $j> -\frac{1}{2}$\\
$-s\leq-j_{3}\leq j$ & \ if & $j<-\frac{1}{2}$%
\end{tabular}
.\right. 
\end{align*}
 
 \subsection{The relation between spin-$1$ discrete modes and ours  } \label{AppA2}

Let $\nabla_{\mu}$ be the covariant derivative of diffeomorphisms. The
Laplace-Beltrami operator is defined as $\nabla^{\mu}\nabla_{\mu}$ and acting
upon a covariant vector field $\overrightarrow{X}$ of components $(X_{\theta
},X_{\varphi})$ has the explicit form%
\[
\nabla^{\mu}\nabla_{\mu}\overrightarrow{X}:=\left(
\begin{array}
[c]{cc}%
\square_{s=0}+\coth^{2}(\theta) & 2\frac{\coth(\theta)}{\sinh^{2}(\theta
)}\partial_{\varphi}\\
-2\coth(\theta)\partial_{\varphi} & \square_{s=0}-2\coth(\theta)\partial_{\theta
}+1
\end{array}
\right)  \left(
\begin{array}
[c]{c}%
X_{\theta}\\
X_{\varphi}%
\end{array}
\right),
\]
where $\square_{s=0}$ is the scalar Laplacian. We have added the subscript
$s=0$ to remind that it can be obtained from the magnetic Laplacian previously
defined, by particularising to $s=0$. The eigenvector
\[
\overrightarrow{X}_{0}=\nabla\Phi,\text{ \ with \ }\Phi:=\left(  \frac
{\sinh(\theta)}{1+\cosh(\theta)}\right)  ^{|j_{3}|}e^{i\text{ }j_{3}\varphi
},\text{ \ \ }j_{3}=\pm1,\pm2,...\text{\ .}%
\]
One can check that indeed%
\[
\square_{s=0}\Phi=0,
\]
and second that%
\[
\nabla^{\mu}\nabla_{\mu}\overrightarrow{X}_{0}=\overrightarrow{X}_{0}.
\]
In words, $\overrightarrow{X}_{0}$ is an eigen-tensor of rank one of the
Laplace-Beltrami operator $\nabla^{\mu}\nabla_{\mu}$ with eigenvalue 1.

More important to our purpose, we have checked that
\[
\left(  -\partial_{\theta}^{2}-\coth^{2}\theta\partial_{\theta}+\frac{1}%
{\sinh^{2}\theta}(|j_{3}|-\cosh\theta)^{2}\right)  X_{0\theta}=X_{0\theta}.
\]
Notice that the operator in the LHS equation above coincides with our
$\square_{s=1}$ \text{if and only if }%
\[
j_{3}>0.
\]
In fact, the equation above for $X_{0\theta}$ implies that $X_{0\theta}$ obeys
our defining equation%
\[
\left(  \square_{s}+\Delta\right)  X_{0\theta}=0,
\]
\text{if and only if}%
\[
0=j\ (or-1=j)<s=1\leq j_{3}\in \mathbb{Z}.
\]
It is then consequence that%
\bea
\text{\textbf{ }}\left\{  X_{0\theta}\right\}  _{j_{3}\in \mathbb{N}}&=&\left\{
\frac{\left(  \cosh\theta-1\right)  ^{j_{3}}}{\sinh^{j_{3}+1}\theta}e^{i\text{
}j_{3}\varphi}=\frac{1}{\sinh{\theta}}\left(  \tanh{\frac{\theta}{2}}\right)
^{j_{3}}e^{i\text{ }j_{3}\varphi}\right\}  _{j_{3}\in \mathbb{N}} ,\nonumber\\&=&\Xi_{j=0}^{(2)}(s=1).
\eea
The remaining $\theta$-components of the vector discrete modes, $\left\{
X_{0\theta}\right\}  _{j_{3}\in-\mathbb{N}}$, do solve our defining equation $\left(
\square_{s}+\Delta\right)  X_{0\theta}=0$ \ \text{if and only if}%

\[
j_{3}\leq-1=s<j=0\in Z.
\]
We have thence proven that $\left\{  X_{0\theta}\right\}  _{j_{3}\neq0\in Z}$
are included in our set of square integrable modes%
\bea
\left\{  X_{0\theta}\right\}  _{j_{3}\in-\mathbb{N}}&=&\left\{  \frac{\left(  \cosh
\theta-1\right)  ^{-j_{3}}}{\sinh^{-j_{3}+1}\theta}e^{i\text{ }j_{3}\varphi
}=\frac{1}{\sinh{\theta}}\left(  \tanh{\frac{\theta}{2}}\right)  ^{-j_{3}%
}e^{i\text{ }j_{3}\varphi}\right\}  _{j_{3}\in-\mathbb{N}}, \nonumber\\&=&\Xi_{j=0}^{(1)}(s=-1).
\eea
and correspond to the two possible unit flux (spin one) ``helicities'' $s=\pm1.$
\section{On 1 loop determinants}
\subsection{Alternative regularization} \label{reg}
In this appendix we report a second approach to regularize the determinant of $O_B$ in the subspace $\X_{j=s-1}(s)$ -- We present the case $s> \frac{1}{2}$, the $s < -\frac{1}{2}$ is analogous --:
\bea
\det_{\X_{j=s-1}(s)}O_B&=&\prod_{k\in\mathbb{Z}}\prod_{j_3=s}^{\infty}(\r(u)+k)^2\nonumber,\\
&=&\prod_{k\in\mathbb{Z}}\left((\r(u)+k)^{2}\right)^{\sum_{j_3=s}1}=\prod_{k\in\mathbb{Z}}\left((\r(u)+k)^{2}\right)^{\sum_{j_3=1}^{\infty}1-\sum_{j_3=1}^{s-1}1}\nonumber,\\
&=&\prod_{k\in\mathbb{Z}}\left((\r(u)+k)^{2}\right)^{\z(0)-(s-1)}=\prod_{k\in\mathbb{Z}}\left|(\r(u)+k)\right|^{-2s+1}.
\eea
Where we use the basic definition of Riemann $\z$ function $\z(t)=\sum_{n=1}^{\infty}\frac{1}{n^t}$ and the value $\z(0)=-\frac{1}{2}$.

\subsection{Vector multiplet}
\label{VecBZ}
In this appendix, we prove that the index of a vector multiplet coincides with the index of matter multiplet with R-charge $q_R=2$.

The quadratic actions coming out of the localizing terms \eqref{QVvec2BZ} and \eqref{QVvecFBZ}, along the complex path \eqref{GammaV} and after imposing gauge fixing condition \eqref{axial}, are
\begin{eqnarray}
\mathcal{L}^{B}_{quadratic}&:=&(i\,\delta \tilde{D} )^2+(\mathcal{D}_t \delta A_1)^2+ (\mathcal{D}_t \delta A_2)^2, \\
\mathcal{L}^{F}_{quadratic}&:=&i \,\delta \bar{\lambda}^\dagger_2\, \overleftarrow{\hat{\mathcal{D}}}_t \, \delta \lambda_2,
\end{eqnarray}
where
\begin{eqnarray}
i\,\delta \tilde{D} &:=&i\,\delta D+\delta F_{1 2}+ \delta \hat{\mathcal{D}}_3 \sigma .
\end{eqnarray}
The $\delta \tilde{D}$ integrates trivially. The functional spaces to integrate over the vector and ghost degrees of freedom are
\begin{eqnarray}
(\delta A_{1}, ~\delta A_{2}, ~ \delta \sigma, ~\delta \bar{c}, ~ \delta c) \rightarrow (\Xi_{(s-1)},~ \Xi_{(s-1)}, ~\Xi_{(s)},~ \Xi_{(s)},~ \Xi_{(s)}). \label{b37}
\end{eqnarray}
The integration of $\delta A_1$ and $\delta A_2$ is
\begin{eqnarray}
\prod_{k} \prod^{|s-1|-1}_{j=0} \prod^{\infty}_{j_3=j+1} |(\rho(u)+k )| \times  \prod_{k}\prod^{|s-1|-1}_{j=0}  \prod^{\infty}_{j_3=j+1} |(\rho(u)+k )|. \label{ApAmBZ}
 \end{eqnarray}
 In obtaining \label{ApAmBZ} we have used $\sqrt{(\rho(u)+k )^2}=|(\rho(u)+k )|$. In our contour of integration $(\rho(u)+k )$ is real.
 The functional space to integrate the gaugini degres of freedom is
 \begin{eqnarray}
(
\delta \lambda_2,~ \delta\bar{\lambda}_2^\dagger )
\rightarrow (
 \Xi_{(s-1)}, ~ \Xi_{(s-1)}).  \label{b43}
\end{eqnarray}

The integration of $\delta \lambda_2$ and $\delta \bar{\lambda}^\dagger_2$, multiplied by the integration of $\delta \bar{c}^\dagger$ and $\delta c$, following from the $BRST$ action \eqref{BRSTAc}, gives
\begin{eqnarray}
\prod_{n} \prod^{|s-1|-1}_{j=0} \prod^{\infty}_{j_3=j+1} |(\rho(u)+k )|  \times  \prod_{n}\prod^{|s|-1}_{j=0}  \prod^{\infty}_{j_3=j+1} |(\rho(u)+k )| . \label{llcc}
 \end{eqnarray}
As already mentioned, we will not integrate over the zero modes $\delta \lambda_1$ and $\delta \bar{\lambda}_1$ in order not to obtain vanishing results. 

The super-determinant to compute is
\be
\frac{\eqref{llcc}}{\eqref{ApAmBZ}}. \label{SuDet}
\ee
The result \eqref{SuDet} is a divergent quantity. To regularize these objects we use the zeta-regularization procedure but only after co-homological cancellations are performed.

  If $s>\frac{1}{2}$ the only contribution to the quotient \eqref{SuDet}, comes from the integration of the ghosts degrees of freedom  $(\bar{c},c)$ with quantum number $j=s-1$. The divergent contribution from this functional space is
 \be
\prod^{s-1}_{j=s-1} \prod^{\infty}_{j_3=j+1} |(\rho(u)+k )|. \label{ccDet}
 \ee
  Those degrees of freedom  live in $\Xi_{j=s-1}(s)$ and are coupled to $s$ units of flux. As a formal object \eqref{ccDet} is equal to
\be
\sqrt{\underset{\Xi_{j=s-1}(s)}{\det}(O_{B})},
\ee   
as can be straightforwardly checked from taking the product between equations \eqref{EqRefe1} and \eqref{EqRefe2} and particularizing the result to $j=s-1$.   

   We have already computed the zeta regularized determinant of the operator $O_B$ on $\Xi_{j=s-1}(s)$. In this space and for a given $S^1$ KK mode $k$, $O_B$ has a unique eigenvalue: $(\rho(u)+k )^2$ and the square root of its zeta-regularized determinant is
\be
\sqrt{\underset{\Xi_{j=s-1}(s)}{\det}(O_{B})}=|(\rho(u)+k)|^{-s+\frac{1}{2}}. \label{EqRef3}
\ee
The value of the parameter $s$ is
\be
s=\frac{-\rho(\mathfrak{m})}{2},
\ee
because the ghosts have $q_R=0$.

From \eqref{EqRef3}, after taking the products over roots and KK modes and after regularization we obtain
\bea
Z_{1-loop}^{vector}(\mathbb{H}_2,\mathfrak{m})&=&{\displaystyle\prod\limits_{\rho^*}}
\left[  |C_{reg}\sin\left(  \frac{\rho(u)}{2}\right)| \right]  ^{\frac{\rho(\mathfrak{m})+1}{2}}, \\
&=&{\displaystyle\prod\limits_{\rho^*}}\left[
 |C_{reg}\sin\left(  \frac{\rho(u)}{2}\right)|\right]  ^{\frac{-\rho(\mathfrak{m})+1}{2}},
 \label{1loopV}
\eea
where
\be
C_{reg}=-2 i. \label{def}
\ee
Equality \eqref{1loopV} proves the statement made below equation \eqref{Statement}, that is, independence of the GNO conditions. In the second equality in \eqref{1loopV}, we have
performed the inversion of roots $\rho \rightarrow -\rho$.

 For $s < \frac{1}{2}$ the same result \eqref{1loopV} is obtained by following analog steps.

\section{Conventions: 4d $ \mathcal{N}=2$ gauged supergravity}
In this appendix we summarize our conventions for 4d $\mathcal{N}=2$ gauged supergravity. The construction of black holes  reported in section \ref{hypbhsec}, was implemented in a Mathematica file. If the reader is interested in the file, please write an email to us. If there is interest, we are more than happy to share it.

The 4d gamma matrices
\bea
\gamma^1=\left(
\begin{array}{cccc}
 i & 0 & 0 & 0 \\
 0 & -i & 0 & 0 \\
 0 & 0 & i & 0 \\
 0 & 0 & 0 & -i
\end{array}
\right), ~ \gamma^2=\left(
\begin{array}{cccc}
 0 & 0 & 0 & i \\
 0 & 0 & -i & 0 \\
 0 & -i & 0 & 0 \\
 i & 0 & 0 & 0
\end{array}
\right), ~ \gamma^3=\left(
\begin{array}{cccc}
 0 & -i & 0 & 0 \\
 -i & 0 & 0 & 0 \\
 0 & 0 & 0 & -i \\
 0 & 0 & -i & 0
\end{array}
\right), \nonumber\\
\gamma^4=\left(
\begin{array}{cccc}
 0 & 0 & 0 & -i \\
 0 & 0 & i & 0 \\
 0 & -i & 0 & 0 \\
 i & 0 & 0 & 0
\end{array}
\right).~~~
\eea

\be
\gamma_{a b}=\frac{1}{2}[\gamma_a,\gamma_b], ~~ \gamma_5=i \gamma^4\gamma^1\gamma^2\gamma^3.
\ee

The $SU(2)_R$ $R$-symmetry invariant tensors
\be
\epsilon_{AB}=\epsilon^{A B}=\left(
\begin{array}{cc}
 0 & 1 \\
 -1 & 0
\end{array}
\right).
\ee

The $SU(2)_R$  generators
\bea
\sigma^1_{AB}&=&\left(
\begin{array}{cc}
 1 & 0 \\
 0 & -1
\end{array}
\right), ~ \sigma^2_{AB}=\left(
\begin{array}{cc}
 -i & 0 \\
 0 & -i
\end{array}
\right), ~ \sigma^3_{AB}=\left(
\begin{array}{cc}
 0 & -1 \\
 -1 & 0
\end{array}
\right), \\
\sigma^{1 AB}&=&\left(
\begin{array}{cc}
 -1 & 0 \\
 0 & 1
\end{array}
\right), ~ \sigma^{2 AB}=\left(
\begin{array}{cc}
 -i & 0 \\
 0 & -i
\end{array}
\right), ~ \sigma^{3 AB}=\left(
\begin{array}{cc}
 0 & 1 \\
 1 & 0
\end{array}
\right).
\eea

The $\sigma^{I ~ B}_{A}$ with $I=1,2,3$ are the Pauli matrices.

The ordering of coordinates is
\be
\left(1,2,3,4\right) ~ \leftrightarrow ~ \left(r,~\theta,~\varphi,~ t\right).
\ee

For hyperbolic solutions:
\be
F^\Lambda_{\mu \nu}=\left(
\begin{array}{cccc}
 0 & 0 & 0 & 0 \\
 0 & 0 & \frac{1}{2} \sinh (\theta ) p_{\Lambda } & 0 \\
 0 & -\frac{1}{2} \sinh (\theta ) p_{\Lambda } & 0 & 0 \\
 0 & 0 & 0 & 0
\end{array}
\right),
\ee
\be
{{F}_{\mu,\nu}^{- \Lambda }}=\left(
\begin{array}{cccc}
 0 & 0 & 0 & \frac{i p_{\Lambda }}{4 h(r)^2} \\
 0 & 0 & \frac{1}{4} \sinh (\theta ) p_{\Lambda } & 0 \\
 0 & -\frac{1}{4} \sinh (\theta ) p_{\Lambda } & 0 & 0 \\
 -\frac{i p_{\Lambda }}{4 h(r)^2} & 0 & 0 & 0
\end{array}
\right).
\ee

\subsection{Parametrization in terms of scalar}

In this subsection we post  a series of useful parametrizations in terms of physical scalars $z^1,~z^2,~z^3$. The results reported in this subsection, are consistent with the $BPS$ equations obtained for the choice $+$ in equation \eqref{KSEq}.
\be
\mathcal{K}=-\log \left(\frac{8 \sqrt{z^1 z^2 z^3}}{\left(z^1+z^2+z^3+3\right){}^2}\right),
\ee
\be
\mathcal{F}=\left(\bar{\mathcal{F}}\right)^*=-2 i \sqrt{\frac{z^1 z^2 z^3}{\left(z^1+z^2+z^3+3\right){}^4}},
\ee

\be
\mathcal{F}_\Lambda=\left(\frac{i \sqrt{z^1 z^2 z^3}}{z^1+z^2+z^3+3},\frac{i \sqrt{\frac{z^2
   z^3}{z^1}}}{z^1+z^2+z^3+3},\frac{i \sqrt{\frac{z^1
   z^3}{z^2}}}{z^1+z^2+z^3+3},\frac{i \sqrt{\frac{z^1
   z^2}{z^3}}}{z^1+z^2+z^3+3}\right),
\ee

\be
\mathcal{F}_{\Lambda \Sigma}=\left(\bar{\mathcal{F}}_{\Lambda \Sigma}\right)^*=\left(
\begin{array}{cccc}
 \frac{1}{2} i \sqrt{z^1 z^2 z^3} & -\frac{1}{2} i \sqrt{\frac{z^2 z^3}{z^1}} &
   -\frac{1}{2} i \sqrt{\frac{z^1 z^3}{z^2}} & -\frac{1}{2} i \sqrt{\frac{z^1
   z^2}{z^3}} \\
 -\frac{1}{2} i \sqrt{\frac{z^2 z^3}{z^1}} & \frac{1}{2} i \sqrt{\frac{z^2 z^3}{{z^1}^3}}
   & -\frac{i z^3}{2 \sqrt{z^1 z^2 z^3}} & -\frac{i z^2}{2 \sqrt{z^1 z^2 z^3}} \\
 -\frac{1}{2} i \sqrt{\frac{z^1 z^3}{z^2}} & -\frac{i z^3}{2 \sqrt{z^1 z^2 z^3}} &
   \frac{1}{2} i \sqrt{\frac{z^1 z^3}{{z^2}^3}} & -\frac{i z^1}{2 \sqrt{z^1 z^2 z^3}} \\
 -\frac{1}{2} i \sqrt{\frac{z^1 z^2}{z^3}} & -\frac{i z^2}{2 \sqrt{z^1 z^2 z^3}} &
   -\frac{i z^1}{2 \sqrt{z^1 z^2 z^3}} & \frac{1}{2} i \sqrt{\frac{z^1 z^2}{{z^3}^3}}
\end{array}
\right),
\ee

\be
\bar{f}^\Lambda_j=\left(
\begin{array}{cccc}
 \frac{1}{8 \sqrt{2} \sqrt[4]{{{z^{1}}^{ 5}} z^2 z^3}} & -\frac{3}{8 \sqrt{2} \sqrt[4]{z^1 z^2
   z^3}} & \frac{z^2}{8 \sqrt{2} \sqrt[4]{{{z^{1}}^{ 5}} z^2 z^3}} & \frac{z^3}{8 \sqrt{2}
   \sqrt[4]{{{{z^{1}}^{5}}} z^2 z^3}} \\
 \frac{1}{8 \sqrt{2} \sqrt[4]{z^1 {z^{2}}^{5} z^3}} & \frac{z^1}{8 \sqrt{2} \sqrt[4]{z^1
   {z^{2}}^{5} z^3}} & -\frac{3}{8 \sqrt{2} \sqrt[4]{z^1 z^2 z^3}} & \frac{z^3}{8 \sqrt{2}
   \sqrt[4]{z^1 {z^{2}}^{5} z^3}} \\
 \frac{1}{8 \sqrt{2} \sqrt[4]{z^1 z^2 {z^{3}}^{5}}} & \frac{z^1}{8 \sqrt{2} \sqrt[4]{z^1 z^2
   {z^3}^5}} & \frac{z^2}{8 \sqrt{2} \sqrt[4]{z^1 z^2 {z^{3}}^5}} & -\frac{3}{8 \sqrt{2}
   \sqrt[4]{z^1 z^2 z^3}}
\end{array}
\right),
\ee

\be
\mathcal{N}_{\Lambda \Sigma}= i \left(
\begin{array}{cccc}
 -\sqrt{z^1 z^2 z^3} & 0 & 0 & 0 \\
 0 & -\sqrt{\frac{z^2 z^3}{{z^1}^3}} & 0 & 0 \\
 0 & 0 & -\sqrt{\frac{z^1 z^3}{{z^2}^3}} & 0 \\
 0 & 0 & 0 & -\sqrt{\frac{z^1 z^2}{{z^3}^3}}
\end{array}
\right),
\ee

\be
g_{z \bar{z}}=\left(
\begin{array}{ccc}
 \frac{3}{16}\frac{1}{{z^1}^2} & -\frac{1}{16} \frac{1}{ z^1 z^2} & -\frac{1}{16 }\frac{1}{z^1 z^3} \\
 -\frac{1}{16} \frac{1}{ z^1 z^2} & \frac{3}{16}\frac{1}{{z^2}^2} & -\frac{1}{16}\frac{1}{ z^2 z^3} \\
 -\frac{1}{16}\frac{1}{ z^1 z^3} & -\frac{1}{16} \frac{1}{ z^2 z^3} & \frac{3}{16}\frac{1}{ {z^3}^2}
\end{array}
\right).
\ee

\bibliographystyle{JHEP}
\bibliography{Localization}

\end{document}